\documentclass[fleqn,usenatbib]{mnras}
\usepackage{comment}
\usepackage{gensymb}
\usepackage{refcount}
\usepackage{url}
\usepackage{newtxtext,newtxmath}

\usepackage[T1]{fontenc}
\usepackage{ae,aecompl}
\usepackage{xcolor}
\usepackage{threeparttable}

\usepackage{graphicx}	
\usepackage{amsmath}	
\usepackage{amssymb}	
\makeatletter
\newcommand\footnoteref[1]{\protected@xdef\@thefnmark{\ref{#1}}\@footnotemark}
\makeatother

\title{MeerKAT-16 H\,{\sc i} observation of the dIrr galaxy WLM.}
\author[Ianjamasimanana R. et al.]{
Roger Ianjamasimanana$^{1,2}$\thanks{E-mail: ianjamasimanana26@gmail.com; ianja@starscientist.org},
Brenda Namumba$^{1}$,
Athanaseus J. T. Ramaila$^{2}$,
\newauthor
Anna S. Saburova$^{3, 4}$,
Gyula I. G. J{\'o}zsa$^{2,1,5}$,
Talon Myburgh$^{1,2}$,
Kshitij Thorat$^{6}$,
\newauthor
Claude Carignan$^{7}$,
Eric Maina$^{1}$,
W. J. G. de Blok$^{8,9,10}$,
Lexy A. L. Andati$^{1}$,\newauthor
Benjamin V. Hugo$^{1,2}$,
Dane Kleiner$^{11}$,
Peter Kamphuis$^{12}$,
Paolo Serra$^{11}$, \newauthor
Oleg M. Smirnov$^{1,2}$,
Filippo M. Maccagni$^{11}$,
Sphesihle Makhathini$^{1}$,\newauthor
D\'{a}niel Cs. Moln\'{a}r$^{11}$,
Simon Perkins$^{2}$,
Mpati Ramatsoku$^{1,11}$,
Sarah V. White$^{1}$
\\
$^{1}$Department of Physics and Electronics, Rhodes University, PO Box 94, Makhanda, 6140, South Africa\\
$^{2}$South African Radio Astronomy Observatory, 2 Fir Street, Black River Park, Observatory, Cape Town, 7925, South Africa\\
$^{3}$Sternberg Astronomical Institute, Moscow M.V. Lomonosov State University, Universitetskij pr., 13, Moscow, 119234, Russia\\
$^{4}$Institute of Astronomy, Russian Academy of Sciences, Pyatnitskaya st., 48, 119017 Moscow, Russia\\
$^{5}$Argelander-Institut f\"ur Astronomie, Auf dem H\"ugel 71, D-53121 Bonn, Germany\\
$^{6}$Department of Physics, University of Pretoria, Hatfield, Pretoria, 0028, South Africa\\
$^{7}$Department of Astronomy, University of Cape Town, Private Bag X3,Rondebosch 7701, South Africa\\
$^{8}$Netherlands Institute for Radio Astronomy (ASTRON), Postbus 2, 7990 AA Dwingeloo, the Netherlands\\
$^{9}$Department of Astronomy, University of Cape Town, Private Bag X3, Rondebosch 7701, South Africa\\
$^{10}$Kapteyn Astronomical Institute, University of Groningen, Postbus 800, 9700 AV Groningen, The Netherlands\\
$^{11}$INAF - Osservatorio Astronomico di Cagliari, Via della Scienza 5, I-09047 Selargius (CA), Italy\\
$^{12}$Ruhr-Universit\"at Bochum, Faculty of Physics and Astronomy, Astronomical Institute, 44780 Bochum, Germany\\
}

\date{Accepted 2020 July 01. Received 2020 July 01; in original form 2020 June 15.}

\pubyear{2020}

\begin{document}
\label{firstpage}
\pagerange{\pageref{firstpage}--\pageref{lastpage}}
\maketitle

\begin{abstract}

We present observations and models of the kinematics and the distribution of the
neutral hydrogen (H\,{\sc i}) in the isolated dwarf irregular galaxy, Wolf-Lundmark-Melotte (WLM).
We observed WLM with the Green Bank Telescope (GBT) and as part of
the MeerKAT Early Science Programme, where
16 dishes were available. The H\,{\sc i} disc of WLM extends out to a major axis diameter of
30\arcmin~(8.5 kpc), and a minor axis diameter of 20\arcmin~(5.6 kpc) as measured by the GBT.
We use the MeerKAT data to model WLM using the TiRiFiC software suite,
allowing us to fit different tilted-ring models and select the one that best
matches the observation. Our final best-fitting model is a flat disc with a vertical thickness,
a constant inclination and dispersion, and a radially-varying surface brightness with
harmonic distortions. To simulate
bar-like motions, we include second-order harmonic distortions in velocity in the tangential and the vertical directions.
We present a model with only circular motions included
and a model with non-circular motions. The latter describes the data better.
Overall, the models reproduce the global distribution and the kinematics of the gas, except for some faint emission
at the $2\sigma$ level. We model the mass distribution of WLM with a pseudo-isothermal (ISO) and a Navarro-Frenk-White (NFW) dark matter halo models.
The NFW and the ISO models fit the derived rotation curves within the formal errors,
but with the ISO model giving better reduced chi-square values. The mass distribution in WLM is
dominated by dark matter at all radii.
\end{abstract}




\section{Introduction}
\indent
The circular rotation speed of stars or gas in galaxies around their centre of mass is mainly set by the balance between
the galaxy's centrifugal force and its gravitational force \citep{1973A&A....26..483R}. Therefore, the shape of the rotation
curve of a galaxy can be used to model its mass distribution.
When the circular velocities of the ionized gas in the spiral galaxy M31 were
measured (at optical wavelength) as a function of the radial distance from the galaxy's centre in the 1970s by \citet{1970ApJ...159..379R},
it was realised that the measured rotation curve did not follow what was expected from the distribution of
the visible matter in the galaxy. In fact, instead of following a Keplerian decline, the rotation curve rose sharply
in the inner disc and remained flat at larger distance. A flat rotation curve was later observed in a number of spiral
galaxies at even larger distance from the centre than previously achieved using the hyperfine 21 cm line emission of
neutral hydrogen (H\,{\sc i}) gas
\citep{1973A&A....26..483R, 1978ApJ...226..770P, 1981AJ.....86.1825B, 1985ApJ...295..305V, 2001ARA&A..39..137S, 2006ApJ...641L.109C, 2008AJ....136.2648D}.
This result strengthened the idea that the observed matter in galaxies
cannot account for the total gravitational force that holds matter in galaxies.
Therefore, there must be invisible matter in the halo of spiral galaxies that contributes to the total orbital
speed of the gas and the stars. This unseen mass is known as dark matter, whose very first observational evidence came
from the study of the Coma cluster in the 1930s by Fritz Zwicky. Understanding the role and the properties of dark matter
is at the heart of modern cosmology. The decomposition of the galaxy's rotation curve in terms of the contribution
from visible matter (stars and gas) and dark matter has become a powerful tool to model the mass distribution in both spiral and dwarf
galaxies \citep{1985ApJ...294..494C, 2006ApJ...641L.109C, 2008AJ....136.2648D, 2011AJ....141..193O, 2012AAS...21910603O, 2014MNRAS.439.2132R, 2019MNRAS.490.3365N}.
Dwarf galaxies are dominated by dark matter even inside their optical discs \citep{1989ApJ...347..760C, 2011AJ....141..193O}.
Therefore, their inferred central dark matter density distributions are less affected by the
uncertainties due to the assumed contribution of baryonic matter to the total dynamical mass and the mass-to-light ratio of the stellar
discs \citep{2017MNRAS.466.4159I}. However, their kinematics are usually affected by non-circular motions and therefore,
analysing their mass distributions requires careful kinematical modelling \citep{2008AJ....136.2761O}.\\
\indent Several models are proposed in the literature to explain the distribution of
dark matter in galaxies. The most popular ones are the Navarro-Frenk-White model
\citep[NFW,][]{1996ApJ...462..563N} and the pseudo-isothermal model \citep[ISO,][]{1987PhDT.......199B}.
The NFW model was derived from N-body simulations of collisionless cold
dark matter (CDM) haloes and is characterised by a steep power law mass-density distribution (known as the \textit{cusp} model).
The ISO model is motivated by observations and is characterised by a shallow inner mass density profile (the \textit{core} model).
Despite being dark-matter dominated, the rotation curves of dwarf and low surface brightness galaxies (LSB) have been found
to be inconsistent with the cosmologically motivated NFW cuspy profile \citep[but see][]{2019MNRAS.482..821O, 2020MNRAS.491.4993K}.
Instead, they are best fitted by the ISO model. This is known as the core-cusp problem; an extensive review on this can
be found in \citet{2010AdAst2010E...5D}. In this paper, we use both the NFW and the ISO models to model the dark matter distribution of the dwarf irregular
galaxy, Wolf-Lundmark-Melotte (WLM, DDO 221, UGCA 444), using
H\,{\sc i} as mass and kinematics tracers. We map the H\,{\sc i} gas in WLM using the MeerKAT radio telescope. This paper presents
the first MeerKAT data in its 32k mode. To estimate the completeness of the emission recovered by MeerKAT,
we compare the flux recovered by MeerKAT with the flux recovered
by observations using the 100 m Robert C. Byrd Green Bank Telescope (GBT).
We model the distribution and kinematics of the H\,{\sc i} in
WLM using sophisticated 3D modelling techniques implemented in the software
package {\tt {TiRiFiC}}\footnote{\url{http://gigjozsa.github.io/tirific/}}\citep{2007AA...468..731J}.

The shapes of the rotation curves of galaxies can be influenced by several factors; inherent in the data itself
\citep[beam smearing, pointing offsets, non-circular motions, see][]{2008AJ....136.2761O} and also in the method
applied (1D vs 2D vs 3D method). One of the widely used approaches to derive a rotation curve consists of extracting a
velocity that best represents the circular motion of the gas along each line of sight. By doing so,
one gets a map of the representative velocities along all lines of sight (i.e., a velocity field).
The extracted velocity field can then be fitted with a tilted ring model to obtain the rotation curve.
This approach is known as the 2D velocity field method and is implemented in many software packages,
such as in the GIPSY task ROTCUR \citep{1987PhDT.......199B, 2007ApJ...664..204S}, DiskFiT \citep{2015arXiv150907120S},
and most recently the software 2DBAT \citep{2018MNRAS.473.3256O}.
As mentioned in \citet{2008AJ....136.2761O}, the 2D velocity-field approach requires the projection of
a 3D data cube to an infinitely thin disc. While this makes the computation faster,
it has some limitations. These include beam smearing, the presence of non-circular motions, the inability
to model the vertical thickness of the disc and to simulate the case where the sight-line cross the disc multiple times,
the most extreme case of this being edge-on galaxies \citep{2008AJ....136.2761O}. Beam smearing tends to lower the true rotation velocity,
whereas the assumption of a thin disc may overestimate the rotation curve at the outer radius
\citep{1978PhDT.......195B, 1997AstL...23..522B, 2004MNRAS.352..787K, 2009A&A...493..871S}. In addition,
the presence of non-circular motions, which manifests itself as kinks in the velocity fields, makes the derivation of
rotation curves uncertain; especially for dwarf galaxies where the kinematics can be severely affected by non-circular
motions \citep{2008AJ....136.2761O, 2008AJ....136.2720T, 1978PhDT.......195B}. For this analysis, we opt for the 3D approach due to,
its great flexibility to make a complex model, it is less affected by beam smearing effects,
and its ability to work even in the presence of strong non-circular motions. The 3D method consists of directly
fitting a tilted-ring model to the data cube \citep{2007AA...468..731J, 2015MNRAS.452.3139K, 2015MNRAS.451.3021D}
in order to extract kinematic information such as the rotation velocity, the velocity dispersion and the
surface brightness profile. We use the derived rotation curve from our tilted-ring model to
model the luminous and the dark matter distribution in the galaxy. This analysis will contribute to
our understanding of the dark matter distribution in dwarf galaxies.

We organise our paper as follows. In Section~\ref{sec:sample-selection}, we describe the properties of WLM.
In Section~\ref{sec:observations}, we present the observation set-up. In Section~\ref{sec:data-processing}, we
present the data reduction procedures. In Section~\ref{sec:optical}, we
contrast our observations with archival optical data. In Section~\ref{sec:3dmodel},
we describe our kinematical modelling methods and present the kinematic parameters.
In Section~\ref{sec:massmodels}, we present the mass modelling of the galaxy. In Section~\ref{sec:discussion}, we discuss
the results, and in Section~\ref{sec:conclusions}, we give a summary.
\section{Properties of WLM}\label{sec:sample-selection}
WLM is a dwarf irregular (dIrr) galaxy in the Local Group. It was first discovered
by \citet{1909AN....183..187W}, and later identified as a nebula by Lundmark (unpublished) and
\citet{1926MNRAS..86..636M} when analysing Franklin-Adams Charts Plates, thus the name
Wolf-Lundmark Melotte or, in short, WLM. A tip of the red giant branch (TRGB) analysis
placed it at a distance of 0.97 Mpc \citep{2013AJ....145..101K}, close enough
to have allowed detailed studies of its stellar population and star formation history.
Using \textit{Hubble Space Telescope (HST)} data, \citet{2000ApJ...531..804D}
reported that WLM formed half of its stellar population earlier than
9 Gyr ago. This was followed by a gradual decrease in star formation
until 1-2.5 Gyr ago after which a rise in activity was observed.
WLM star formation continues until the present epoch but is confined in
what appears to be a bar in the centre of the galaxy \citep{2000ApJ...531..804D}.
Despite having a metallicity 50\% lower than the reported
carbon monoxide (CO) detection threshold, \citet{2015Natur.525..218R}
found 10 CO clouds with an average radius of 2 pc and an average virial mass of
10$^3$ M$_{\odot}$ in WLM using the Atacama Large Millimetre Array (ALMA).
\citet{2015Natur.525..218R} also noted that WLM formed stars efficiently despite its
low CO content. The basic properties of WLM are
presented in Table~\ref{tab:gal_prop}.
\begin{table*}
    \centering

    \begin{tabular}{llrlc} 
       \multicolumn{5}{c}{WLM}\\
       \hline
       \hline
	Parameters & Symbols & Values & Units & Ref.\\
       \hline
       	Morphological Type & & dIrr & & 1\\
	Optical centre & RA &00$\rm{^{h}}$01$\rm{^{m}}$ 58.1$\rm{^{s}}$& & 2\\
	(J2000)& Dec &-15$\degree$ 27$\arcmin$ 40$\arcsec$& & 2\\
	Systemic Velocity & $v_{\rm sys}$ & $-122\,\pm\,4$& $\mathrm{km}\,\mathrm{s}^{-1}$ & 3\\
 	Distance & $d$ & $0.97$ & $\mathrm{Mpc}$ & 2 \\
	B-band magnitude & $M_\mathrm{B}$ & $-13.56$ & mag & 4\\
	Stellar mass & $M_{\star}$ &  $4.3\times10^{7}$& $M_\odot$ &  4 \\
        Star formation rate & $SFR(\mathrm{FUV})$ & $0.008 $ & $M_\odot\,\mathrm{yr}^{-1}$ & 5 \\
	Optical diameter & $D_{25}$ &$1.98$&kpc& 5\\
	Optical position angle & $PA_\mathrm{opt}$ & $4$ & \degree  & 4\\
	\ion{H}{i} diameter & $D_\ion{H}{i}$ & $15.\!\!\arcmin 8\,\pm\, 1.\!\!\arcmin 2$ & &3\\
	\ion{H}{i} diameter & $D_\ion{H}{i}$ & $4.3\,\pm\,0.4$ & $\mathrm{kpc}$ & 3\\
	&$D_\ion{H}{i}$/D$_{25}$&2.17& &3\\
	Total \ion{H}{i} Mass& $M_\ion{H}{i}$ &$5.5\, \pm\, 0.6\cdot 10^7$& $M_\odot$ &  3\\
	Rotation velocity & $v_\mathrm{rot}(r_\ion{H}{i})$ &$29\,\pm\,5$& $\mathrm{km}\,\mathrm{s}^{-1}$ & 3\\
	Dyn. Mass & $M_\mathrm{dyn}(r_\ion{H}{i})$ & $4.0\,\pm\,1.4\cdot 10^8$&$M_\odot$ & 3\\
	\ion{H}{i} line width & $W_{50}$ & $54\,\pm\, 11 $ &
	$\mathrm{km}\,\mathrm{s}^{-1}$ & 3\\
    & $W^\mathrm{c}_{50}$ & $55\,\pm\, 10 $ & $\mathrm{km}\,\mathrm{s}^{-1}$ & 3\\
    & $W_{20}$ & $85\,\pm\, 14 $ & $\mathrm{km}\,\mathrm{s}^{-1}$ & 3\\
    & $W^\mathrm{c}_{20}$ & $84\,\pm\, 13 $ & $\mathrm{km}\,\mathrm{s}^{-1}$ & 3\\
    \hline
    \end{tabular}
     \caption{Fundamental properties of WLM galaxy. (1) \citet{2012AJ....144....4M}, (2) \citet{2013AJ....145..101K},
     (3) This work, (4) \citet{2014MNRAS.445..881C} (5) \citet{2018AJ....156..109M}.}
    \label{tab:gal_prop}
 \end{table*}

WLM has previously been observed in H\,{\sc i}.
Using the GBT, \citet{2011AJ....142..173H} mapped its H\,{\sc i} disc and measured
an H\,{\sc i} diameter of~28\arcmin~at $\rm{10^{19}~cm^{-2}}$. This is about
2.4 times the optically defined Holmberg diameter and is in good agreement with the previous VLA H\,{\sc i}
observation by \citet{2007AJ....133.2242K}. \citet{2007AJ....133.2242K} found that the H\,{\sc i} distribution of WLM is not
uniform, but has a hook-like structure of high column density in the centre, encompassing roughly 20\%
of the total H\,{\sc i} mass. The origin of this feature is not clear but is most likely due to active
star formation in the centre which
blows-out or ionizes the gas. According to \cite{2007AJ....133.2242K} the \ion{H}{i} velocity
field of WLM is asymmetric, with the northern part appearing to be warped.

WLM appears to have evolved in isolation, free from significant external
encounters. This is because WLM lies far away from its nearest neighbour, the Cetus dwarf spheroidal galaxy.
They are separated from each other by a distance of ~250 kpc \citep{2011EAS....48...59L}.
Note that the size of WLM is approximately 8~kpc as traced by H\,{\sc i} emission \citep{2007AJ....133.2242K, 2011AJ....142..173H}.
A search for possible H\,{\sc i} companions around WLM by \citet{2004MNRAS.351..333B} was unsuccessful.
\citet{2012ApJ...750...33L} classified WLM as among the five \textit{least tidally disturbed and one of the most isolated galaxies
within 1 Mpc of the Milky Way}. Thus, WLM is an interesting target for at least three main reasons.
First, it is a gas-rich dwarf galaxy with a very extended HI disc, allowing to trace its gravitational potential out to very large radius.
Second, being a gas-rich dwarf irregular galaxy, it falls within a mass range where cold gas accretion is still expected to occur \citep{2014AA...561A..28S}.
Lastly, its isolated nature enables the description of the radial and the vertical velocity structure of the disc without being hampered by the
presence of debris or irregular kinematics.

\section{Observations}\label{sec:observations}
\subsection{MeerKAT}
The MeerKAT \ion{H}{i} observations of WLM were conducted with the L-band (900-1670 MHz) receivers of
the MeerKAT-16 telescope \citep{2016mks..confE...1J},
composed of 16, 13.5~metre dishes, with a maximum baseline of 8~km as part of the MeerKAT Early Science observations.
The target was observed for about 5.65 hours, including overheads. A total bandwidth of 856 MHz centred at 1283.98 MHz was
recorded and was divided into 32k channels with a channel width of 26.12 kHz each and two linear polarisations.
The target field was observed for a total time of 3.15 hours in multiple scans of 13.5 minutes at 16 seconds integration time.
Moreover, 2.5 hours were spent on three calibrator sources: J1938-6341 and J0408-6544 for bandpass calibration, and J0003-1727
for gain (amplitude and phase) calibration.

To achieve the required sensitivity, the 16 MeerKAT antennas were arranged such that 12 antennas formed part of a
compact configuration, and the remaining 4 antennas were at a distance of more than 1000 m from the core,
resulting in an extended beam of up to 60\arcsec. The few chosen extended baselines allowed a better estimate of the bandpass and phase solutions.
The quasi-random distribution of antennas allowed us to attain imaging of suitable quality, despite the proximity of WLM to the equator.
In addition, the MeerKAT primary beam FWHM at 1.4 GHz is significantly larger than the H\,{\sc i} diameter of WLM,
allowing us to map its extended emission without mosaicing.

\begin{table}
	\centering
	\caption{MeerKAT HI observations of WLM galaxy.}
	\label{tab:obsinfo}
	\begin{tabular}{ll} 
		\hline
		\hline
		Property & Value\\
		\hline
		Number of antennas & 16\\
		Total observation time (hr) & 5.65\\
		Frequency range (MHz) & 900 - 1670\\
		Central Frequency (MHz) & 1285\\
		Bandpass calibrator 1 & J1938-6341 [19:39:25 -63:42:45]\\
		Bandpass calibrator 2 & J0408-6544 [04:08:20 -65:44:09]\\
		Gain calibrator & J0003-1727 [00:03:22 -17:27:14]\\
		Number of channels & 4098\\
		Channel width (kHz) & 26.123\\
		Central frequency (MHz) & 1420.9893\\

		\hline
	\end{tabular}
\end{table}
\subsection{GBT}
We observed WLM with the GBT for 10 hours, including overheads,
from April to May 2016. We combined this observation
with a previous 4.2-hour older observation, giving a total observing time of 14.2 hours.
We used the L-band (1.15-1.73 GHz) spectral line mode of
the VErsatile GBT Astronomical Spectrometer (VEGAS) backend to map 4
square degrees around WLM in a \textit{basket-weave} fashion. The data was taken in
frequency-switching mode, using the edges of the maps as OFF position to improve the
sensitivity. We recorded four integrations from each edge of the maps, giving a total of
8 OFF integrations. The calibrator 3C48 was observed at the beginning and end of each run.
The total bandwidth was 100 MHz, with a spectral resolution
of 3.1 kHz, and centred at a rest frequency of 1420 MHz.
The noise level was $\sim$0.025 K ($\sim$11 $\mathrm{mJy~beam^{-1}}$). We convert the
brightness temperature $T$ in K to $\mathrm{Jy~beam^{-1}}$ using the following equation:
\begin{equation}
    T = 1.36\dfrac{\lambda^{2}}{\theta^{2}}S,
\end{equation}
where $\lambda$ is the wavelength in cm (21.1), $\theta$ is the beam in arcsec (522.18)
and $S$ is the flux in $\mathrm{Jy~beam^{-1}}$:
\begin{equation}
    S~(\mathrm{Jy~beam^{-1}}) = 0.455\times T~(K).
\end{equation}
Therefore, the noise is $\sim$11 $\mathrm{mJy~beam^{-1}}$.
\section{Data Processing}\label{sec:data-processing}
\subsection{MeerKAT}
We reduced the MeerKAT data using the Containerised Automated Radio Astronomy
Calibration pipeline\footnote{\url{https://caracal.readthedocs.io}}~\citep[CARACal, formerly
known as MeerKATHI,][]{jozsacal}, being developed in an international collaboration, with main contributions coming from Rhodes University,
the South African Radio Astronomy Observatory (SARAO), and the
Italian National Institute for Astrophysics (INAF). CARACal has initially been developed to reduce MeerKAT data,
but its scope is intentionally to be able to reduce
data from any radio interferometers. CARACal performs different data reduction steps (flagging, calibration, imaging)
using standard radio astronomy
data reduction tools such as CASA \citep{2007NRAON.113...17B}, AOFlagger \citep{2010MNRAS.405..155O}, WSClean
\citep{2014MNRAS.444..606O}, CubiCal \citep{2018MNRAS.478.2399K} etc. in containerised environments provided by a
Python-based scripting framework called {\tt{Stimela}}\footnote{\url{https://github.com/ratt-ru/Stimela}}.
The advantages of using CARACal include reproducibility of results, highly customisable, and the ability
to use a wide variety of data analysis tools in a seamless manner.
\subsubsection{Flagging and calibration}
We perform the data reduction within CARACal using a chunk of 50 MHz around the WLM central H\,{\sc i} emission frequency
with a width of 26.12 KHz, resulting in a data set of 1915 channels.
Prior to cross-calibration of the data, the known RFI-prone channels are flagged using a baseline-independent
flagging mask obtained from SARAO. To remove emission from the Milky Way, the channels corresponding
to the Milky Way's emission frequencies are flagged using the CASA task {\tt{flagdata}}. This is the default
setting of CARACal and is appropriate for a galaxy far away from the Galactic plane like WLM.
After flagging, cross-calibration is performed. This includes delay calibration, as well as bandpass and gain calibrations.
The CASA task {\tt{SETJY}} is used to determine the flux density of the flux calibrator J1938-6341.
The CASA task {\tt{BANDPASS}} is used to correct for instrumental delays. We estimate a maximum delay correction of 1.5 ns,
which is standard for MeerKAT. To estimate the time-dependent antenna-based gain and to solve for the phase, the CASA tasks {\tt{gaincal}} and {\tt{bandpass}}
are used, respectively.

After cross-calibration, the data gets processed through a set of four self-calibration steps.
In each step, the data go through a WSClean imaging and calibration
using CubiCal. The clean components produced by WSClean are used to make a sky model for the calibration.
To avoid the selection of artefacts in the mask, we start with shallow cleaning and then iteratively go deeper by lowering the {\tt{auto-mask}}
and the {\tt{auto-threshold}} parameters in CARACal until all sources are detected and the residuals are noise-like. To compute image statistics
such as the dynamic range and moments of distribution of residuals in the successive self-cal steps, we enabled the image-based quality assessment
tool {\tt{aimfast}}. These quantities are used as stopping
criteria for the self-calibration processes. Our self-calibration strategy improves the dynamic range of the images and allows
a better model of the continuum sources.
\subsubsection{Continuum subtraction and imaging}
To examine the astrometry and the flux density fidelity of the continuum images of the WLM field, we compare the positions and
flux densities of the sources in the field with the equivalent quantities in the NRAO Very Large Array (VLA)
\footnote{\label{note1}The VLA and GBT are facilities of the National Radio Astronomy Observatory. NRAO is a facility of
the National Science Foundation operated under cooperative agreement by Associated Universities, Inc.}
Sky Survey \citep[NVSS,][]{1998AJ....115.1693C} for the same sources.

\begin{figure}
 \includegraphics[scale=0.4]{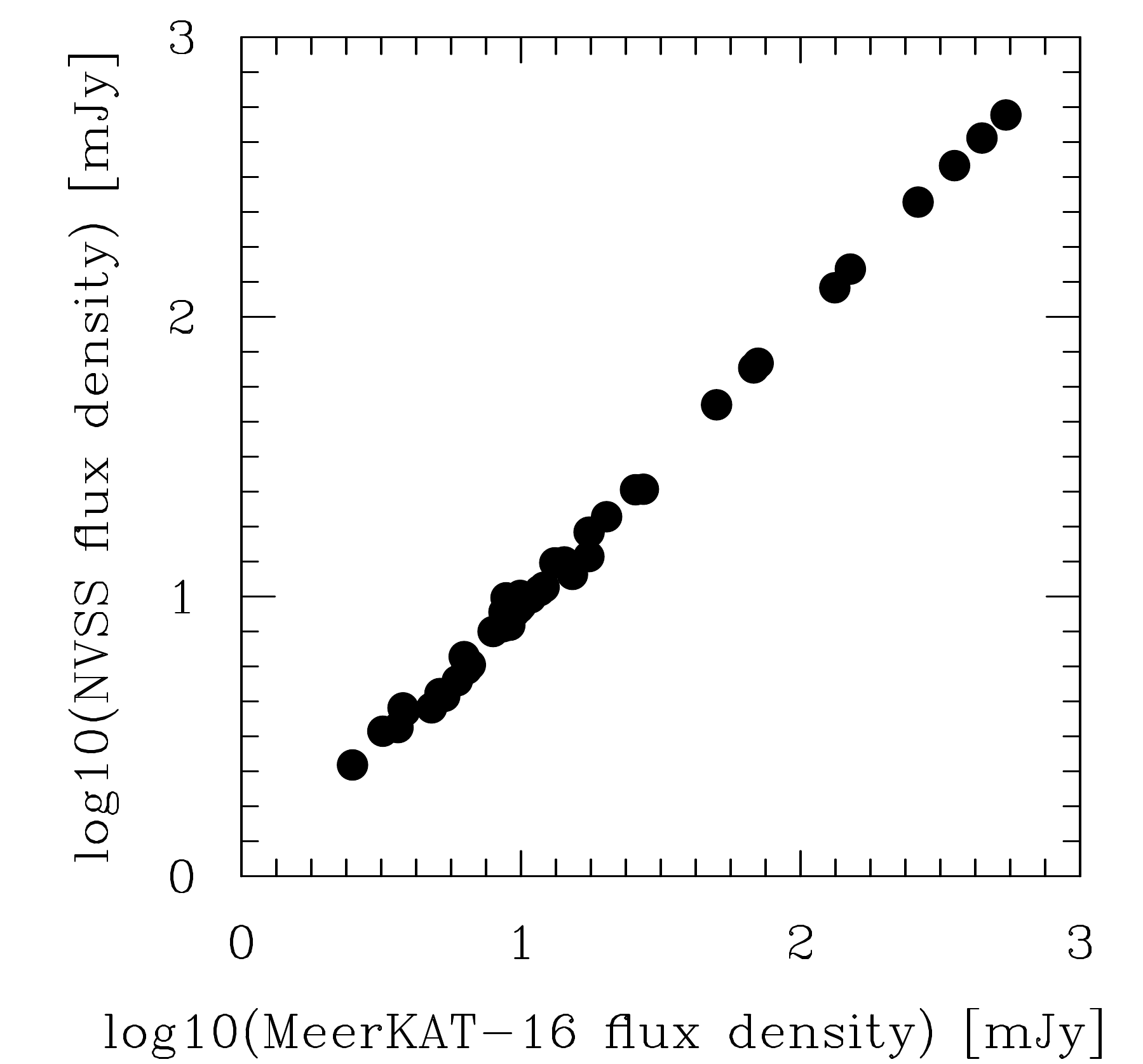}
 \caption{Comparison of the flux densities of matched sources in the MeerKAT-16 WLM field and the NVSS catalogue.}
 \label{fig:fluxcompare}
\end{figure}

\begin{figure}
 \includegraphics[scale=0.52]{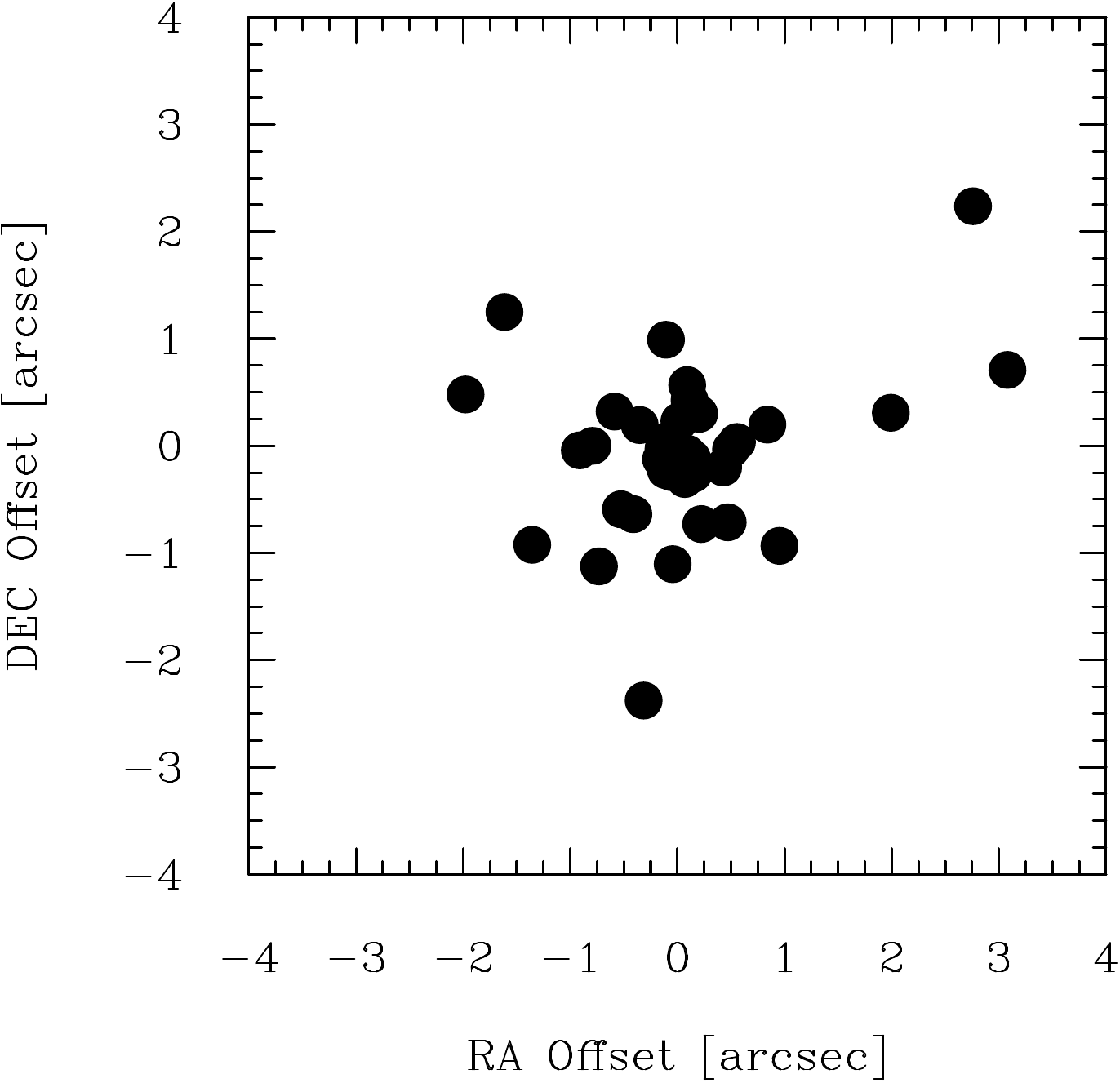}
 \caption{Positional offsets in RA and DEC between sources in our MeerKAT-16
 WLM field and the corresponding matched sources in the NVSS catalogue.}
 \label{fig:astrometry}
\end{figure}

We use the PyBDSF source finder \citep{2015ascl.soft02007M} on our continuum images (with the default settings, which we find satisfactory)
to construct a source catalogue for the WLM field. From this catalogue, only the sources with the code "S" are chosen, from which we found 64 sources.
These sources are single-Gaussian sources that are the only sources in their respective "islands" \citep{2015ascl.soft02007M}.
We compare the position and the integrated flux densities of these sources with those in the catalogue of the NVSS.
The results are presented in Fig.~\ref{fig:fluxcompare} and Fig.~\ref{fig:astrometry}. There is an excellent agreement between the flux
densities recovered by our MeerKAT-16 observation and those recovered by the VLA NVSS survey.
The mean positional offsets in RA and DEC are 0.05 and -0.07 arcsec, with standard deviations of 0.9 and 0.7 arcsec, respectively.
We find no systematic trends in RA and DEC and conclude that the WLM field source positions well match those of the NVSS catalogue.

The continuum subtraction was done in two steps.
First, the continuum model was subtracted from the corrected data.
Then, the CASA task {\tt{uvconstsub}} was used to remove any remaining faint extended sources by
fitting a polynomial of order 3 to the line-free channels. The cleaning was done with WSClean using the Cotton-Schwab
major iteration mode, a clean mask obtained from SoFiA~\citep{2014ascl.soft12001S}, and an {\tt{-auto-threshold}} parameter of 0.75.
In addition, a Briggs weighting with a robust parameter of 0 and a Gaussian taper of 8$\arcsec$ were used,
resulting in a final resolution of $35.12\arcsec~\times~11.7\arcsec$ (163 $\times$ 56 pc). The rms noise is 2.5~$\mathrm{mJy~beam^{-1}}$.
The expected theoretical noise is 1.8~$\mathrm{mJy~beam^{-1}}$. This has been calculated using the following radiometer formula:
\begin{equation}
    rms~=~nfac~\dfrac{SEFD}{e_{sys}~\sqrt{n_{pol}~N_{ant}~(N_{ant}-1)~t_{int}~\Delta_{\nu}}}
\end{equation}
where $nfac$ is a factor to multiply the expected rms noise from natural weighting; here we use a value 1.5 for a robust weighting of 0.
$SEFD$ is the system equivalent flux density ($SEFD$~=~443~Jy), $e_{sys}$ is the system efficiency ($e_{sys}$ = 1), $n_{pol}$
is the number of polarisation ($n_{pol}=2$), $N_{ant}$ is the number of antenna ($N_{ant}$ = 16), $t_{int}$ is the
total on-source integration time ($t_{int} = 3.15~h$), and $\Delta_{\nu}$ is the channel width (26 kHz).
To also image and model the faint extended emission in our galaxy,
we use the MIRIAD\footnote{\url{https://www.atnf.csiro.au/computing/software/miriad/}} task {\tt{CONVOL}}
to convolve the data cube with a Gaussian FWHM of 60\arcsec. The convolved cube has an rms noise per channel of 5~$\mathrm{mJy~beam^{-1}}$,
which corresponds to a column density limit of about $\mathrm{5\times10^{19}~cm^{-2}}$ (3$\sigma$
detections over 2 channels of $\mathrm{5.5~km~s^{-1}}$ width). We use the convolved cube for our tilted-ring fitting analysis.
\subsection{GBT}
The standard data reduction was done with {\tt{AIPS}}\footnote{\url{http://www.aips.nrao.edu}}~and
the GBTIDL\footnote{\url{http://gbtidl.nrao.edu/}}~routine \citep{2006ASPC..351..512M}.
The GBTIDL data reduction steps for frequency-switched spectral line data
are described in detail in \citet{2018ApJ...865...36P},
and here we follow the same procedure as they adopted. We smoothed the data from their native
resolution of 3.1 kHz to a channel width of 24.4 kHz (5.2 $\mathrm{km~s^{-1}}$) using a boxcar function.
The rms noise in the final cube was 32 $\mathrm{mJy~beam^{-1}}$. This corresponds to a column density
of 4 $\times \mathrm{10^{18}~cm^{-2}}$ ($3\sigma$ detections over 2 channels of 5.2 $\mathrm{km~s^{-1}}$ width, and a beam size of $\sim$ 9$\arcmin$).
\section{H\,{\sc i} and optical data}\label{sec:optical}
\subsection{Total flux and global H\,{\sc i} profile}
\begin{figure}
 \includegraphics[scale=0.55]{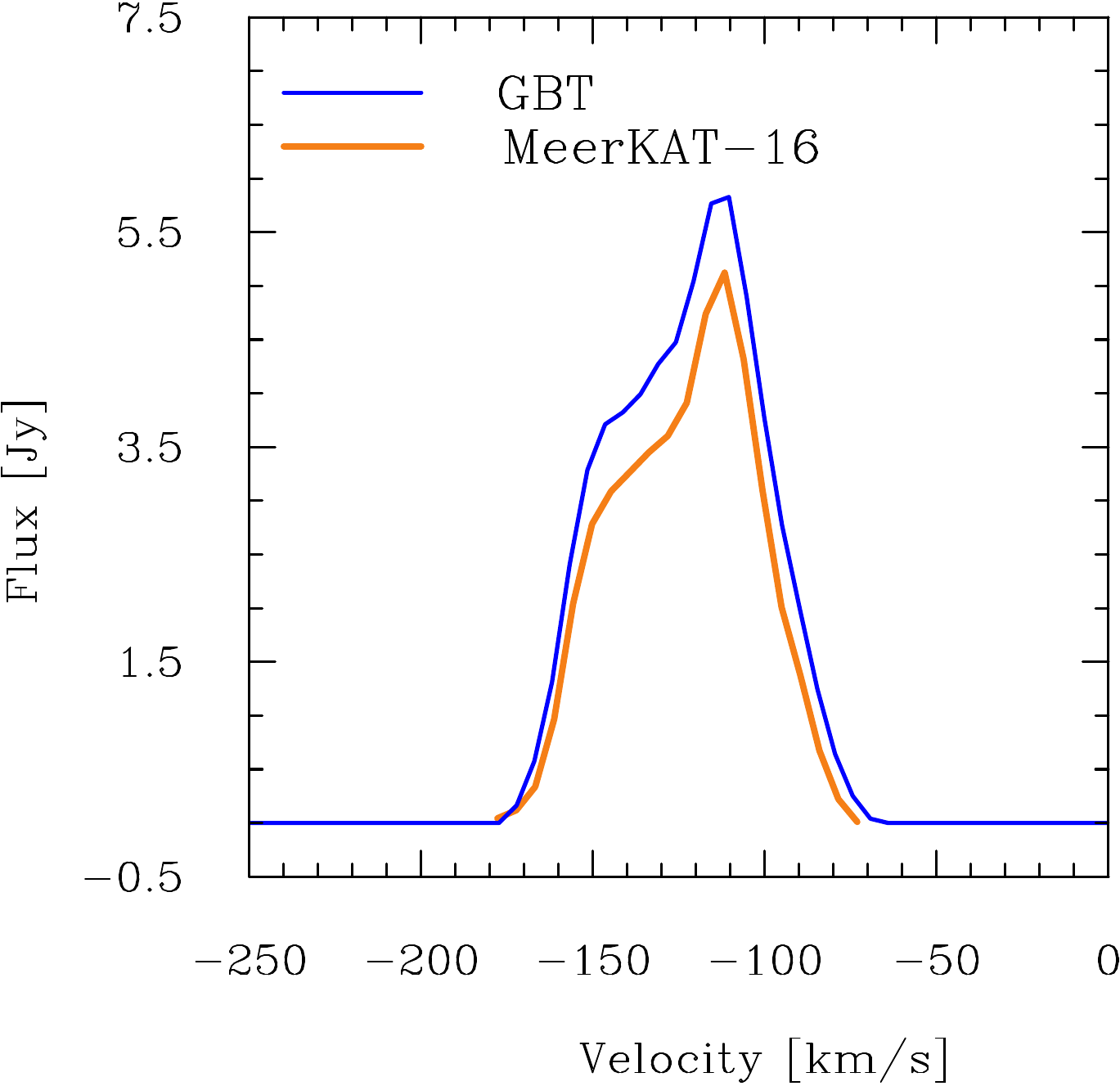}
 \caption{Global H\,{\sc i} velocity profiles of WLM from the GBT data (blue) and the MeerKAT data (red).}
 \label{fig:globalprof}
\end{figure}

We present the global H\,{\sc i} emission profile of WLM in
Fig.~\ref{fig:globalprof}, derived from our MeerKAT cube convolved with a 60\arcsec~Gaussian beam.
Note that the spectrum has been corrected for primary beam attenuation. As reported in previous
observations \citep{2004AJ....128.1219J, 2007AJ....133.2242K}, the
total profile of WLM shows a strong asymmetry, which is confirmed by our observations.
The physical origin of this is unclear
given the isolated nature of WLM. Using the conventional
zeroth moment of the spectrum, we recover a total
flux of 249 Jy~km~s$^{-1}$. We convert this to H\,{\sc i} mass using
the following equation: \begin{equation}
    \dfrac{M_{\rm{HI}}}{M_{\odot}} = 2.36\times10^{5}\Bigg(\dfrac{\rm{S_{HI}}}{\rm{Jy~km~s^{-1}}}\Bigg)\Bigg(\dfrac{D}{\rm{Mpc}}\Bigg)^2,
\end{equation}
where $\rm{S_{HI}}$ is the integrated
flux in Jy~km~s$^{-1}$ and D is the distance in Mpc.
Note that this equation does not take into account the optical depth effects. Adopting a distance of 0.97 Mpc \citep{2013AJ....145..101K}, we find a total H\,{\sc i} mass of
$5.5\times10^{7}~M_{\odot}$. To check if we recover most of the H\,{\sc i} flux in WLM, we compare our MeerKAT-16 observation with the single dish data from the GBT. From the GBT observations, we derived a total flux of about 310 Jy~km~s$^{-1}$, resulting in a
total H\,{\sc i} mass of $6.9\times10^{7}~M_{\odot}$.
Thus, there is a $\sim$ 20$\%$ difference between the flux recovered by the GBT and the one recovered by the MeerKAT-16 observation.
The H\,{\sc i} mass of WLM has also been
reported in the literature. \citet{2004AJ....128...16K}
used the Parkes 64m telescope and reported an H\,{\sc i} mass of $4.9\times10^{7}~M_{\odot}$.
Using the Australia Telescope Compact Array (ATCA), \citet{2004AJ....128.1219J}
obtained an H\,{\sc i} mass of $3.3\times10^{7}~M_{\odot}$. The VLA observations of \citet{2007AJ....133.2242K} recovered
an H\,{\sc i} mass of $6.4\times10^{7}~M_{\odot}$, whereas the Parkes Multibeam measurements of \citet{2004MNRAS.351..333B} reported a value of $6.7\times10^{7}~M_{\odot}$.
These last two values roughly agree with our GBT measurement. Note that these values have been corrected to the adopted distance of 0.97 Mpc used in this paper.

\subsection{Moment maps and optical data}
\textbf{Moment zero}: to map the total intensity of the H\,{\sc i} emission along each lines-of-sight, we calculate the zero-th moment of
the line profiles using:
\begin{equation}
    I_\mathrm{H\,{\textsc i}} (\mathrm{Jy~beam^{-1}~km~s^{-1}})= \displaystyle\sum_{i=1}^{n} I_{i} (x, y)~\Delta_{v},
\end{equation}
where $n$ is the number of channels, $I_{i}$ represents the emission at a specific channel $i$, and $\Delta_{v}$ is the channel separation.
Iterating over all non-blanked pixels (x, y) gives the total intensity map. The integrated intensity is proportional to the column density,
$N_{\mathrm{H\,{\textsc i}}}~\mathrm{(cm^{-2})}$, given by:
\begin{equation}
    N_{\mathrm{H\,{\textsc i}}} (cm^{-2})= \dfrac{1.823 10^{18} \times 6.07 10^{5}~I_{\mathrm{H\,{\textsc i}}}} {(B_{min} \arcsec \times B_{maj} \arcsec)},
\end{equation}
where $I_{\mathrm{H\,{\textsc i}}}$ is the integrated intensity (moment zero map), $B_{min}$ and $B_{max}$ are the minor and major axes of the beam, respectively.
We further multiply $N_{\mathrm{H\,{\textsc i}}}$ to the cosine of the galaxy's inclination to project it to face-on value. Here we use an inclination of 77 degree as motivated
by our tilted-ring modelling described in the kinematic-modelling section. The column density $N_{\mathrm{H\,{\textsc i}}}$ in $\mathrm{cm^{-2}}$
can be converted to surface density in $M_{\odot}~\mathrm{pc^{-2}}$ using:
\begin{equation}
\Sigma_{\mathrm{H\,{\textsc i}}} (M_{\odot}~\mathrm{pc^{-2}}) = 8.01325~10^{-21} N_{\mathrm{H\,{\textsc i}}} (\mathrm{cm^{-2})}
\end{equation}
\textbf{Moment one}: we calculate the moment one of the H\,{\sc i} spectrum to map the velocity field of the galaxy as follows:
\begin{equation}
 \langle v \rangle = \dfrac{\displaystyle\sum_{i=1}^{n} I_{i}(x, y)~v_{i}}{\displaystyle\sum_{i=1}^{n} I_{i}(x, y)}
\end{equation}
\textbf{Moment two}: we use the intensity-weighted mean of the deviation around the mean velocity (moment two) as a rough estimate of
the disordered motions in the galaxy. Pixels with moment two values lower than the width of one channel are blanked in the final moment two map.
\begin{equation}
    \sigma = \sqrt{\dfrac{\displaystyle\sum_{i=1}^{n} I_{i} (x, y)~(v_{i}-\langle v \rangle )^2}{\displaystyle\sum_{i=1}^{n} I_{i} (x, y)}}
\end{equation}
\subsubsection{GBT}
\begin{figure}
 \includegraphics[width=\columnwidth]{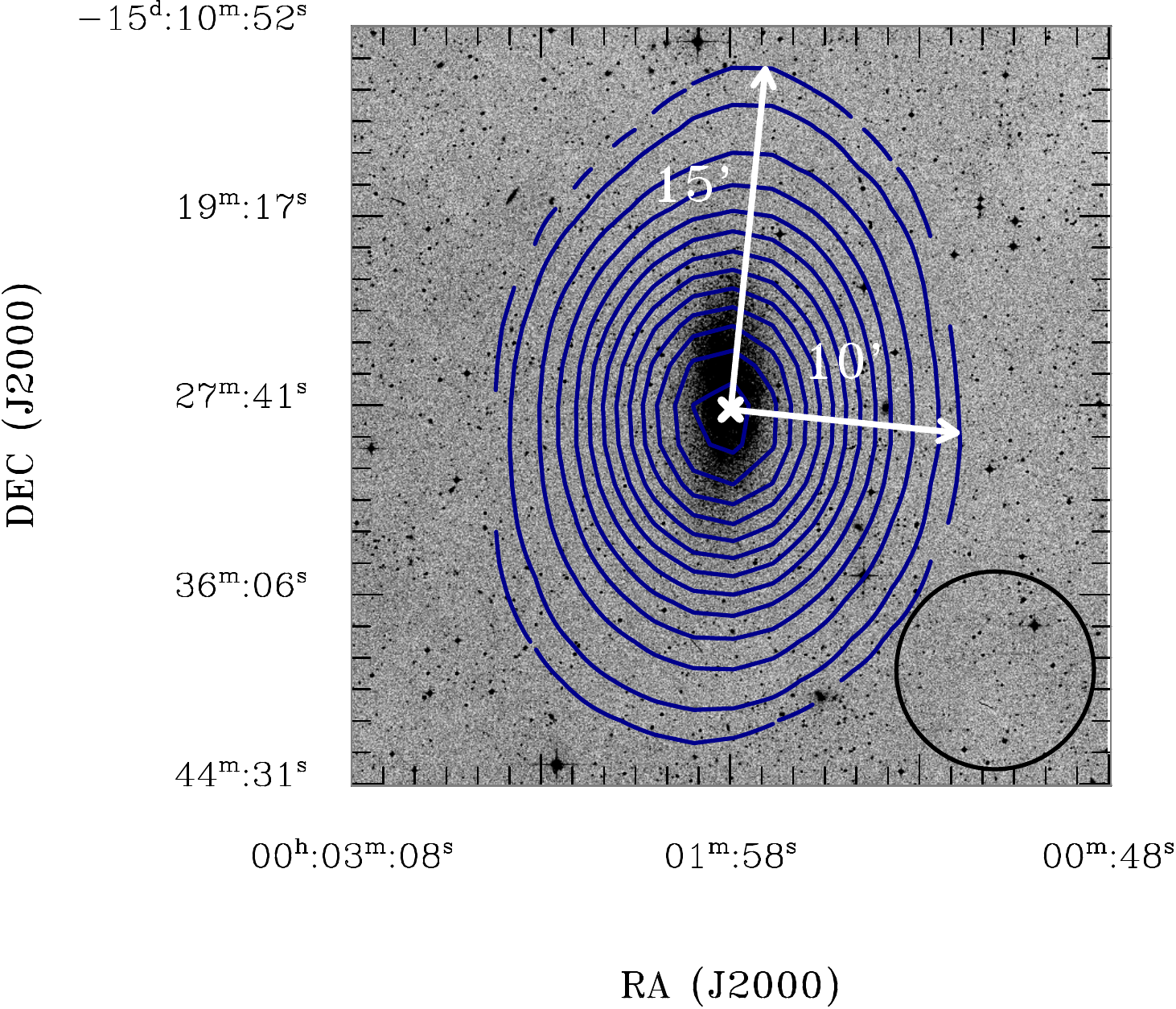}
 \caption{GBT column density map of WLM, overplotted on an archival DSS (red, IIIa-F) optical image.
 Contour levels are (0.62, 1.25, 2.50, 3.74, 4.99, 6.24, 7.49, 8.73, 9.98, 11.23, 12.48 13.73, 14.98,
 16.22, 17.47, 18.72) $\times~10^{19}~\mathrm{cm^{-2}}$. The white cross represents the kinematical centre of the model derived from
 this study.
 The beam of the GBT observation is shown as the black circle.}
\label{fig:gbt_coldens}
\end{figure}

For the GBT, the moment zero map is of particular interest to estimate the extent of the H\,{\textsc i} disc. We use SoFiA
to obtain the moment zero map, and from which we derive a column density map. Pixels below the 3$\sigma$ rms noise level are blanked.
In Fig.~\ref{fig:gbt_coldens}, we overplot the column density map of WLM onto an archival optical image
from the Digitised Sky Survey (DSS\footnote{\url{https://archive.eso.org/dss/dss}}). As shown by the white arrows in the Figure, the H\,{\textsc i}
extends out to a major axis diameter of $\sim$ 30$\arcmin$ ($\sim$8.5 kpc), and a minor axis diameter of $\sim$ 20$\arcmin$ (5.6 kpc).
These values have been derived at a column density level of $6.2~\times10^{18}~\mathrm{cm^{-2}}$. Our measured H\,{\textsc i} diameter
is slightly larger than the extent derived by \citet{2011AJ....142..173H}, who also used the GBT. However it is
in agreement with the value obtained by \citet{2007AJ....133.2242K}.
\subsubsection{MeerKAT}
\begin{figure}
 \includegraphics[width=\columnwidth]{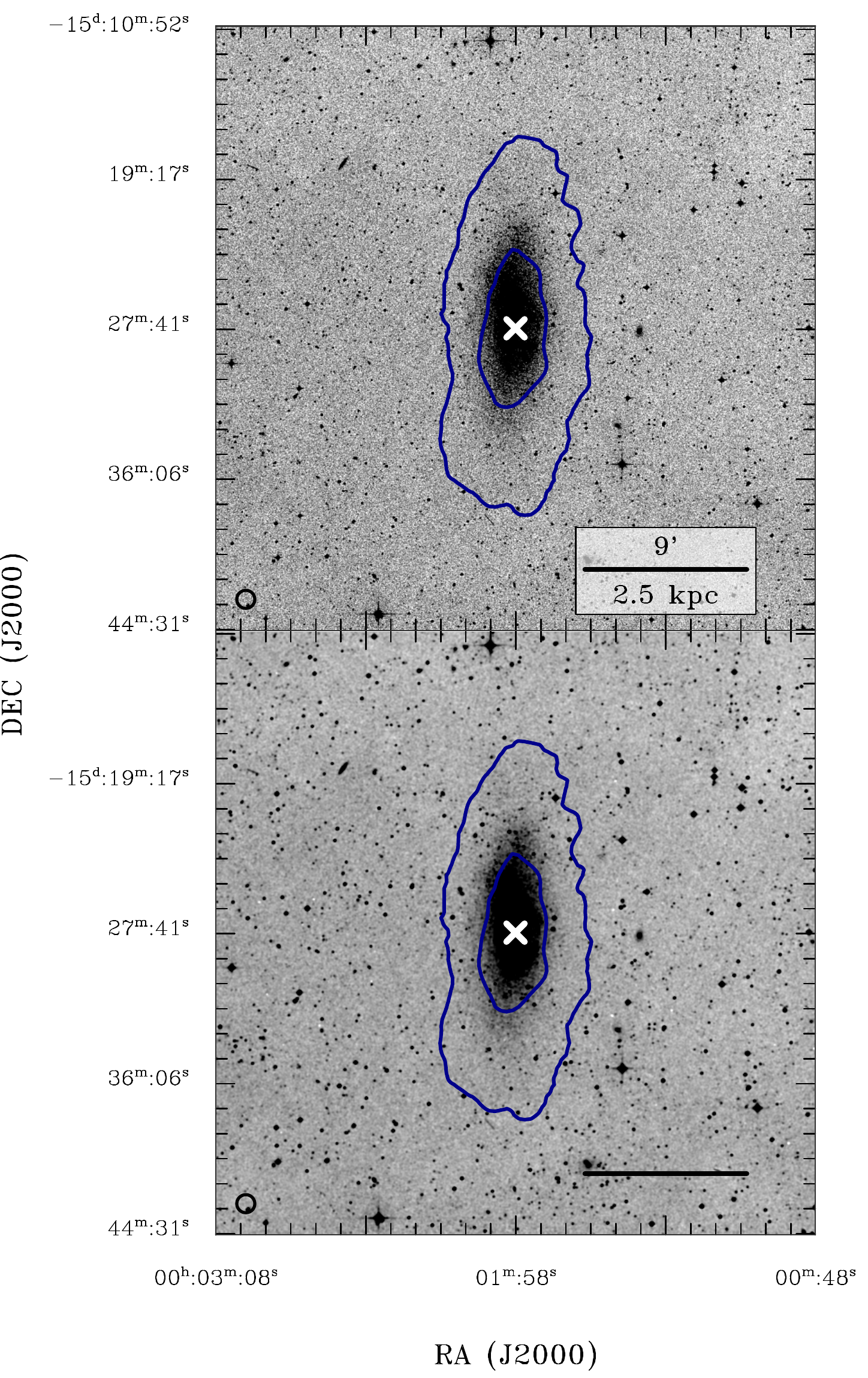}
 \caption{H\,{\sc i} surface density map of WLM overlaid on top of the DSS (red, IIIa-F) images of WLM.
 Top: the DSS image is shown at its native resolution. Bottom: the DSS image has been
 convolved with a 2D-Gaussian of $FWHM\,=\, 6^{\prime\prime}$ in each direction.
 Blue contours denote the H\,{\sc i} surface density at $1,9\,$-$\,M_\odot\,\mathrm{pc}^{-2}$-levels.
 The white crosses represent the kinematical centre of the model. The beam of the MeerKAT-16 observation is shown as
 black circles in each panels.}
\label{fig:WLM_optical_dss_r_norm_con_6}
\end{figure}

We use the MIRIAD task {\tt{MOMENT}} to derive the moment maps of the MeerKAT data.
In Fig.~\ref{fig:WLM_optical_dss_r_norm_con_6}, we overplot
the surface density map of the low-resolution data cubes onto the DSS optical images of WLM.
We show the optical image at its native resolution and after a convolution with a
2D-Gaussian with a FWHM of 6\arcsec. The outermost contour shows the H\,{\sc i} surface brightness level at 1~$M_{\odot}~\mathrm{pc^{-2}}$,
thus it delimits the H\,{\sc i} diameter quoted in
Table~\ref{tab:gal_prop}. As observed in many gas-rich dwarf galaxies, the H\,{\sc i} extends
much farther than the optical disc. We derive an H\,{\sc i}-to-optical diameter
($D_{HI}/D_{25}$) ratio of 2.17. The overall H\,{\sc i} distribution is asymmetric, with the
southern side slightly twisted towards the east.

\begin{figure*}
    \begin{tabular}{c c}
         \includegraphics[scale=0.48]{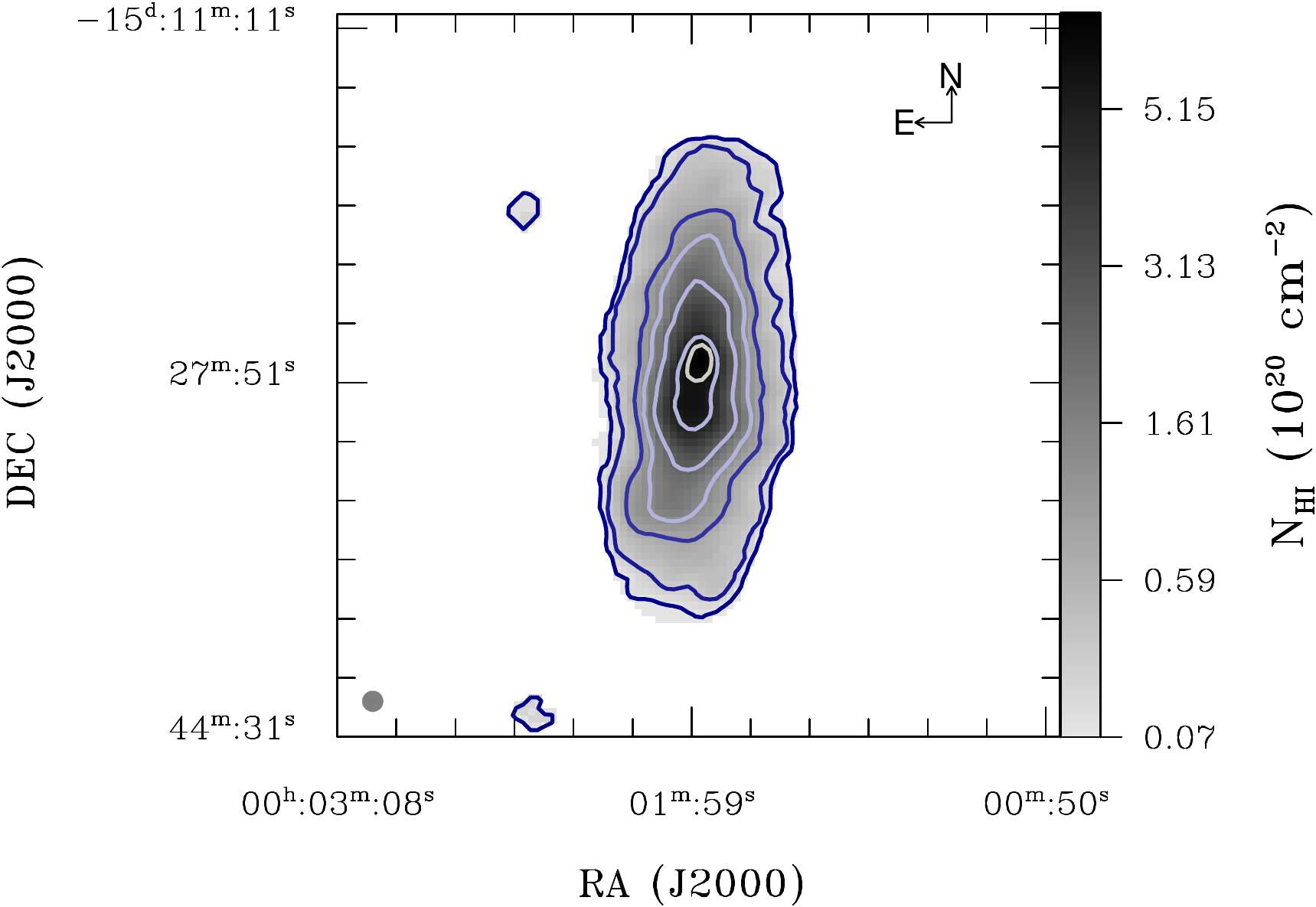}&
          \includegraphics[scale=0.48]{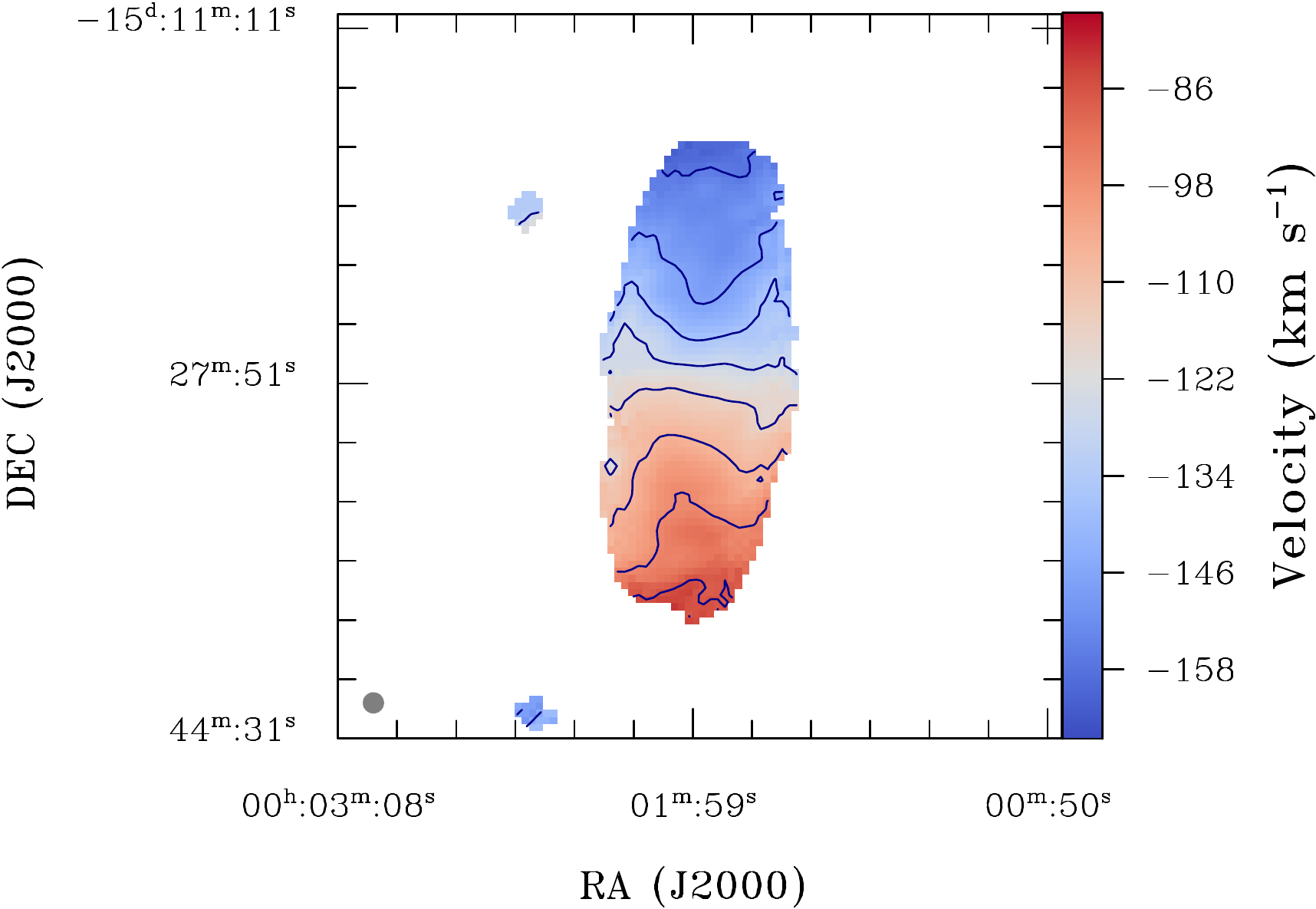}\\
          \includegraphics[scale=0.48]{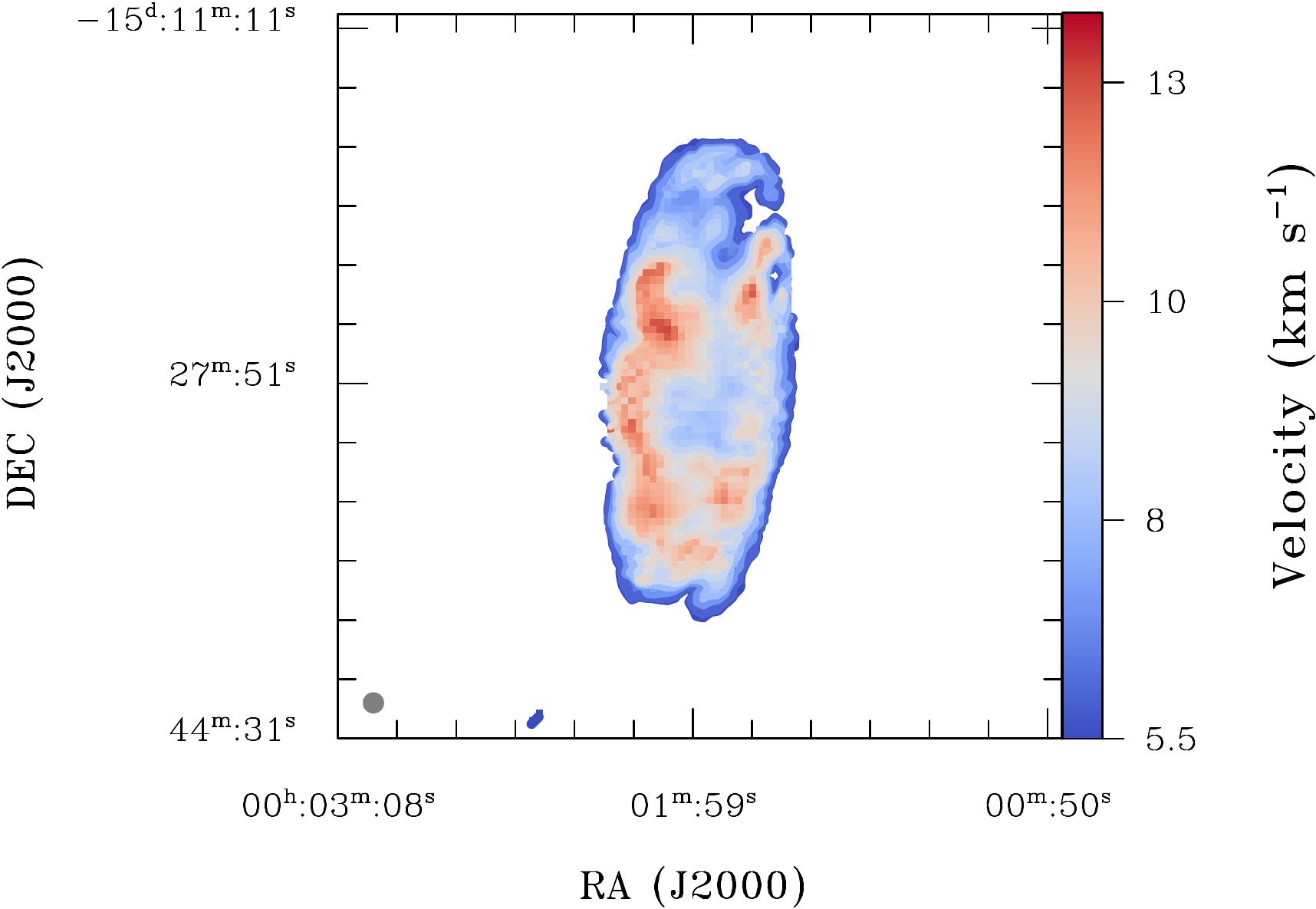} &
         \includegraphics[scale=0.43]{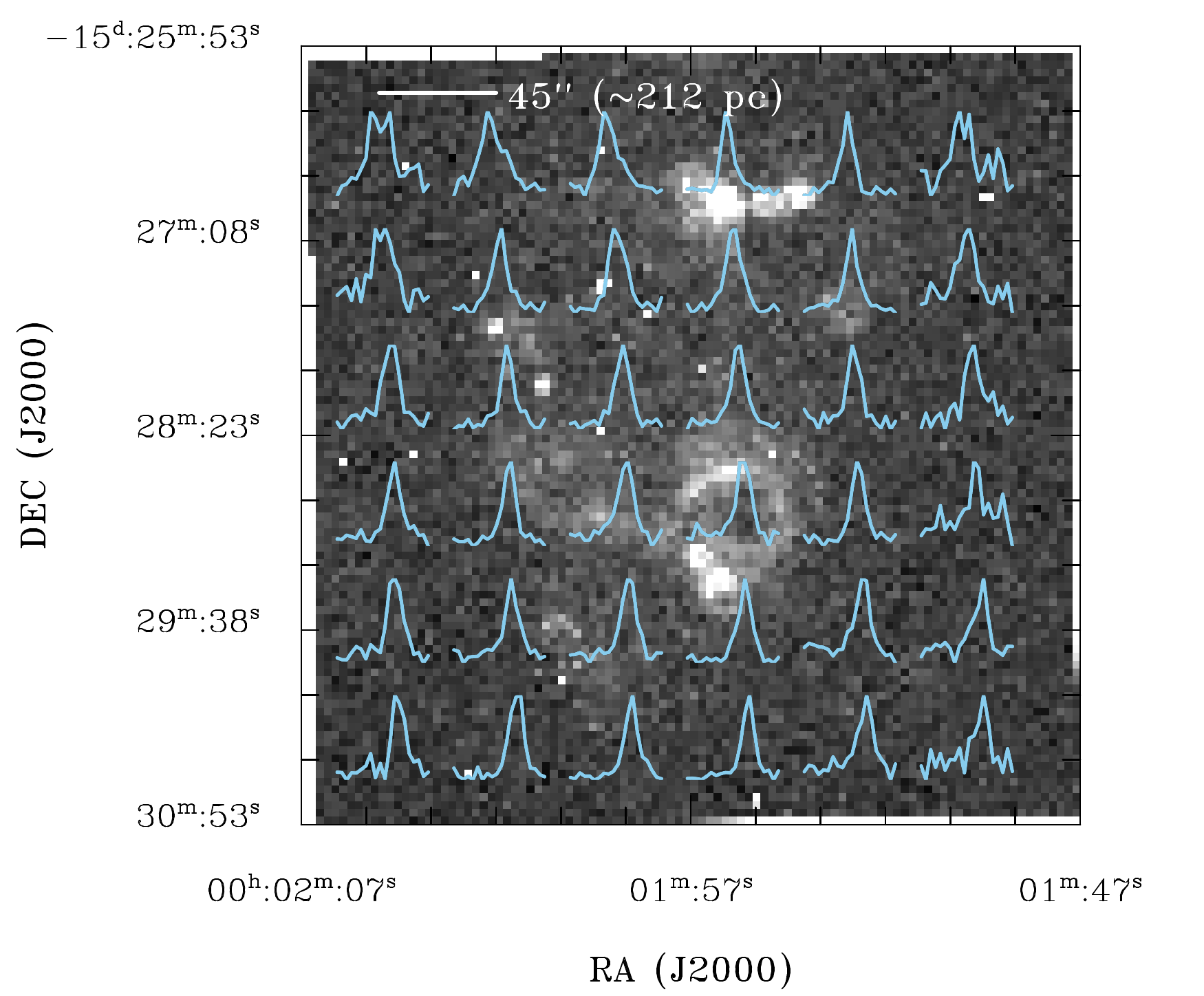}
    \end{tabular}
     \caption{Top left: column density map; contour levels are (0.9, 2.8, 8.4, 14.0, 25.2, 50.5, 59.0)~$\times~10^{19}~\rm{cm}^{-2}$.
     Top right: first moment map; contour levels are spaced by 10 $\mathrm{km~s^{-1}}$ and start
     from -35 $\mathrm{km~s^{-1}}$ to 35 $\mathrm{km~s^{-1}}$ relative to the systemic velocity $v_{sys}$ = -121.62 $\mathrm{km~s^{-1}}$.
     Bottom left: second moment map. The beam ($60\arcsec$) is represented by the gray circle at the bottom left of the first three panels.
     Bottom right: H\,{\sc i} spectra from the high-resolution data cube of WLM ($35\arcsec \times 12\arcsec$).
     The spectra are separated from each other by a distance of 45\arcsec~in the x and y direction. The background gray-scale image shows
     $\rm{H_{\alpha}}$ emission from the SPITZER/LVL archive.}
    \label{fig:mommaps}
\end{figure*}

We present the column density map, the first moment map (velocity field), and the second moment map in Fig.~\ref{fig:mommaps}.
The H\,{\sc i} column density distribution appears to be smooth in the outer disc.
However, in the central part, there are a few high column density peaks
whose location closely matches that of the H$\alpha$ emission observed as part of the Spitzer Local Volume Legacy project \footnote{\label{archive}\url{https://irsa.ipac.caltech.edu/data/SPITZER/LVL/galaxies/WLM.html}}\citep{2009ApJ...703..517D}.
The hook-like morphology observed by
\citet{2007AJ....133.2242K} is not recovered by our observations but the location of the high-column density peaks roughly matches
that of the hook. The velocity field shows twisted iso-velocity contours in both the Northern and the Southern sides. While the twisted, curved iso-velocity
contours in the Northern side were seen previously in high resolution observations \citep{2004AJ....128.1219J, 2007AJ....133.2242K}, the ones
in the Southern side did not show up due to the limited sensitivity of these observations. They indicate a warp-like morphology but their origin are not clear given that WLM is an isolated galaxy. The inner iso-velocity contours are mostly parallel to the minor axis, indicating a solid-body rotation.
Signatures of differential rotation, indicated by curved iso-velocity contours, are found slightly further away from the centre.
As reported by \citet{2007AJ....133.2242K}, here also we find that the inner approaching halve of the galaxy show steeper velocity gradient than the inner receding side. For the second moment map, there are regions with enhanced second moment values in the Southern side,
corresponding to the location of the warp suggested by the moment-1 velocity field. The inner North-East side also has regions of higher
second moment values. This corresponds to the region where we find a few double-peaked profiles. The central and the far Northern side have
lower second moment values than the rest of the pixels. Regions with enhanced second moment values have been attributed to turbulence effects
caused by, e.g., star formation feedback or magneto rotational instability \citep{2009AJ....137.4424T}. Investigating the origin of
this patterns of high second moment values requires pixel-by-pixel analysis of the shapes of the H\,{\sc i} spectrum, which is beyond
the scope of this analysis.

WLM has experienced an initial burst of star formation, and continues to form stars
until the present epoch \citep{2019MNRAS.490.5538A}. Therefore, we may expect to see expanding shells of gas produced by
past supernova explosions in WLM. We look for possible signatures of such shells in the star forming disc of WLM. For this, we use
the high-resolution data cube, which has a spatial resolution of $163~\times~56$ pc. Supernova-driven shells, which appear as
H\,{\sc i} holes with size above our linear resolution limit are common in spirals and in dwarf galaxies
\citep{2008A&A...490..555B, 2011AJ....141...23B, 2011ApJ...738...10W}. Thus, if expanding shells with size above our linear resolution
are present in WLM, we would expect to see their imprints on the H\,{\sc i} kinematics.
We use the $\mathrm{H_{\alpha}}$ image from the IRAC/LVL\footnoteref{archive} archive to trace the location of star formation.
Signatures of expanding shells include the presence of double peaked H\,{\sc i} spectra, and enhanced velocities at the radius of the shell.
At the bottom right of Fig.~\ref{fig:mommaps}, we show individual H\,{\sc i} spectra in the central region of WLM where the $\mathrm{H_{\alpha}}$ emission
occurs. Virtually all profiles inside the region are single peaked, except in the North-Eastern
side where we find a few double peaked profiles. This location
also corresponds to the region where we find high second moment values as mentioned previously.
In addition to looking at the shapes of the
individual profiles, we used the Karma-kshell\footnote{\url{https://www.atnf.csiro.au/computing/software/karma/user-manual/node9.html}}
tool to search for enhanced velocity signatures around the expected centre of the shells but our findings were inconclusive.
Thus, at our resolution limit of $163~\times~56$ pc, we do not find any convincing evidence that the H\,{\sc i} is currently expanding in single star formation regions in WLM.
\section{3D modelling}\label{sec:3dmodel}
\indent To model the global kinematics and the morphology of WLM, we fit the data with an extended version of the
tilted-ring model of \citet{1974ApJ...193..309R} implemented in the TiRiFiC
software suite. TiRiFiC allows the user to fit a tilted-ring  model directly to the data cube instead of a velocity
field, as is done in classic tilted-ring fitting method. TiRiFiC has the advantage of having a larger set of parameters compared to the
velocity-field based approach. This enables the user to explore different combinations of model parameters
and makes a more complex modelling where necessary.
\subsection{Modelling strategy}
\indent TiRiFiC models a galaxy as a rotating disc of radius
{\tt{R}}, with a
thickness that follows a certain distribution in the vertical direction. The disc is divided into a set of {\tt{TIRNR}} circular,
concentric rings with widths {\tt{RADSEP}}, each centred at
{\tt{(XPOS, YPOS)}},
tilted at an inclination of {\tt{INCL}} with respect to the sky plane and oriented at a position angle defined by
the {\tt{PA}} parameter. Each ring simulates rotating gas with a surface brightness distribution
{\tt{SBR}}, a circular velocity {\tt{VROT}}, a systemic velocity
{\tt{VSYS}}, and a velocity dispersion {\tt{SDIS}}. The user can specify the number of discs to be modelled using the {\tt{NDISKS}} parameter.
Parameters which correspond to a particular disc number are then denoted by the parameter names followed by an underscore and the disc number.
Fixing one or more parameters with radius makes the model
simpler and allows a faster computing time. Thus, as described in \citet{2007AA...468..903J}, \citet{Saburova2013}, \citet{2014AA...561A..28S}, and \citet{Henkel2018}, we start with a simple model and compare the result with the observation. If the simplest possible model does not sufficiently
describe the observations, then we allow one or more parameters to vary with radius and check if that improves the model.
If no significant improvements are found, we set the corresponding parameters to remain fixed. We therefore
model the galaxy with different set of parameter combinations described as follows. We assume that all rings (nodes)
have the same rotation centre and the same systemic velocity.
We further assume that the \ion{H}{i} is optically thin.
We then explore the following parameter set up:
\begin{itemize}
    \item with fixed or radially-varying position angle,
    \item with fixed or radially-varying inclination angle,
    \item with fixed or radially-varying disc-thickness,
    \item with or without global vertical and radial motion,
    \item excluding or including second-order harmonic distortions in tangential and radial motions.
    This has been used in the literature to simulate bar-like motions.
    \citep[e.g.,][]{franx_evidence_1994, schoenmakers_asymmetries_1999, 2007ApJ...664..204S, Saburova2013}.
    \item with or without first-order distortions in surface brightness,
    \item with or without first and second-order distortions in surface brightness.
\end{itemize}
\subsection{Results}
After a visual inspection of the previously described parameter combinations,
our final best-fitting model is described as an optically thin \ion{H}{i} disc with
the following parameters:
\begin{itemize}
    \item one rotation centre and one systemic velocity for all rings,
    \item a constant position angle and inclination,
    \item a constant velocity dispersion,
    \item a fixed disc thickness,
    \item a radially-varying surface brightness with first and second-order harmonic distortions described by
    \begin{equation}
    \begin{aligned}
    \Sigma(r,\theta) \,=\, &SBR(r)+SM1A(r)\cdot \cos(\theta-SM1P(r))+\\
    &SM2A(r)\cdot \cos(2(\theta-SM2P(r))),\\
\end{aligned}
\end{equation}
where $r$ is the radius, $\theta$ is the azimuthal angle, $SM1A/SM2A$ and $SM1P/SM2P$ are the amplitudes and the phases of the distortions.
    \item a radially-varying rotation velocity,
    \item a radially-varying tangential and radial velocity with second order harmonic distortions.
    To keep a minimal number of parameters, the angle between the phase of the  radial and the tangential velocity was kept at an angle of 45\degree.
\end{itemize}

\indent The harmonic distortions in surface brightness are required to model the
asymmetric morphology of the galaxy and to describe the bar-like motions as used in
\citet{2007ApJ...664..204S} and \citet{Saburova2013}. However, we find it necessary to only
include the kinematic distortions from a radius of 120\arcsec.
While adding a warp did not improve our results significantly, a thick disc of 85\arcsec ($\sim$400 pc)
was required to refine the model. In addition, the inclusion of non-circular motions
improved the model and resulted in a more well-behaved rotation curve with less wiggles compared to
the rotation curve from the model with only circular motions.
Therefore, our final best-fitting model is the one with non-circular motions and without a warp, which we use for the mass modelling.
However, for the interested readers, we present below comparison between different models.

\subsubsection{The final flat disc model vs a warp model}

\begin{figure*}
 \includegraphics[scale=0.8]{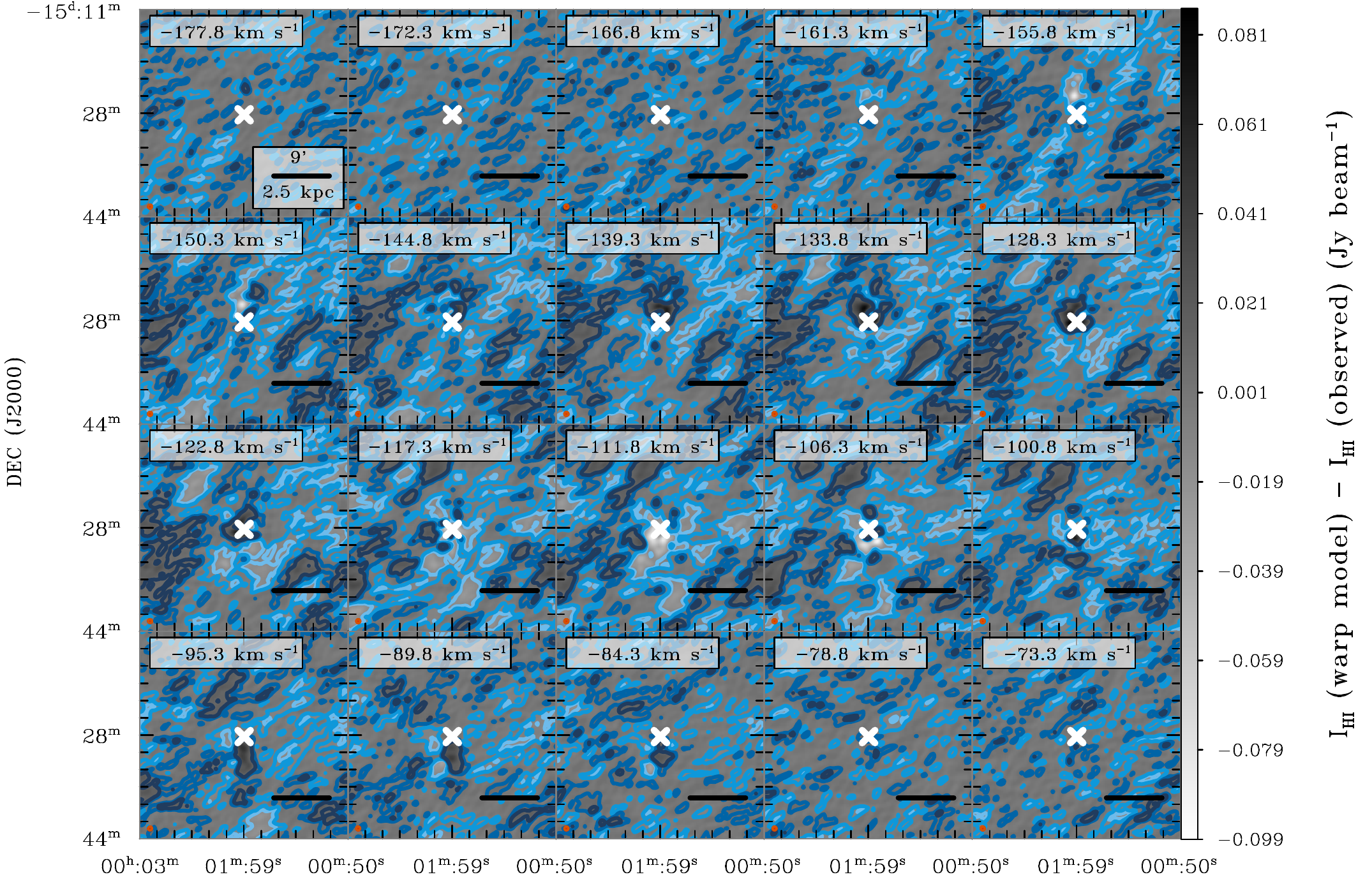}
 \includegraphics[scale=0.8]{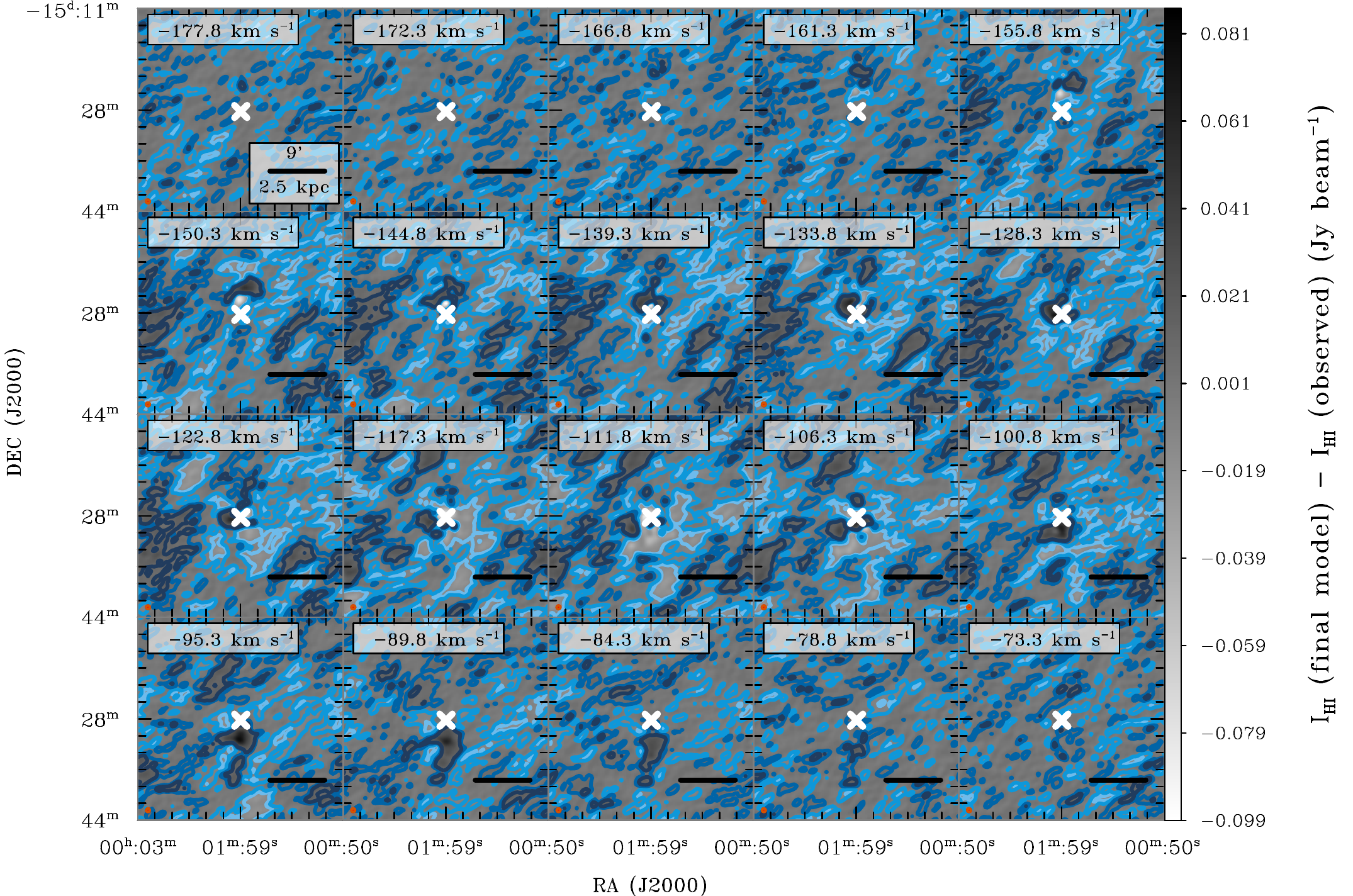}
 \caption{Residual channel maps. Top: a warp model minus the observed data cube; Bottom:
 the final (flat-disc) model minus the observed data cube. Contour levels are
 (-2.2, -1.1, 1.1, 2.2) $\times$ $0.005~\mathrm{Jy~beam^{-1}}$. Light blue contours show negative pixels.}
 \label{fig:warp}
\end{figure*}

In Fig.~\ref{fig:warp}, we show residual channel maps. The top panel shows a warp model (i.e., with a
radially varying inclination and position angle) minus the observed data cube,
whereas the bottom panel shows the final (flat-disc) model minus the observed data cube. In general, the final flat-disc model performs
better in the Northern side compared to the warp model. Inversely, the warp model works best in the Southern side. For example,
the bottom panels of Fig.~\ref{fig:warp} shows elongated excess model emission (positive contours) adjacent
to negative contours at $v~=~(-95.3,~-89.8,~-84.3)~\mathrm{km~s^{-1}}$. This indicates a warp that is not well
recovered by the flat-disc model in the Southern part (as also indicated by the velocity field
shown in Fig.~\ref{fig:mommaps}). Overall though, the final flat-disc model is able to recover
the general distribution of the emission while still keeping a minimal number of model parameters.\\\\

\subsubsection{The final (flat-disc) model with non-circular motions vs the (flat-disc) model with only circular motions}
\begin{figure*}
 \includegraphics[scale=0.8]{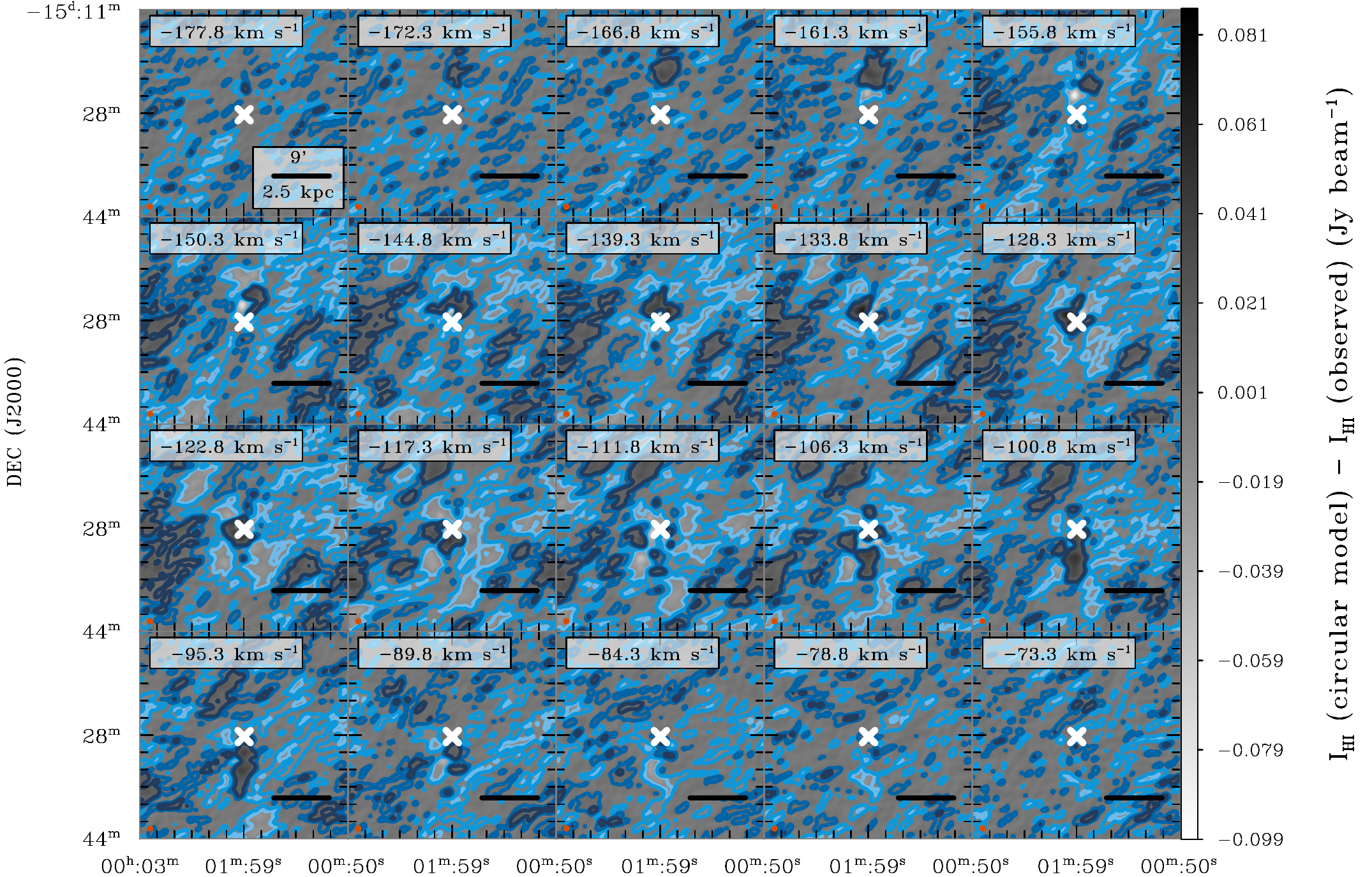}
 \includegraphics[scale=0.8]{residual_datacube_final.pdf}
 \caption{Residual channel maps. Top: the circular (flat-disc) model minus
 the observed data cube; bottom: the final (flat-disc) model minus the observed data cube. Contour levels are
 (-2.2, -1.1, 1.1, 2.2) $\times$ $0.005~\mathrm{Jy~beam^{-1}}$. Light blue contours show negative pixels.}
 \label{fig:final-circular}
\end{figure*}

\begin{figure*}
 \includegraphics[scale=0.68]{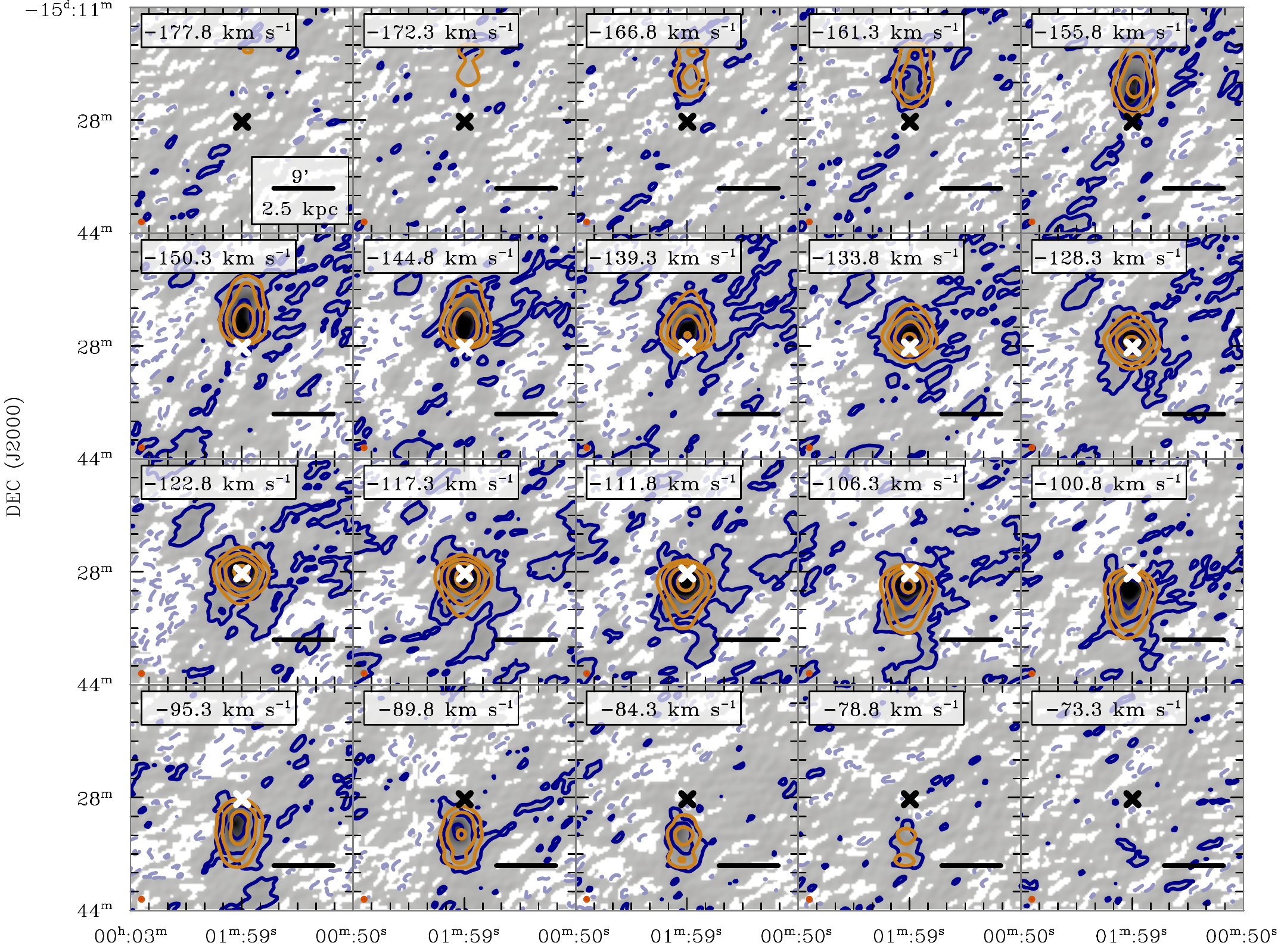}
 \includegraphics[scale=0.68]{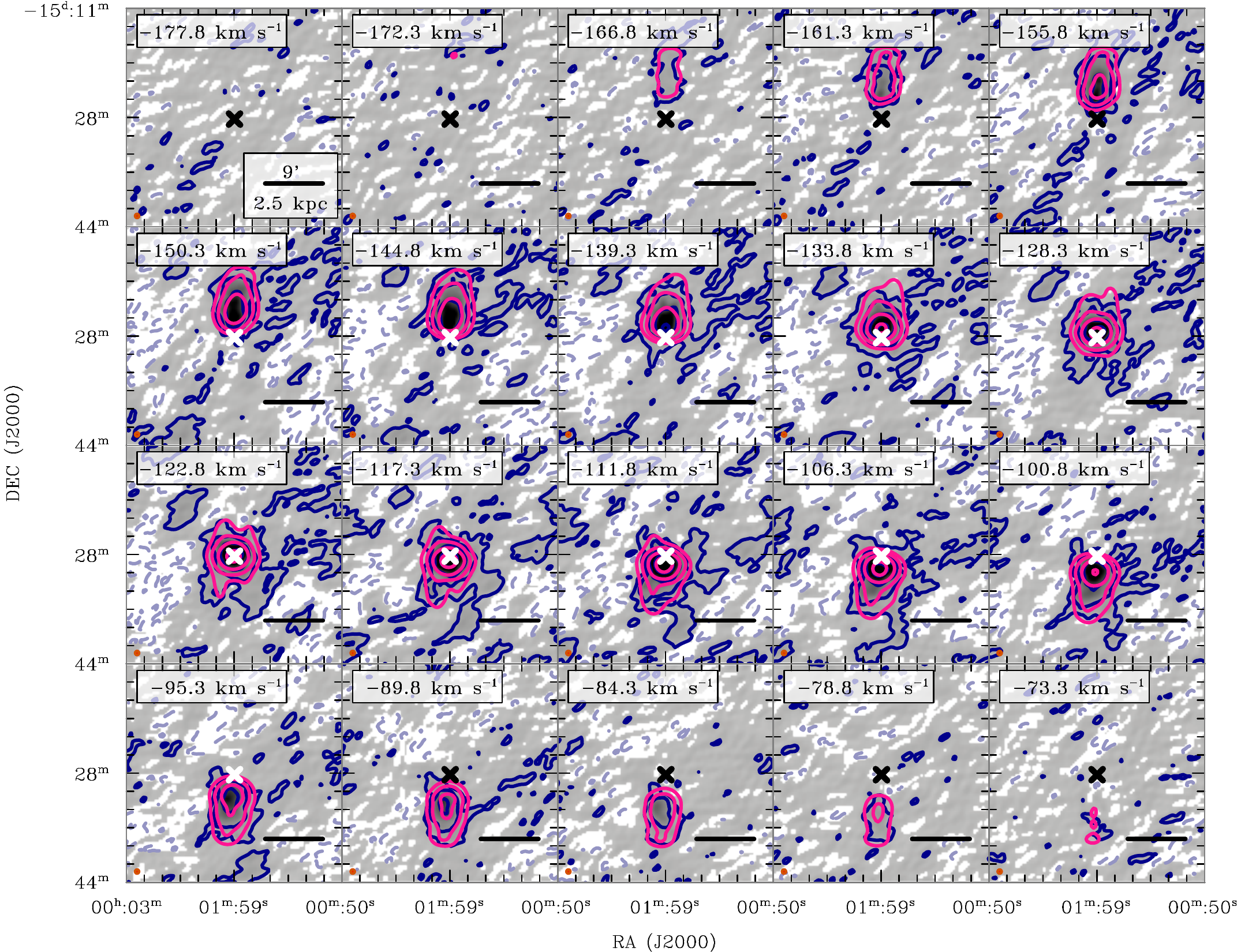}
 \caption{Individual channel maps of the TiRiFiC models overlaid on the observed data shown as blue contours.
 Contours denote the $-2,2,6,18,54\,-\,\sigma_\mathrm{rms}$-levels, where $\sigma_\mathrm{rms}\,=\,5\,\mathrm{mJy}\,\mathrm{beam}^{-1}$.
 Dashed lines represent negative intensities. Top: the {\sc TiRiFiC} model allowing for circular motions only.
 Bottom: our final best-fitting model with non-circular motions. Crosses represent the kinematical centre of the models. The circles
 shown in the lower left side of each panels represent the synthesised beam ($HPBW$).
 A version showing only the observed data cube is presented in Fig.~\ref{fig:WLM_original_data_cube_20_obs} of the Appendix.}
 \label{fig:WLM_circular_data_cube_20}
\end{figure*}

To compare the final model including non-circular motions with the model allowing for circular motions only, we
show residual channel maps of the two models in Fig.~\ref{fig:final-circular}. In addition,
we show a channel-by-channel comparison of the observed data cube with the two model data cubes in Fig.~\ref{fig:WLM_circular_data_cube_20}.
Overall, both models seem to perform equally
better in reproducing the overall distribution of the gas. The discrepancies between the two models are only mostly apparent
at the 2$\sigma$ rms noise level, where the final model is to be preferred.
Note that including the 2$\sigma$ level emission in the fitting does not affect
the modelling results in a systematic way but rather show the relative strength of the models to fit faint extended emission.

\begin{figure*}
 \includegraphics[scale=0.88]{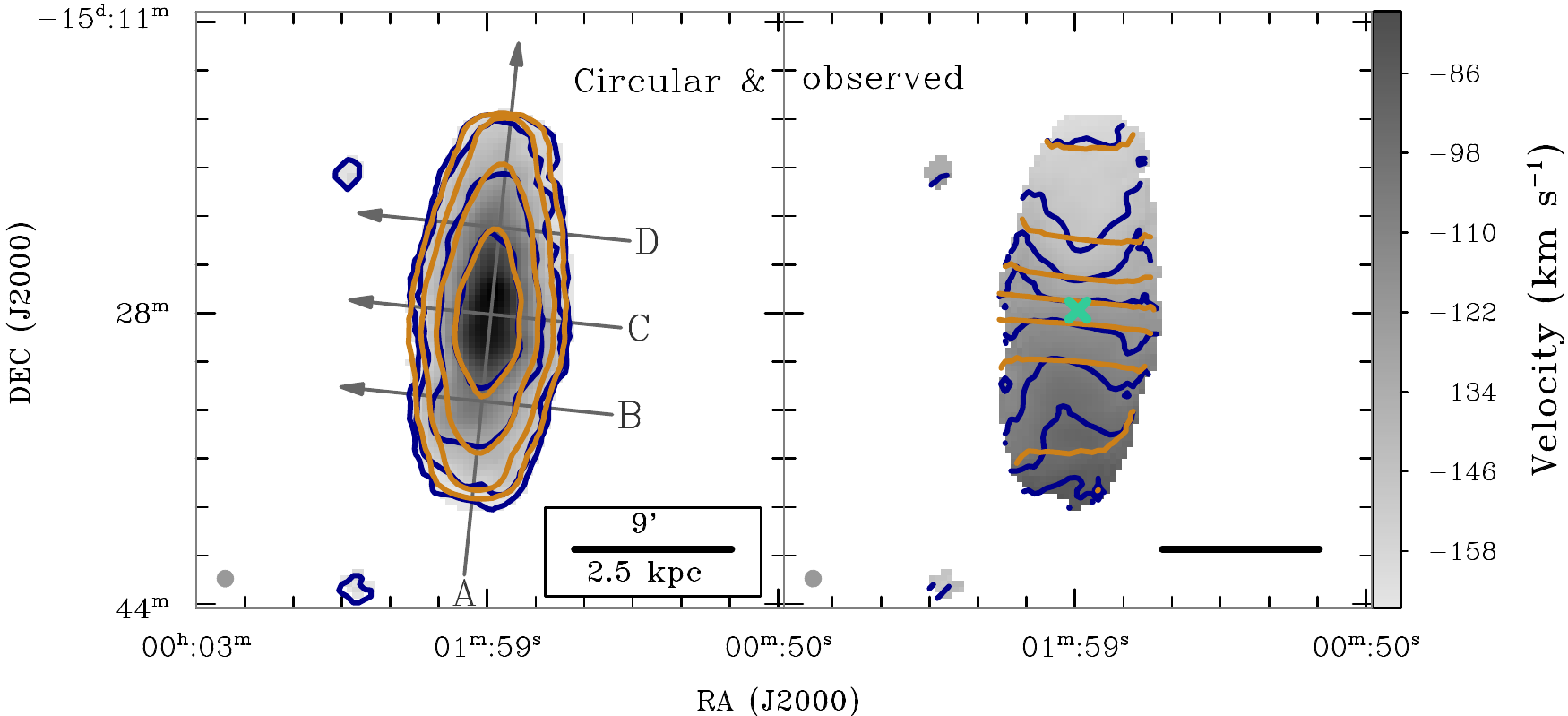}
 \includegraphics[scale=0.88]{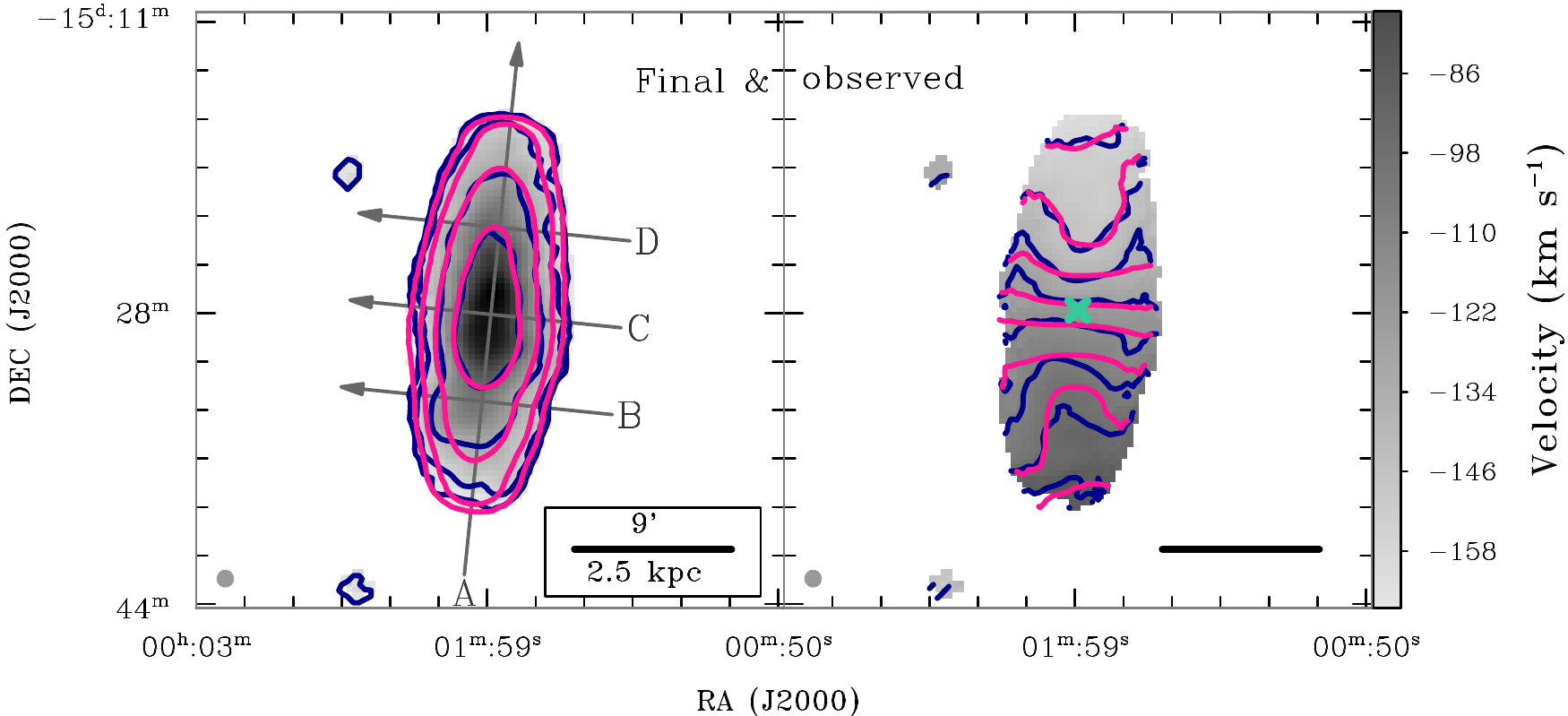}
 \caption{Total-intensity maps (left panels) and moment-1 velocity fields (right panels)
 of the {\sc TiRiFiC} models overlaid onto the observed data (blue contours). Top: the {\sc TiRiFiC} model
 allowing for circular motions only. Bottom: the final {\sc TiRiFiC} model
 with non-circular motions. For the total-intensity maps, contours denote the
 $\frac{1}{3},1.0,3,9,27\,$-$\,M_\odot\,\mathrm{pc}^{-2}$-levels. Arrows indicate the positions of
 slices along which the position-velocity diagrams in Fig.~\ref{fig:WLM_circular_PV-diagrams} and
  Fig.~\ref{fig:WLM_final_PV-diagrams} were taken. For the velocity fields, contours are spaced by
  10 $\mathrm{km}\,\mathrm{s}^{-1}$ and range from -35  $\mathrm{km}\,\mathrm{s}^{-1}$ to 35 $\mathrm{km}\,\mathrm{s}^{-1}$
  relative to the systemic velocity $v_\mathrm{sys}\,=\,-122\,\mathrm{km}\,\mathrm{s}^{-1}$. The crosses or the
  intersection between arrows A and C represents the kinematical centre of the model. The circles to the lower left of each panels show the synthesised beam ($HPBW$).}
\label{fig:WLM_circular_mom0_mom1_vf}
\end{figure*}

In Fig.~\ref{fig:WLM_circular_mom0_mom1_vf}, we compare the moment maps from the TiRiFiC models with the moment maps from
the observed data cubes. The total intensity map is well reproduced
by both models. However, the model with circular motions clearly fails to recover the curved iso-velocity contours in the velocity field.
This is a farther evidence regarding the strength of the model with non-circular motions over the model with only circular motions.

\begin{figure*}
 \begin{tabular}{c c}
 \includegraphics[width=0.45\textwidth]{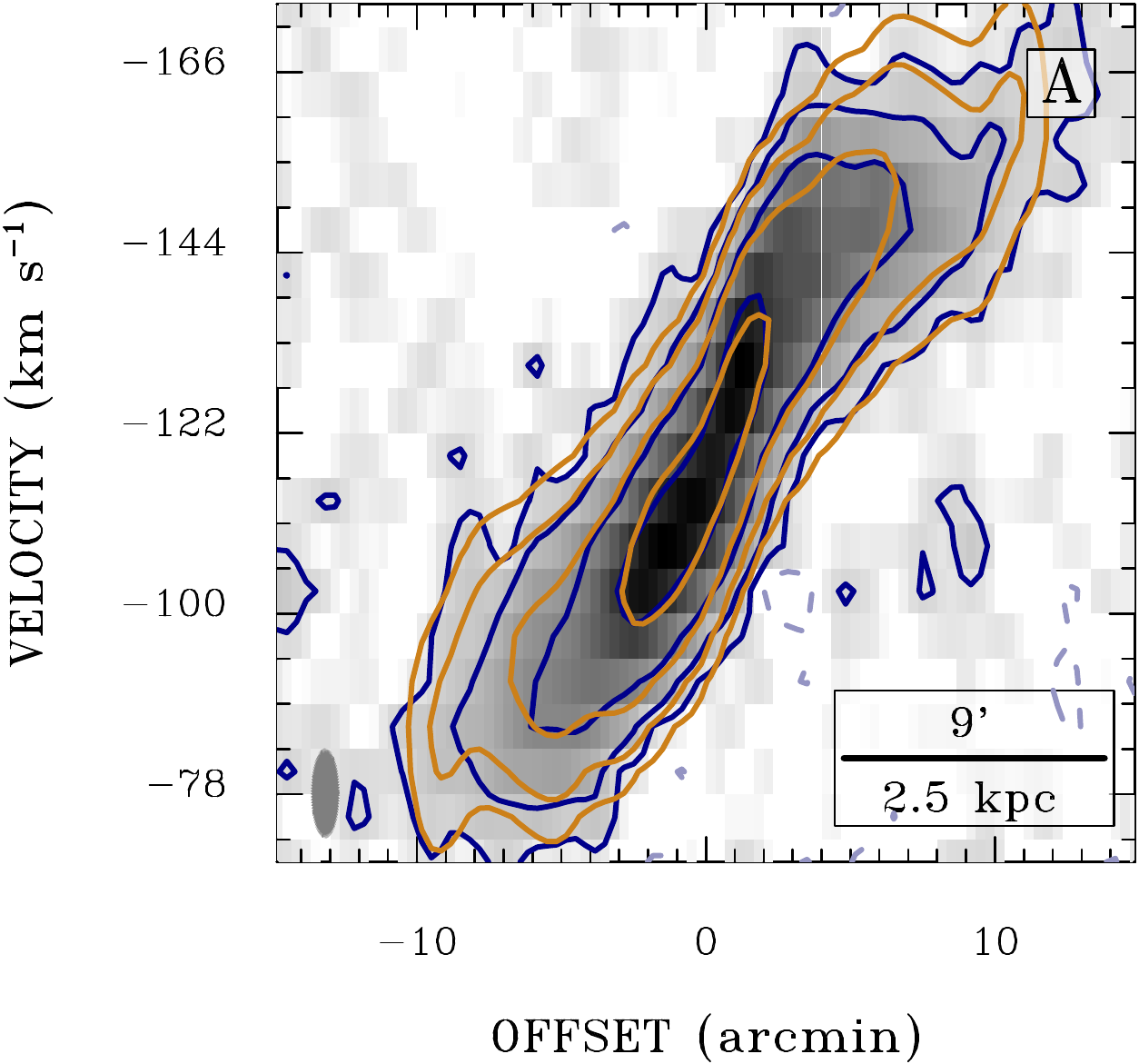}&
 \includegraphics[width=0.45\textwidth]{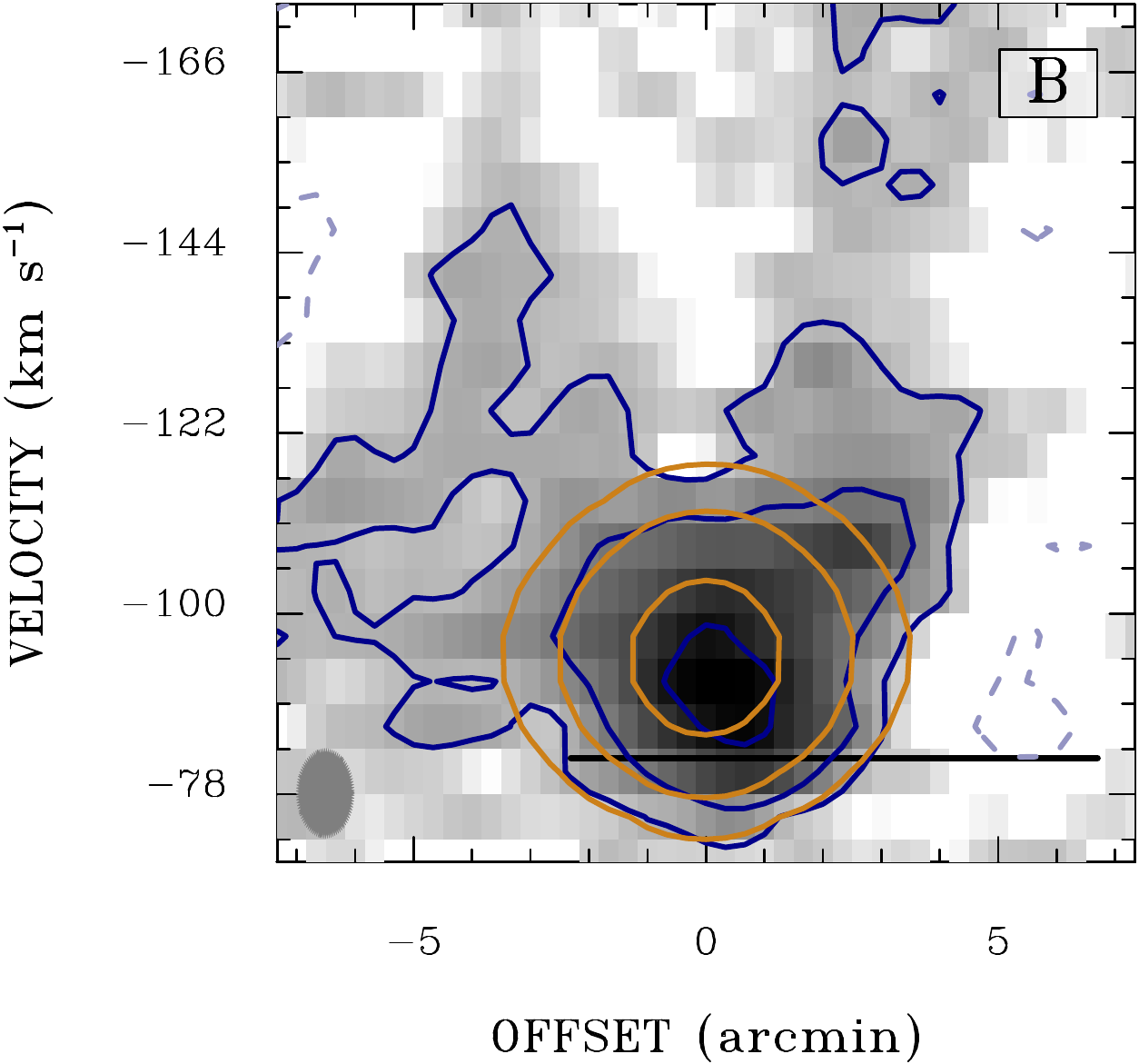}\\\\
 \includegraphics[width=0.45\textwidth]{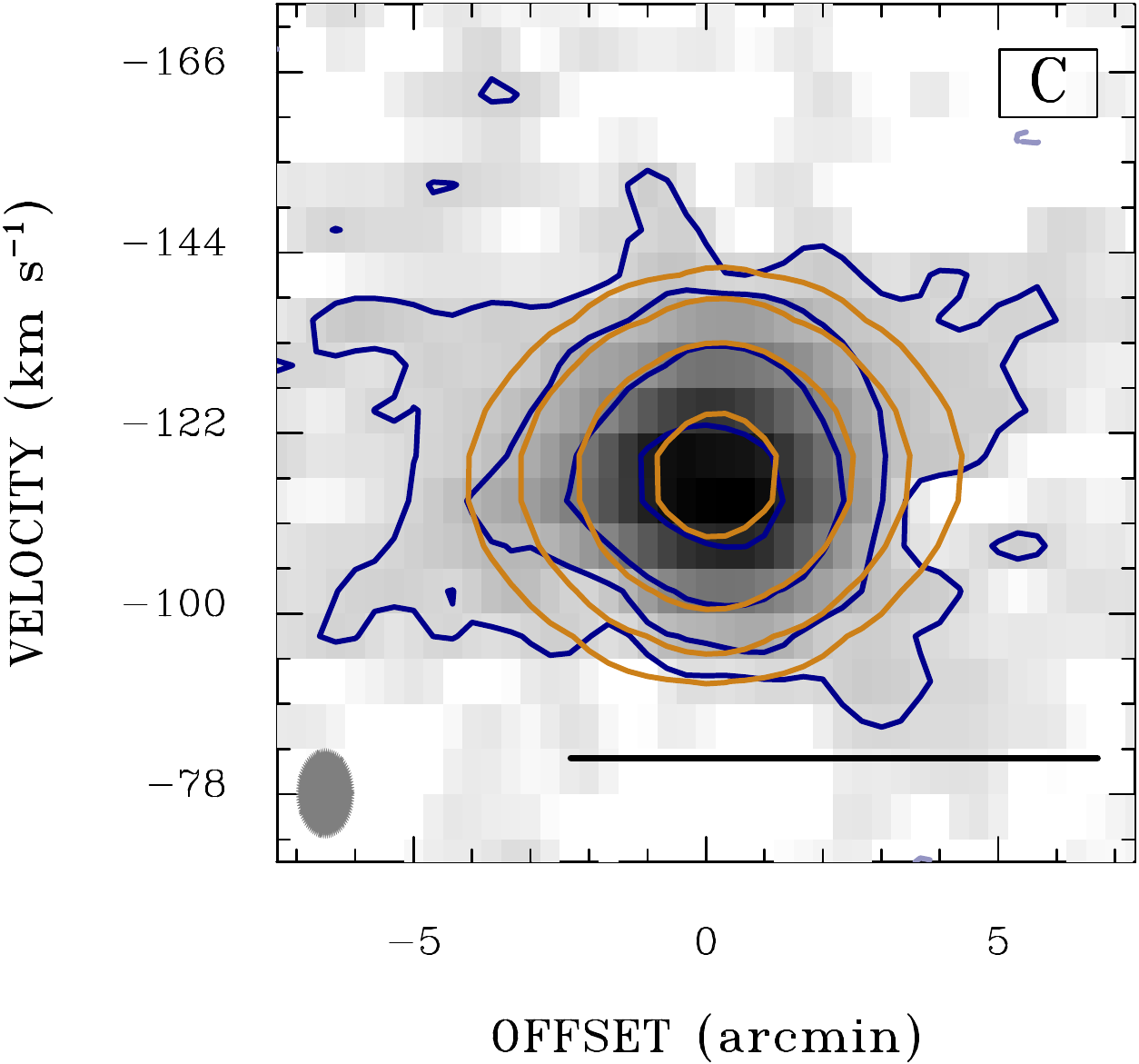}&
 \includegraphics[width=0.45\textwidth]{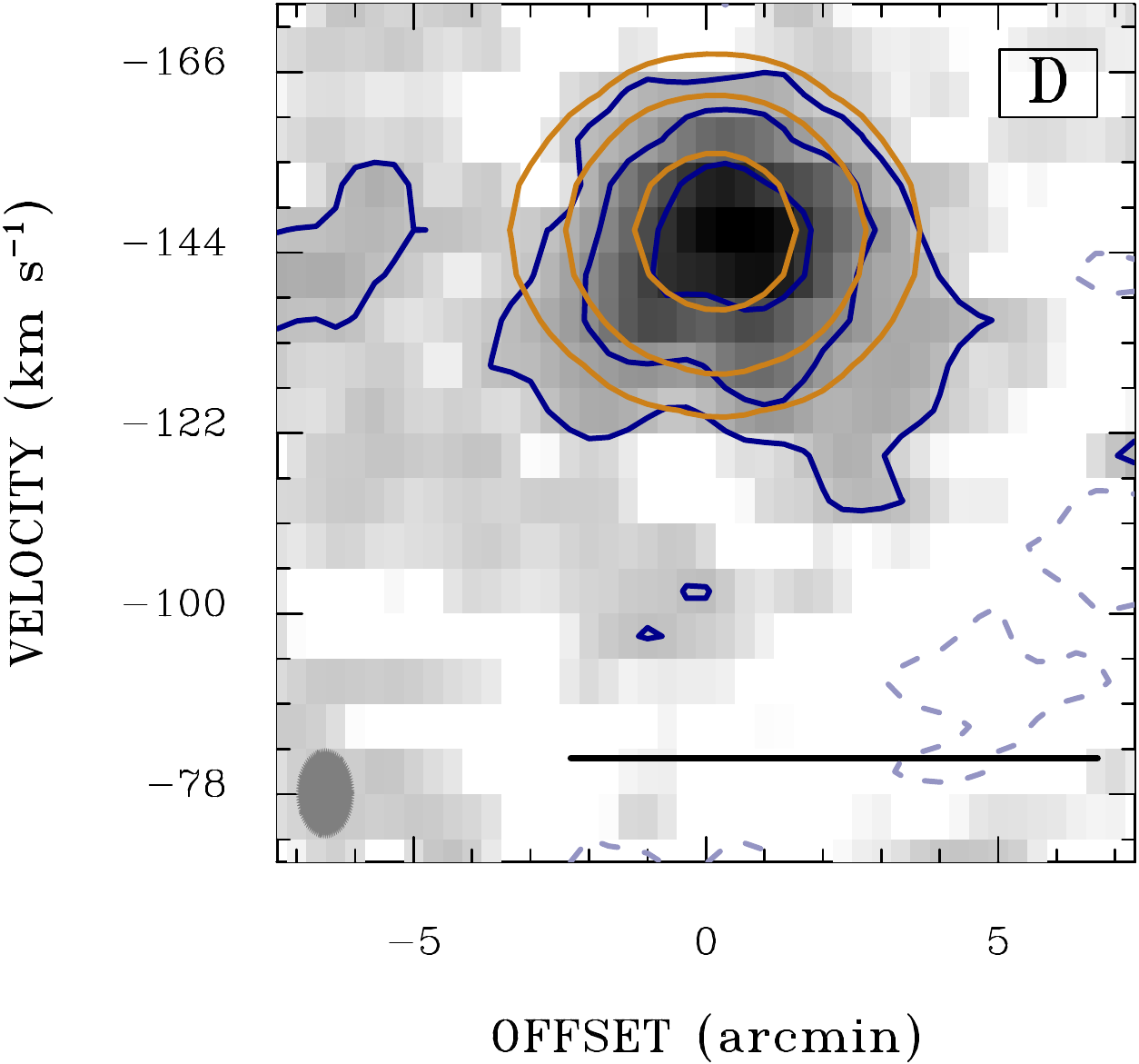}
 \end{tabular}
 \caption{Position-velocity diagrams taken along slices shown in Fig.~\ref{fig:WLM_circular_mom0_mom1_vf}.
 Contours denote the $-2,2,6,18,54\,-\,\sigma_\mathrm{rms}$-levels, where $\sigma_\mathrm{rms}\,=\,5\,\mathrm{mJy}\,\mathrm{beam}^{-1}$.
 Blue: the observed data cube. Strong orange: the {\sc TiRiFiC} model allowing for circular motion only.
 Dashed lines represent negative intensities.
 The ellipse represents the velocity (2 channels) and the spatial resolution ($\sqrt{HPBW_\mathrm{maj}*HPBW_\mathrm{min}}$,
 where $HPBW_\mathrm{maj}$ and $HPBW_\mathrm{min}$ are the major and minor axis half-power-beam-widths).}
\label{fig:WLM_circular_PV-diagrams}
\end{figure*}

\begin{figure*}
 \begin{tabular}{c c}
 \includegraphics[width=0.45\textwidth]{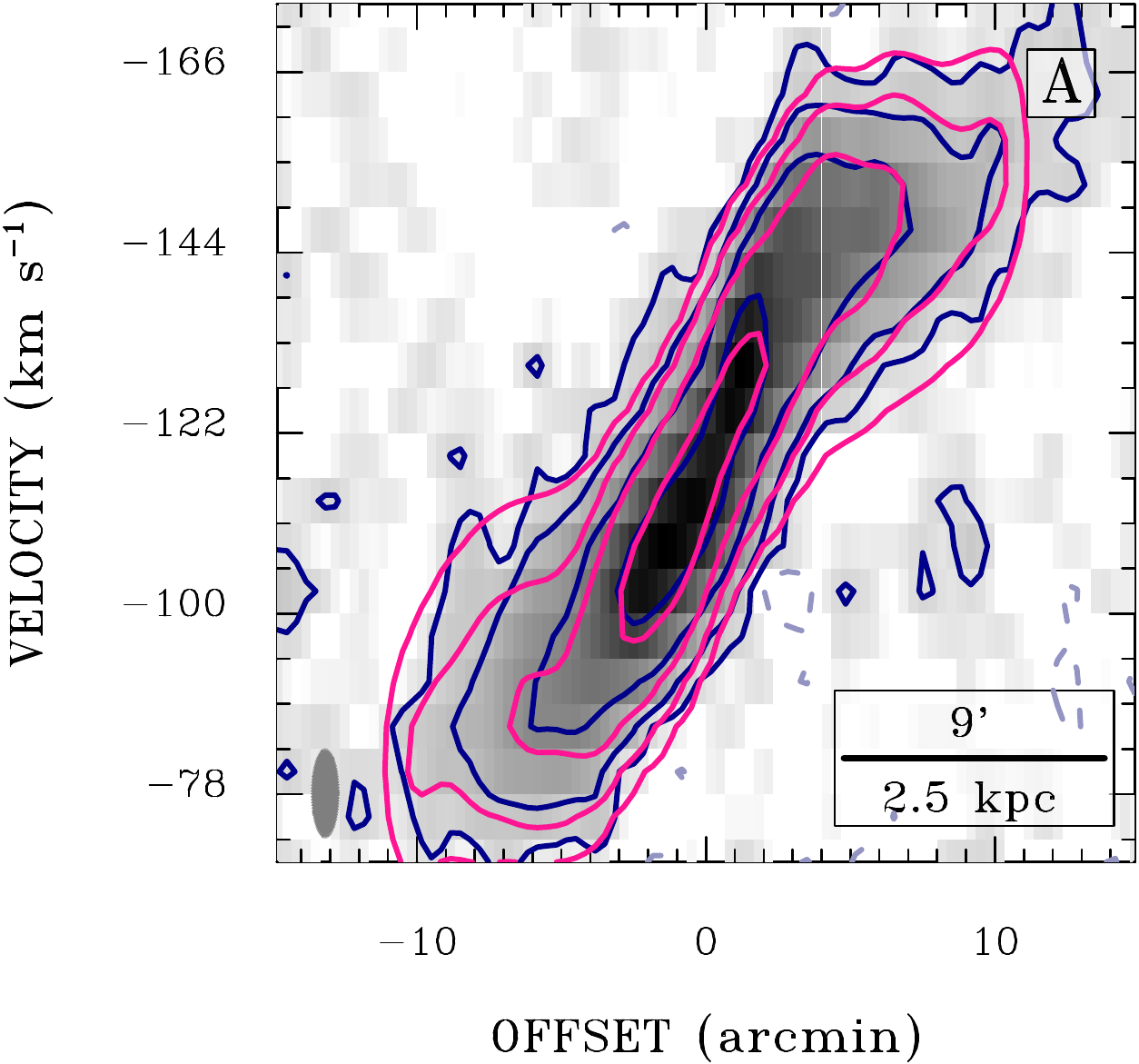}&
 \includegraphics[width=0.45\textwidth]{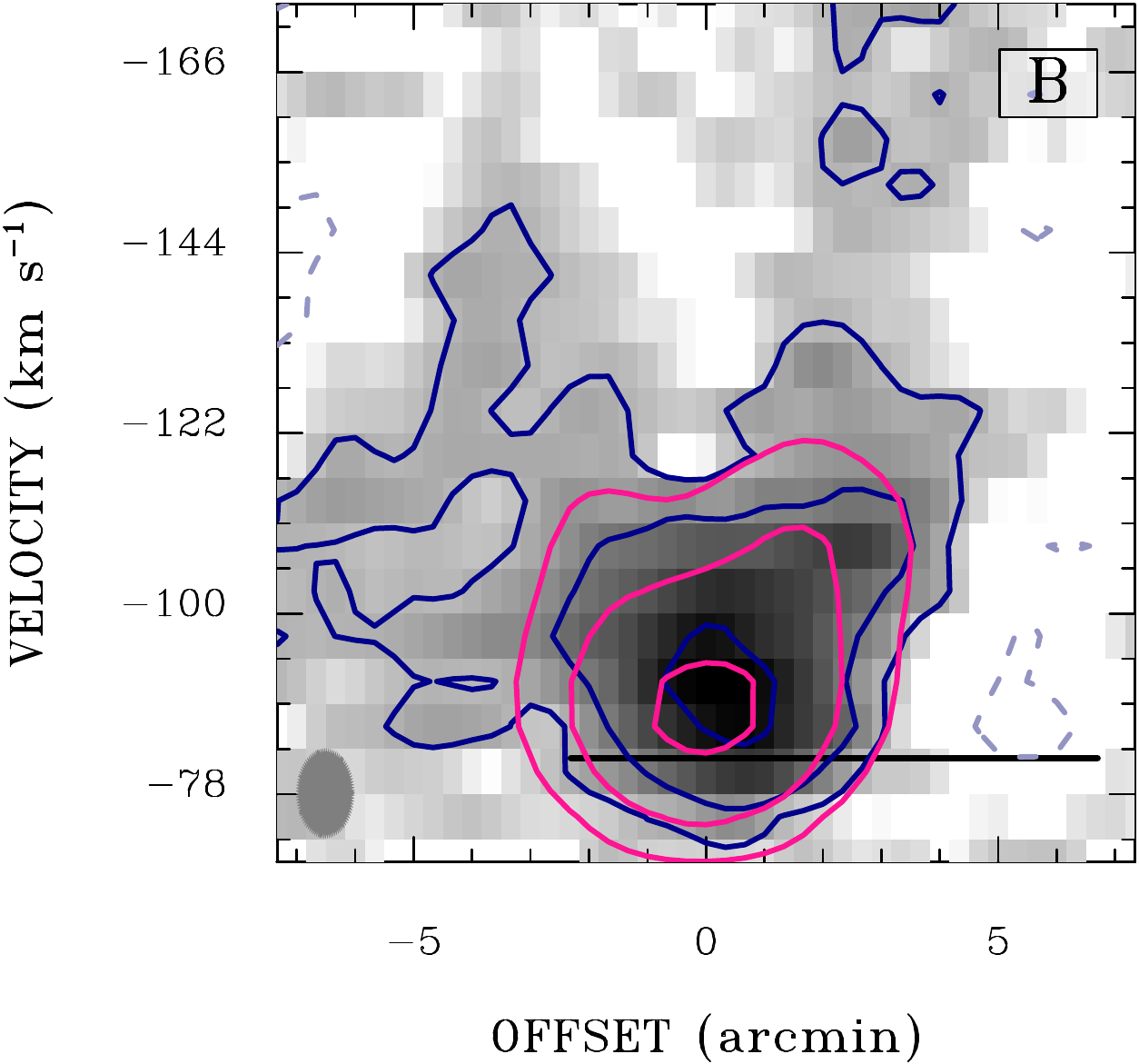}\\\\
 \includegraphics[width=0.45\textwidth]{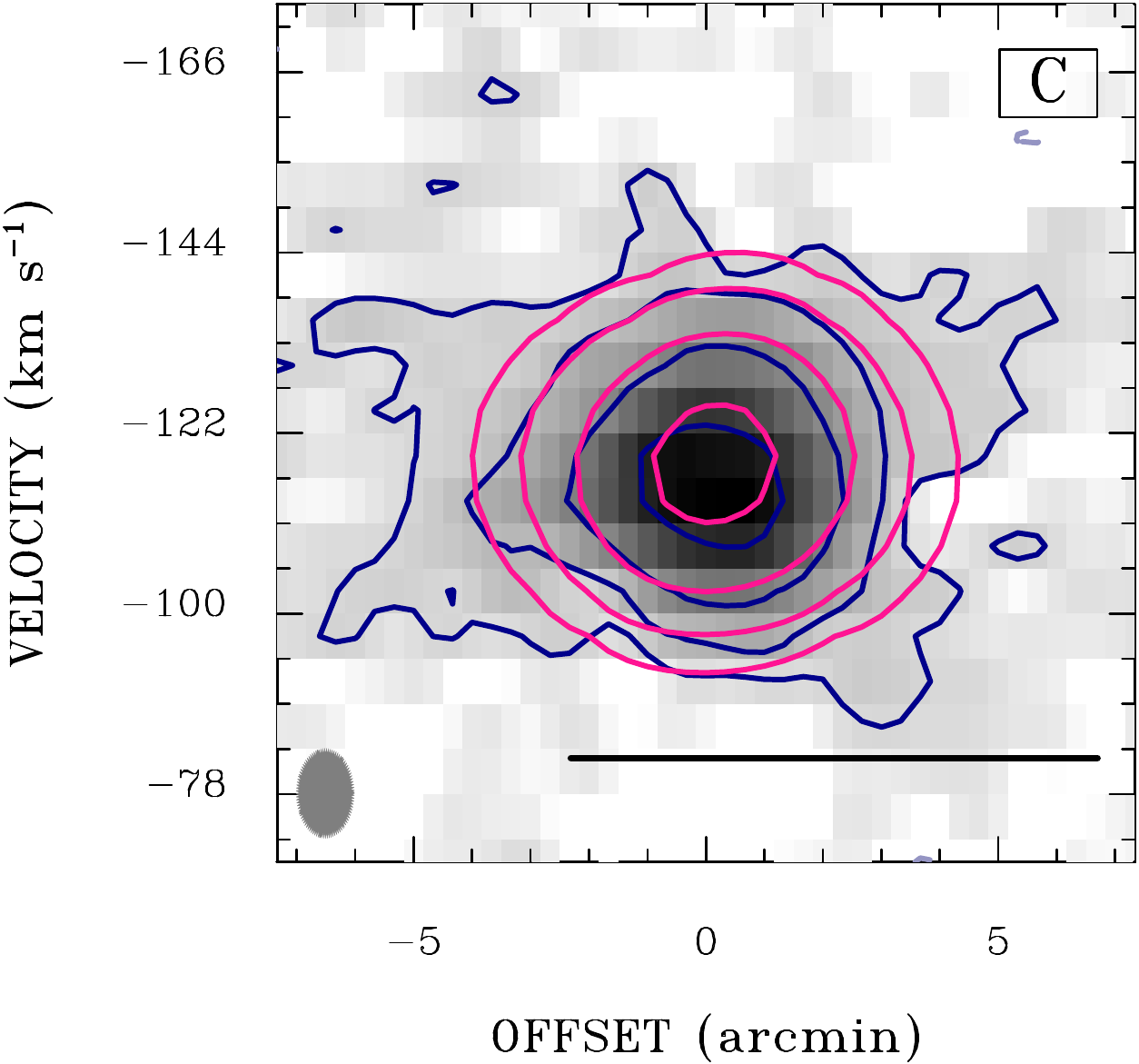}&
 \includegraphics[width=0.45\textwidth]{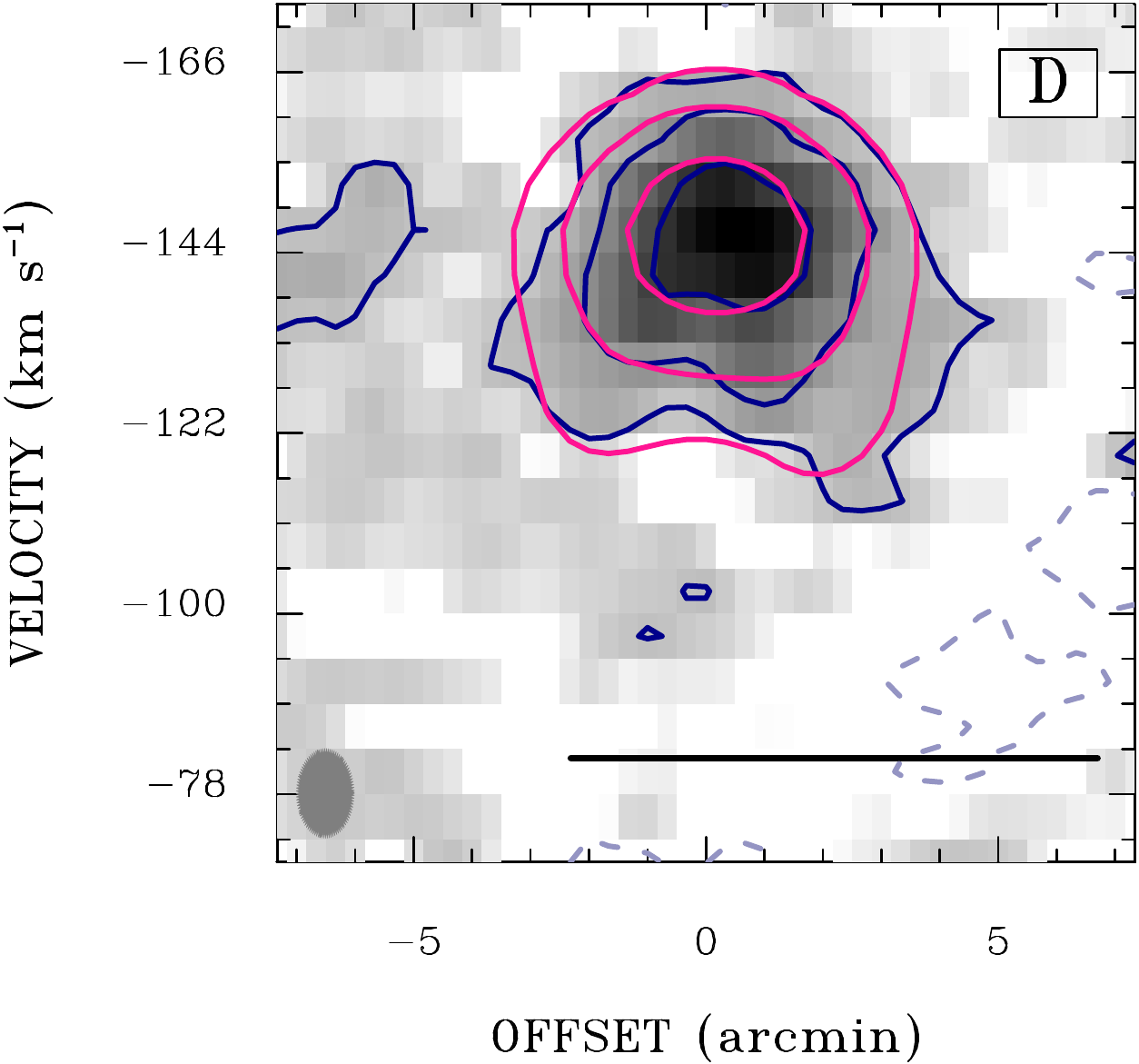}
 \end{tabular}
 \caption{Position-velocity diagrams taken along slices shown in Fig.~\ref{fig:WLM_circular_mom0_mom1_vf}.
 Contours denote the $-2,2,6,18,54\,-\,\sigma_\mathrm{rms}$-levels, where $\sigma_\mathrm{rms}\,=\,5\,\mathrm{mJy}\,\mathrm{beam}^{-1}$.
 Blue: the observed data cube. Pink: the {\sc TiRiFiC} final model. Dashed lines represent negative intensities.
 The ellipse represents the velocity (2 channels) and the spatial resolution ($\sqrt{HPBW_\mathrm{maj}*HPBW_\mathrm{min}}$,
 where $HPBW_\mathrm{maj}$ and $HPBW_\mathrm{min}$ are the major and minor axis half-power-beam-widths).}
\label{fig:WLM_final_PV-diagrams}
\end{figure*}

In Fig.~\ref{fig:WLM_circular_PV-diagrams} and Fig.~\ref{fig:WLM_final_PV-diagrams}, we show
position-velocity (PV) diagrams across cuts parallel to the major and the minor axes of WLM shown in
Fig.~\ref{fig:WLM_circular_mom0_mom1_vf}. Slices B and C show 2$\sigma$-emission that are not reproduced by the models.
Note that we have cleaned the data cube below the noise level. Thus, it is possible that they are real emission, but a follow-up study
is required to confirm this. It is also interesting that for slice B, there are discrepancies between the models
and the data even at the high-flux density contours. Gas at anomalous velocities is often attributed to star formation
or intergalactic gas accretion.
\subsubsection{Final model parameters}
\begin{figure}
 \includegraphics[width=\columnwidth]{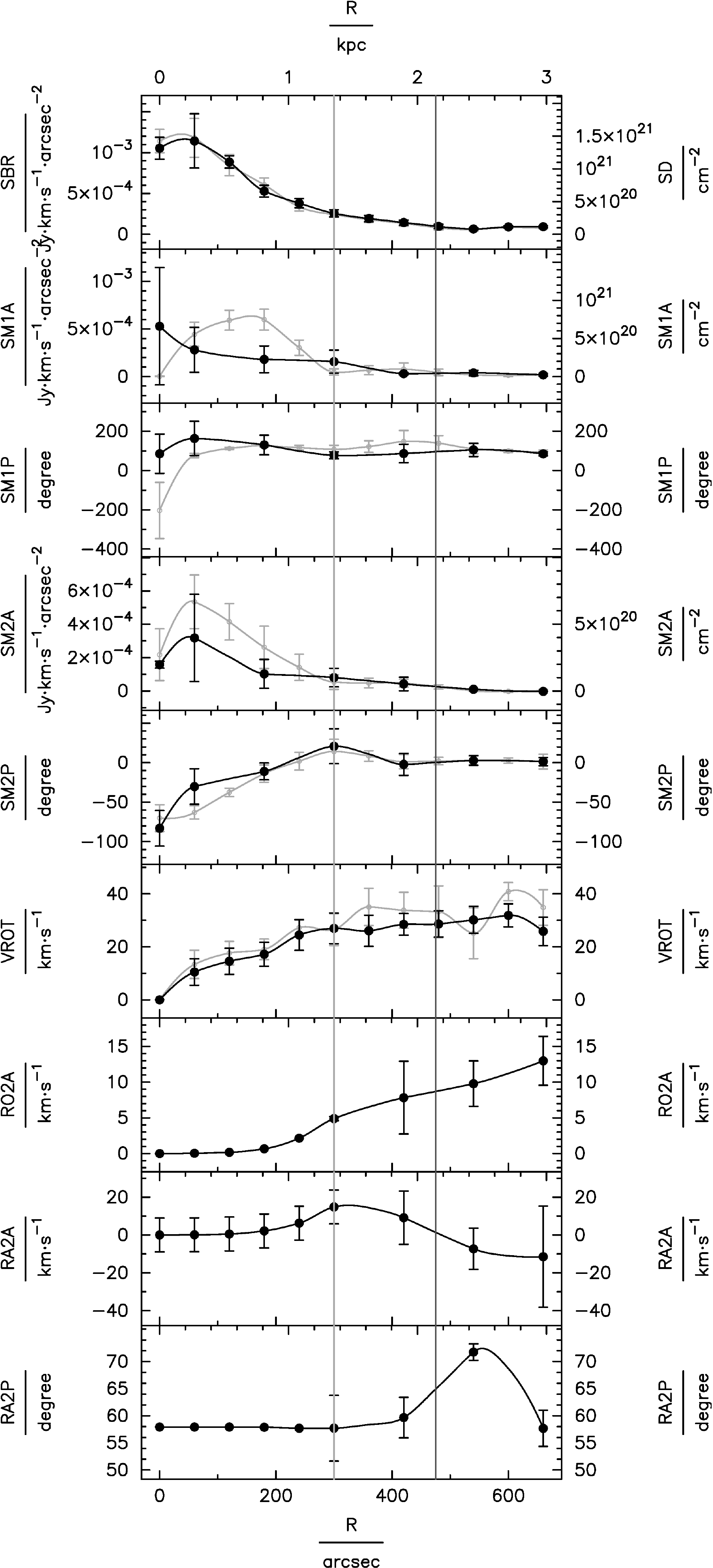}
 \caption{Final {\sc TiRiFiC} model of the \ion{H}{i} disc of WLM.
 SBR/SD: surface brightness or surface column density. The two lines denote $r_{25}$ and $r_\ion{H}{i}$ (at the larger radius).
 SM1/2A: Amplitude of a harmonic distortion in surface brightness/surface column density, first and second-order.
 SM1/2P: phase of a harmonic distortion in surface brightness/surface column density, first and second-order.
 VROT: rotation velocity. RO/RA2A: Amplitude of a second order harmonic distortion in velocity in tangential/radial direction.
 RA2P: Phase (azimuth) of harmonic distortion second order in velocity in radial direction.
 The phase for the distortion in tangential direction is shifted by $45^\circ$ with respect to RA2P.
 The grey line is the result of a fit, in which the rotation curve was left to vary on the nodes shown as dots.}
\label{fig:WLM_modelpars}
\end{figure}

We present the final model parameters that do not vary with radius in Table~\ref{tab:model_prop}. The radially-varying parameters
are presented in Fig.~\ref{fig:WLM_modelpars}. We estimate the errors on the best-fitting parameters using
a bootstrap method. In summary, we fit the data using a
Golden-Section nested intervals fitting algorithm to derive the model parameters. Then, we generate
many synthetic data sets by shifting the model parameters at a single node by a random value. After that,
we fit each sets using the same fitting routine as the real data. Finally,
we calculate the standard deviations of the best-fitting parameters of the synthetic data and we use them as
the standard errors for the model parameters quoted in Table~\ref{tab:model_prop} and shown as errorbars in
Fig.~\ref{fig:WLM_modelpars}. We derived a slowly-rising solid-body rotation curve, which is typical for dwarf galaxies.

\begin{table}
    \centering
 \begin{small}

    \begin{tabular}{llrl} 
       \multicolumn{4}{c}{WLM: TiRiFiC radially invariant parameters}\\
       \hline
       \hline
	Parameter & Symbol & Value & Unit\\
       \hline
    Model centre & RA & $0^\mathrm{h}\, 1^\mathrm{ m}\,58.1^\mathrm{s} \, \pm\, 4^\mathrm{s}$&\\
	(J2000)& Dec & $-15^\mathrm{d}\,27^\mathrm{m}\, 56^\mathrm{s}\, \pm \, 32^\mathrm{s}$ &  \\
	Systemic Velocity & $v_{\rm sys}$ & $-122\,\pm\,4$ & $\mathrm{km}\,\mathrm{s}^{-1}$ \\
	Thickness & $z_0$ & $85\arcsec\,\pm\,9\arcsec$ & \\
	Inclination & $i$ & $77\degree,\pm\,6\degree$ &\\
	Position angle & $pa$ & $ 174.3\degree\,\pm\,0.7\degree$ & \\
	Dispersion & $\sigma_v$ & $7.9\,\pm\,0.4$ & $\mathrm{km}\,\mathrm{s}^{-1}$\\
    \hline
    \end{tabular}
    \end{small}
     \caption{Tilted-ring parameters not varying with radius, final model.}
    \label{tab:model_prop}
 \end{table}
\section{Mass models}\label{sec:massmodels}
In this section, we study the properties of the luminous and the dark matter components in WLM. For the reason given previously,
our mass modelling will be based on the rotation curve from the model that include non-circular motions. The mass modelling result
for the circular model is presented in the Appendix.
We model the observed rotation curve as the dynamical contributions of stars, gas and dark matter halo. For this, we consider two
widely used dark matter halo models, described in the next section. The quadratic sum of
the rotation curves of these three components are then compared with observations to gauge the strength of the
assumed models.
\subsection{Asymmetric drift correction}
Random motions in dwarf galaxies like WLM can be significant, and therefore contribute to
a non-negligible fraction of the pressure support against gravity. This has the effect of lowering the observed rotation
velocity, especially in the outer disc where the gas density is low \citep{2012AAS...21910603O}.
Thus, a correction for this dynamical effect must be applied to derive a more reliable rotation curve. This is called
\textit{asymmetric drift correction}. Following \citet{2013AJ....146...48C}, we calculate the asymmetric drift correction using
the following equation:
\begin{equation} \label{eq2}
    V_{corr}^{2} = V_{obs}^{2} -2\sigma \dfrac{\delta\sigma}{\delta ln~R} - \sigma^{2}\dfrac{\delta ln~\Sigma}{\delta ln~R}
\end{equation}
, where $V_{corr}$ is the corrected velocity, $V_{obs}$ is the observed velocity, $\sigma$ is the velocity dispersion, $R$ is the radius,
and $\Sigma$ is the gas surface density. From our tilted-ring fitting result, we use a constant velocity dispersion of 7.9 $\mathrm{km~s^{-1}}$. Thus, Equation~\ref{eq2} reduces to
\begin{equation} \label{eq3}
    V_{corr}^{2} = V_{obs}^{2} - \sigma^{2}\dfrac{\delta ln~\Sigma}{\delta ln~R}
\end{equation}
We correct the rotation curve for the asymmetric drift effect using Equation~\ref{eq3} before proceeding to the mass modelling.
We present in Table~\ref{tab:rotcur} the rotation curve after correcting for the asymmetric drift effect.
\begin{table}
\begin{threeparttable}
    \centering
 \begin{small}

    \begin{tabular}{ccc} 
       \multicolumn{3}{c}{WLM: Rotation curve}\\
       \hline
       \hline
 RAD  &  VROT  &  ERR \\
 (kpc) & ($\mathrm{km~s^{-1}}$) & ($\mathrm{km~s^{-1}}$)\\
\hline
0.00  &  0.00  &  0.00\\
0.29  &  11.55  &  4.99\\
0.58  &  17.09  &  4.96\\
0.87  &  19.16  &  4.54\\
1.16  &  26.64  &  5.81\\
1.45  &  28.71  &  5.79\\
1.75  &  28.24  &  5.87\\
2.04  &  31.66  &  4.11\\
2.33  &  31.97  &  4.95\\
2.62  &  26.73  &  5.13\\
2.91  &  31.65  &  4.34\\
3.20  &  25.56  &  5.36\\
    \hline
    \hline
    \end{tabular}
      \small
      \textbf{Note.}~~RAD (kpc): Radius.
      VROT ($\mathrm{km~s^{-1}}$): Rotation curve after asymmetric drift correction. ERR ($\mathrm{km~s^{-1}}$): Error in VROT.
    \end{small}
     \caption{WLM rotation curve after asymmetric drift correction.}
    \label{tab:rotcur}
\end{threeparttable}
 \end{table}
\subsection{Contribution of the gas component}
\begin{figure*}
 \begin{tabular}{l l}
 \includegraphics[scale=0.45]{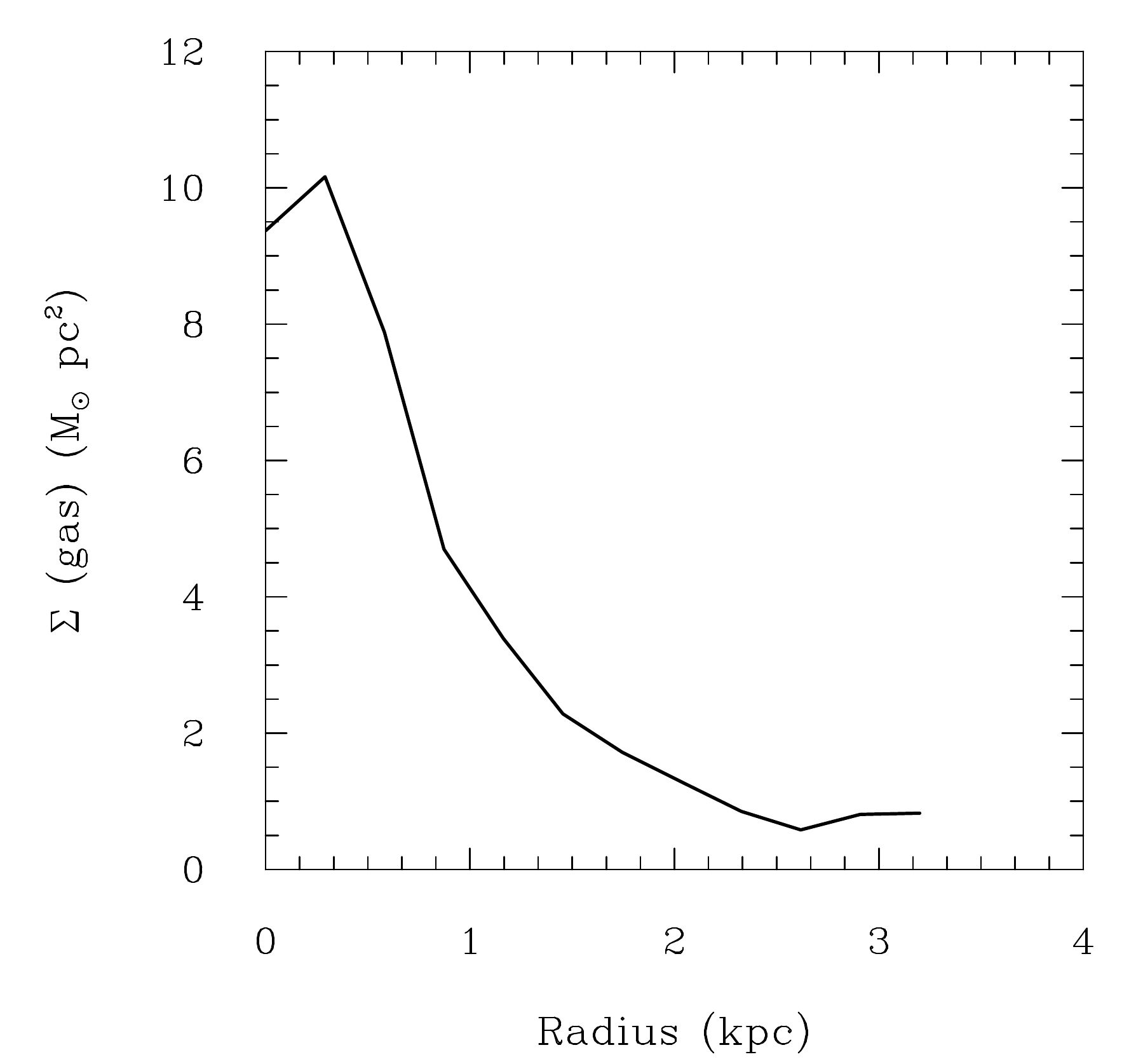} &
 \includegraphics[scale=0.45]{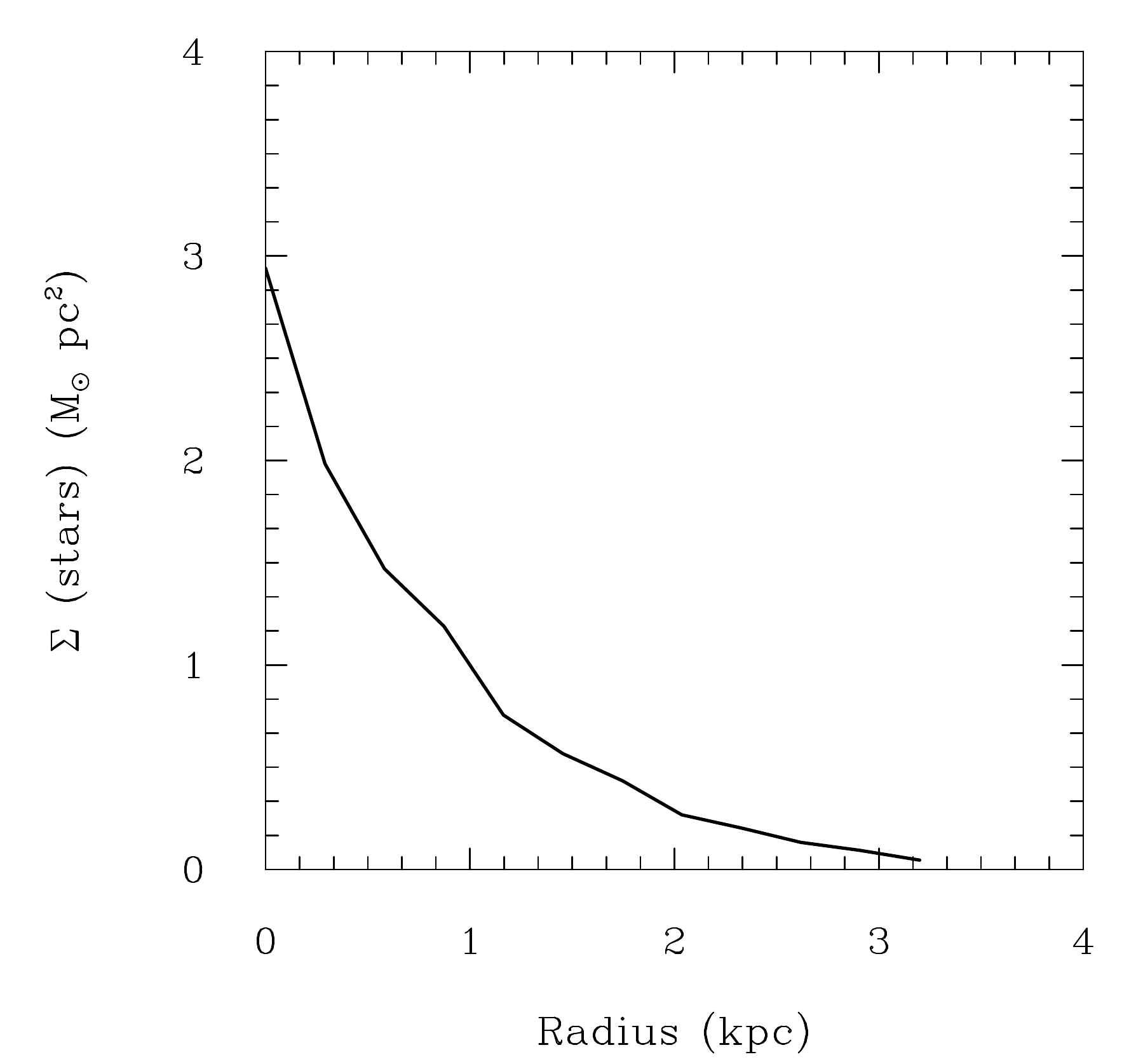}
  \end{tabular}
 \caption{Left: H\,{\sc i} surface density profile of WLM. Right: Stellar surface density profile of WLM.}
 \label{fig:gasstellardensity}
\end{figure*}

To model the contribution of the gas component to the total rotation curve, we first convert the
surface brightness profiles obtained from TiRiFiC (see Section \ref{sec:3dmodel}) to mass surface density profiles.
Then, we scale the derived surface density profiles by a factor of 1.4 to take into account the presence of Helium and other metals.
We show the gas surface densities in Fig.~\ref{fig:gasstellardensity}. Finally, we use the obtained gas density profiles
to model the contribution of the
gas component to the total rotation velocities using the {\tt{GIPSY}} task {\tt{ROTMOD}}.

\subsection{Contribution of the stellar component}
To model the contribution of the stellar components, we use the 3.6$\mu$m \textit{Spitzer IRAC} stellar surface brightness
profile from \citet{2015AJ....149..180O}.
The \textit{Spitzer} 3.6$\mu$m light traces the old stellar populations
containing the bulk of the stellar mass and is therefore an effective measure of the stellar mass.
Following \citet{2008AJ....136.2761O}, we convert the stellar surface brightness profiles in $\mathrm{mag~arcsec^{-2}}$
to luminosity profiles in units of L$_{\odot}$ pc$^{-2}$, and then to mass density profiles using the following
expression
\begin{equation}
\Sigma_{\text{disc}}[M_{\odot} pc^{-2}] =  \Upsilon_{\star}^{3.6} \times 10^{-0.4 \times (\mu_{3.6} - C_{3.6})}
\end{equation}
where $\mu_{3.6}$ is the stellar surface brightness profile, $\Upsilon_{\star}^{3.6}$ is
the mass-to-light ratio in the 3.6 $\mu$m \textit{Spitzer}
band, and C$_{3.6}$ is the constant used for conversion from mag arcsec$^{-2}$ to L$_{\odot}$ pc$^{-2}$ and is
calculated as C$_{3.6}$ = M$_{\odot}$ + 21.56. M$_{\odot}$ = 3.24 is the absolute magnitude of the Sun in
the 3.6 $\mu$m \textit{Spitzer} band. We then model the stellar rotation velocities from the
stellar surface densities using the GIPSY task {\tt{ROTMOD}}.

To estimate the stellar mass in galaxies, one needs to assume a stellar mass-to-light ratio, ($\Upsilon_{\star}$), from
stellar population synthesis models. As mentioned in \citet{2015AJ....149..180O}, this assumption
drives the largest uncertainties associated with converting the luminosity profile to the mass density profile.
For this study, we have adopted a mass-to-light ratio based on $3.6$ $\mu$m light from $Spitzer$ IRAC. The
$3.6$ $\mu$m images are less affected by dust compared to optical images and mostly trace stellar light from old stellar
populations, which are dominant in late-type dwarf galaxies. Here we adopt a $\Upsilon_{\star}^{3.6}$ value of 0.37 from \citet{2015AJ....149..180O},
which is found to be appropriate for late-type dwarf galaxies like WLM.
This value was derived using the empirical relation between galaxy optical colors and $\Upsilon_{\star}$ values
based on stellar population synthesis models \citep{2001ApJ...550..212B}.
The detailed derivation of the mass-to-light ratio is given in \citet{2008AJ....136.2761O}.

\subsection{Contribution of the dark matter component}
For the dark matter distribution, we have assumed two models: the pseudo-isothermal dark matter halo ISO \citep[ISO,][]{1991MNRAS.249..523B}
and the Navarro-Frenk and White halo \citep[NFW,][]{1997ApJ...490..493N}. The ISO halo is an observationally motivated model
with a central constant-density core. The density profile is given as follows:
\begin{equation}
    \rho_{\text{ISO}}(R) =\rho_{0}\Bigg[1 + \Bigg(\frac{R}{R_{c}}\Bigg)^2\Bigg]^{-1},
\end{equation}
where $\rho_{0}$ is the core density and R$_{c}$ is the core radius of a halo. The ISO dark matter halo model is used to describe
the steepness of the inner slope with a power law $\rho \sim$ r$^{\alpha}$, with $\alpha = 0$.
The corresponding rotation velocity is given by:
\begin{equation}
  V_{\text{ISO}} = \sqrt{4 \pi G \rho_{0} R_{c}^{2}\Bigg[1 - \frac{R}{R_{c}} \text{arctan}\Bigg(\frac{R}{R_{c}}\Bigg)\Bigg]}
\end{equation}

The NFW halo, also known as the \textit{universal density profile} \citep{1997ApJ...490..493N},
is the cosmologically motivated dark matter model that describes the cusp-like radial dark matter distribution.
The NFW density profile takes the form
\begin{equation}
    \rho_{\text{NFW}}(R) = \rho_{i}\Bigg[\frac{R}{R_{s}}\Bigg(1 + \frac{R}{R_{s}}\Bigg)\Bigg]^{-2},
\end{equation}
where R$_{s}$ is the characteristic radius of the halo and $\rho_{i}$ is related to the density of the Universe at
the time of the collapse of the dark matter halo. The corresponding rotation velocity is given by:
\begin{equation}
    V_{\text{NFW}}(R) = V_{200}\sqrt{\frac{\ln{1 + cx} - cx/(1 + cx)}{x[\ln{1 + c} - c/(1 + c)]}},
\end{equation}

where c = R$_{200}$/R$_{s}$ is a concentration parameter, R$_{200}$ is the radius where the density contrast with
respect to the critical density of the Universe exceeds 200, roughly the virial radius; the
characteristic velocity, V$_{200}$, is the velocity at that radius \citep{1996ApJ...462..563N}.
The NFW mass density profile is cuspy in the inner parts and can be represented by $\rho \sim$ r$^{\alpha}$, where $\alpha$ = -1.

\begin{figure*}
\begin{tabular}{c c}
\includegraphics[scale=0.45]{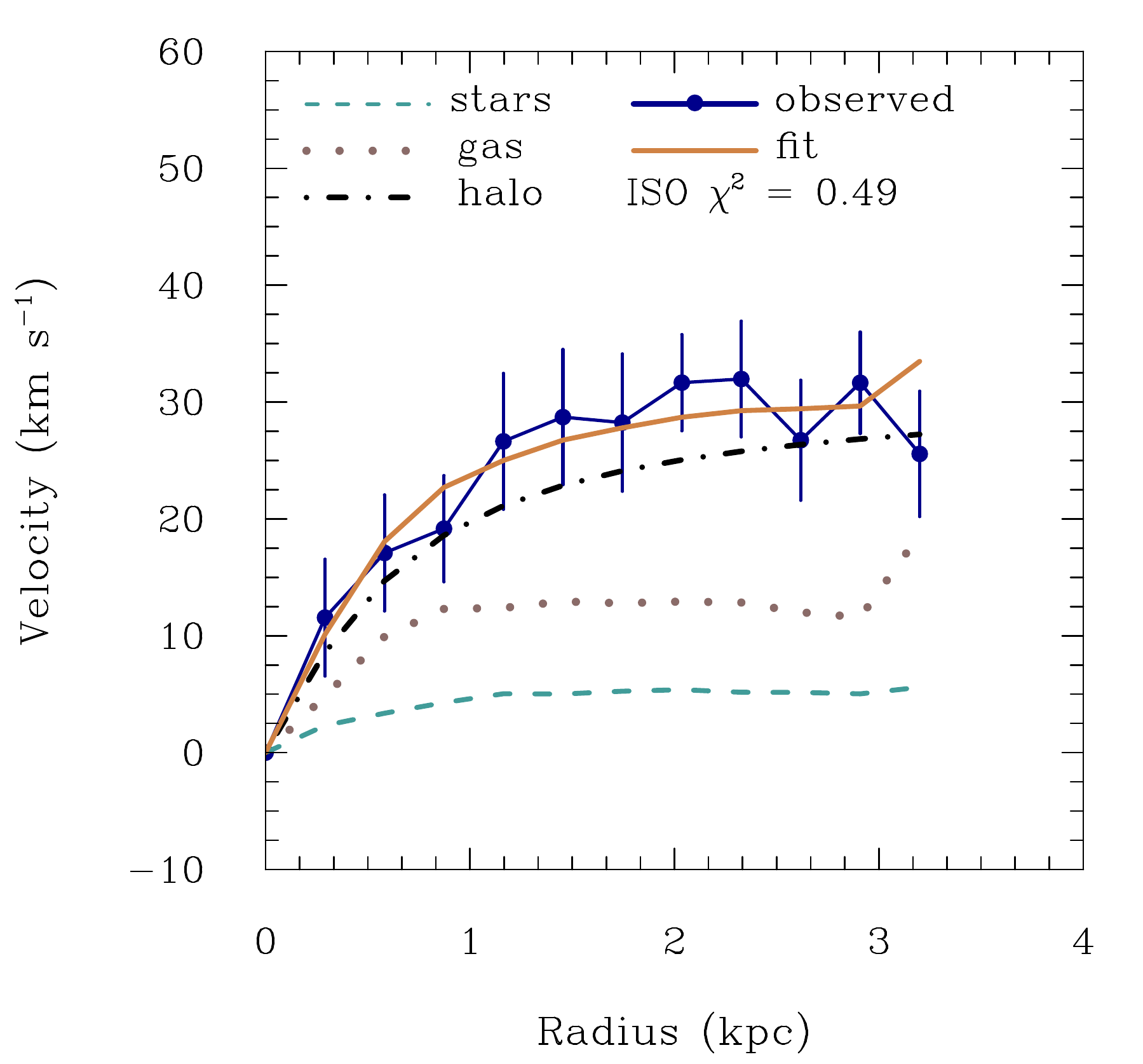} &
\includegraphics[scale=0.45]{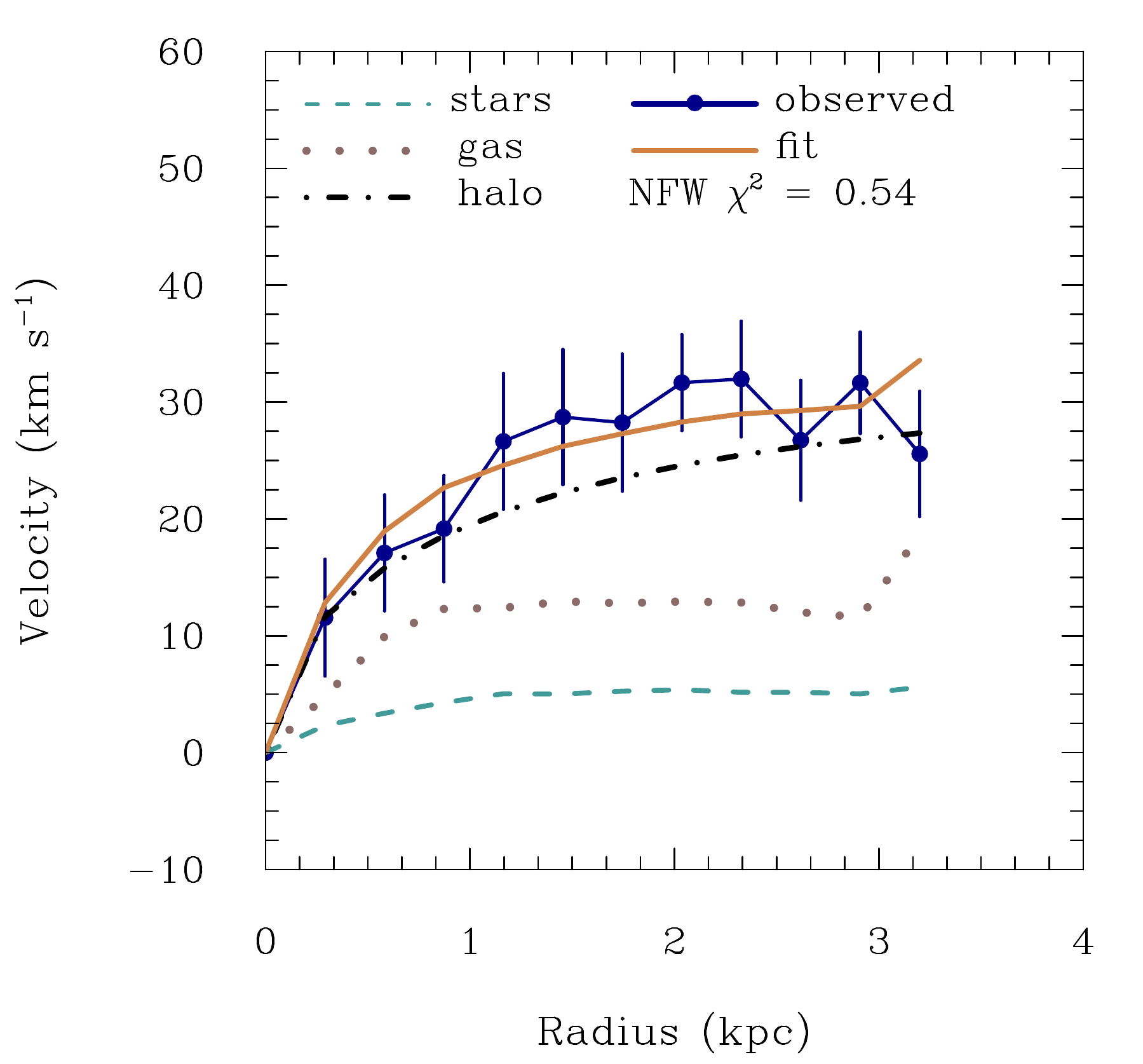}
\end{tabular}
\caption{Mass distribution models of WLM with ISO (left panel) and NFW (right panel) for the model
with non-circular motions.}
  \label{fig:massmodel_noncirc}
\end{figure*}

We use the outputs of {\tt{ROTMOD}} as described previously as inputs for the {\tt{GIPSY}} task {\tt{ROTMAS}}
to decompose the rotation curves of WLM into luminous and dark matter components. We use the inverse square of the
uncertainties in the input rotation curves as weights for the least-squares fit in {\tt{ROTMAS}}.
The quality of the {\tt{ROTMAS}} fits are judged by the calculated $\chi^{2}$ values.
We present the results of our mass modelling in Fig.~\ref{fig:massmodel_noncirc} and Table~\ref{tab:massmodel} (See also
Fig.~\ref{fig:massmodel_circ} and Table~\ref{tab:massmodel_c} for the circular model).

\begin{table}
\begin{threeparttable}
 \begin{small}

    \begin{tabular}{llr} 
       \multicolumn{3}{c}{WLM: Mass models}\\
       \hline
       \hline
	Model & Parameter & $\Upsilon$ = 0.37  \\
       \hline
    ISO & $\rho_{0}$ & 55.38 $\pm$ 32.11 \\
                  & $R_{C}$ & 0.57 $\pm$ 0.22 \\
                  & $\chi^{2}_{red}$ & 0.49 \\
    \hline
    NFW & $R_{200}$ & 35.51 $\pm$ 11.86 \\
                  & $C$ & 8.58 $\pm$ 4.33 \\
                  & $\chi^{2}_{red}$ & 0.54 \\
    \hline
    \end{tabular}
      \small
      \textbf{Note.}~~$\rho_{0}$ [$10^{-3} M_{\odot}/pc^3$]: core density of the ISO model.
      $R_{C}$ [kpc]: Core radius of the ISO model. $C$: concentration parameter for the NFW model.
      $R_{200}$~[kpc]: Radius where the mass density contrast with respect to the critical density of the Universe exceeds 200.
      $\chi^{2}_{red}$: Reduced chi-square.
    \end{small}
     \caption{Mass modelling results of WLM.}
    \label{tab:massmodel}
    \end{threeparttable}
 \end{table}

Both the ISO and the NFW models fits the rotation curve within the errors. However the ISO model
has a lower reduced $\chi^{2}$ (0.49) value than the NFW model (0.54). Note that all our reduced $\chi^{2}$ values are less than one,
which indicate that the estimated errors are likely to be overestimated. The dark matter
component dominates
the gravitational potential at all radii in WLM like
most dwarf irregular galaxies \citep{1988ApJ...332L..33C, 2008AJ....136.2648D,2008AJ....136.2761O,10.1093/mnras/stx2256,10.1093/mnras/sty1056}.
Previous fitting of dwarf galaxies tend to favour the ISO model
over the NFW model \citep{2015AJ....149..180O}. However, recent studies of 11 void dwarf galaxies by \citep{2020MNRAS.491.4993K} reported
that the average slope of the dark matter density profiles in these galaxies are consistent with what is expected from
an NFW model when using a 3D-based approach. In contrast, their 2D-based fitting approach,
similar to the method used by \citet{2015AJ....149..180O}, gave a slope in line with an ISO model.
This difference may be caused by projection effects in the 2D-based approach. Within the errors, our result is not
in conflict with the one by \citet{2020MNRAS.491.4993K}. More data is needed to make a
robust conclusion on the general, preferred model of the dark matter distribution in dwarfs.
Another effect that can mimic an ISO model is non-circular motions. Although non-circular motions are found in dwarf galaxies,
their effects are not significant enough to hide a cusp-like density profile \citep{2008AJ....136.2761O, 2009A&A...505....1V}.

\section{discussion}\label{sec:discussion}
The rotation curve and the mass models for WLM have previously been studied using
both 2D and 3D approaches, most recently by \citet{2015AJ....149..180O, 2016MNRAS.462.3628R, 2017MNRAS.466.4159I}.
The study by \citet{2015AJ....149..180O}
assumed an infinitely-thin disc, whereas that of \citet{2017MNRAS.466.4159I}
assumed an H\,{\sc i} layer of 100 pc. While the assumption of a thin disc might be
reasonable for spiral galaxies,
it may not be the case for dwarf galaxies.
Our derived rotation curve has a lower rotation amplitude compared to that of \citet{2015AJ....149..180O} and \citet{2017MNRAS.466.4159I}.
We attribute this difference to our best-fitting H\,{\sc i} thickness of 85~$\arcsec$,~about 400~pc.
This is roughly four times higher than what was assumed in \citet{2017MNRAS.466.4159I}.
As mentioned in \citet{2017MNRAS.466.4159I}, in the presence of a thick disc,
assuming a thin disc will underestimate the true rotation velocity at smaller radii,
and overestimate it at larger distance from the centre. \citet{2017MNRAS.466.4159I}
estimated the scale
height of WLM to be about 150~pc in the centre and flares linearly up to 600~pc assuming
a vertical hydrostatique equilibrium. Our best-fitting constant value of 400~pc is
thus within the range of this estimate and also in general agreement
with the study by \citet{2011MNRAS.415..687B} who estimate the vertical scale height of four
dwarf galaxies.

Mass modelling for WLM has been studied by \citet{2015AJ....149..180O}. From their fits,
they derived reduced $\chi^{2}$ values of 0.05 and 0.29 for the ISO and the NFW models, respectively.
Although they got smaller reduced $\chi^{2}$ values than what we obtained, both analyses yield smaller reduced
$\chi^{2}$ values for the ISO than for the NFW model.
\section{Summary}\label{sec:conclusions}
We present H\,{\sc i} observations of the isolated, gas-rich dwarf irregular galaxy, WLM, using the MeerKAT
telescope during its 16 dishes era. For a total observing time of 5.65 hours including overheads,
we reach a total flux density of 249 $\mathrm{Jy~km~s^{-1}}$ and an rms noise per channel of 5 $\mathrm{mJy~beam^{-1}}$,
corresponding to a column density limit of about $\mathrm{5\times10^{19}~cm^{-2}}$ for a 3$\sigma$-detection
over 2 channels of 5.5 $\mathrm{km~s^{-1}}$ width. The MeerKAT moment-1 map revealed clear signatures of
a warp in the Southern side of the galaxy, which was not apparent in previous H\,{\sc i} observations.
We also observe the WLM with the GBT for 14.2 hours, and we derive a total flux of about 310
$\mathrm{Jy~km~s^{-1}}$. The rms noise per channel was 32 $\mathrm{mJy~beam^{-1}}$, which corresponds to
a column density sensitivity limit of 4 $\times~10^{18}~\mathrm{cm^{-2}}$ for a 3$\sigma$-detection
over 2 channels of 5.2 $\mathrm{km~s^{-1}}$ at $\sim$9\arcmin~resolution. The H\,{\sc i} disc
extends out to a major axis diameter of $\sim30\arcmin$, and a minor axis diameter of $\sim20\arcmin$ as measured by
the GBT. We use the MeerKAT data cube to model the galaxy using the tilted-ring fitting techniques
implemented in the {\sc TiRiFiC} software suite.
We reproduce the overall distributions and kinematics of the H\,{\sc i} in WLM using a flat disc model with a
vertical thickness, a constant velocity dispersion and inclination, and a solid-body rotation velocity
with azimuthal distortions. In addition, we find it necessary to only include the harmonic distortions in surface brightness to
recover the asymmetric morphology of the galaxy. Finally, we add second-harmonic distortions in velocity in the
tangential and radial directions to simulate bar-like motions. There are faint 2$\sigma$-emissions that are not
reproduced by our best-fitting model. The data has been cleaned much below this level and thus,
these are potentially real emission. The MeeKAT telescope with its 64 dishes will be a perfect instrument to
confirm their origin and study their kinematics. We decompose the rotation curve
obtained from our tilted-ring model in terms of the contribution from luminous and dark components.
For this, we use the NFW and the ISO dark matter halo models. The two models fit the rotation curve within the formal errors, but
with the ISO model having lower reduced $\chi^{2}$ value than the NFW model. Like most late-type dwarf galaxies in the Local Group,
WLM is dominated by dark matter at all radii. This makes WLM a good candidate to search for the presence of dark matter in galaxies, e.g.,
through indirect detection of dark matter annihilation. An example of such study in WLM has been done by \citet{2019arXiv190810178A}
using a gamma-ray telescope.

\section{Data availability}
The data from this study are available upon request to the corresponding author, Roger Ianjamasimanana.
\section{Acknowledgements}
The MeerKAT telescope is operated by the South African Radio
Astronomy Observatory, which is a facility of the National Research
Foundation, an agency of the Department of Science and Innovation.\\
This work is based upon research supported by the South African
Research Chairs Initiative of the Department of Science and Technology
and National Research Foundation.\\
The financial assistance of the South African Radio Astronomy Observatory (SARAO) towards this research is hereby acknowledged (\url{www.sarao.ac.za}).\\
PK is partially supported by the BMBF project 05A17PC2 for D-MeerKAT.
\\
AS acknowledges the Russian Science Foundation grant 19-12-00281 and the Program of development of M.V.
Lomonosov Moscow State University for the Leading Scientific School ``Physics of stars, relativistic objects and galaxies''.
\\
This project has received funding from the European Research Council (ERC) under the European Union's Horizon 2020
research and innovation programme (grant agreement no. 679627; project name FORNAX) 
%


\bibliographystyle{mnras}
\bibliography{references} 



\appendix

\section{Atlas of data and tilted-ring model comparisons}
In this Appendix we present additional Figures that complement the results presented in the
main body of the paper.
%
\begin{figure*}
 \includegraphics[width=\textwidth]{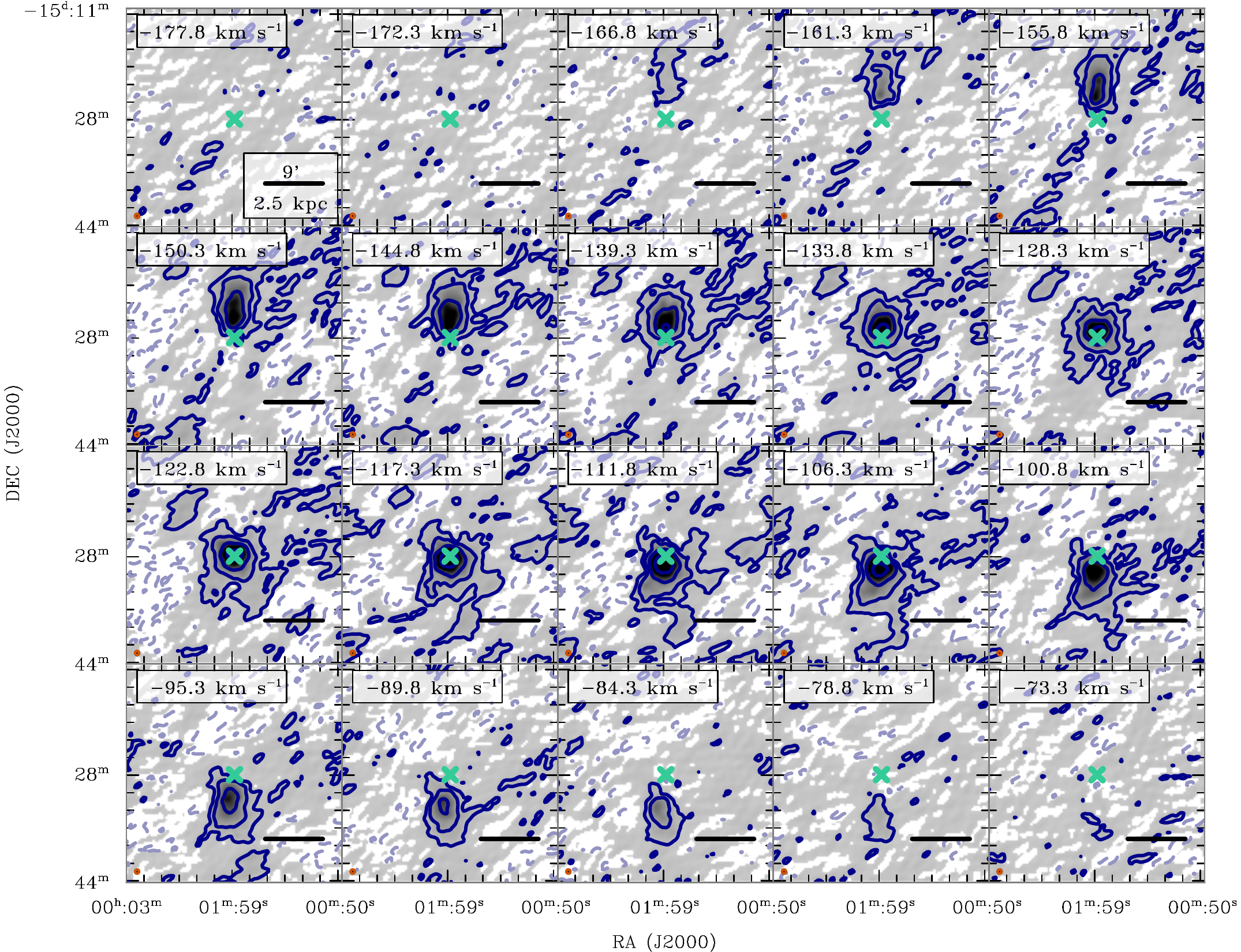}
 \caption{MeerKAT \ion{H}{i} data cube. Contours denote the $-2,2,6,18,54\,-\,\sigma_\mathrm{rms}$-levels,
 where $\sigma_\mathrm{rms}\,=\,5\,\mathrm{mJy}\,\mathrm{beam}^{-1}$. Dashed lines represent negative intensities.
 The crosses represent the kinematical centre of the model,
 the circle to the lower left represents the synthesised beam ($HPBW$).}
 \label{fig:WLM_original_data_cube_20_obs}
\end{figure*}
\begin{figure*}
 \includegraphics[width=\textwidth]{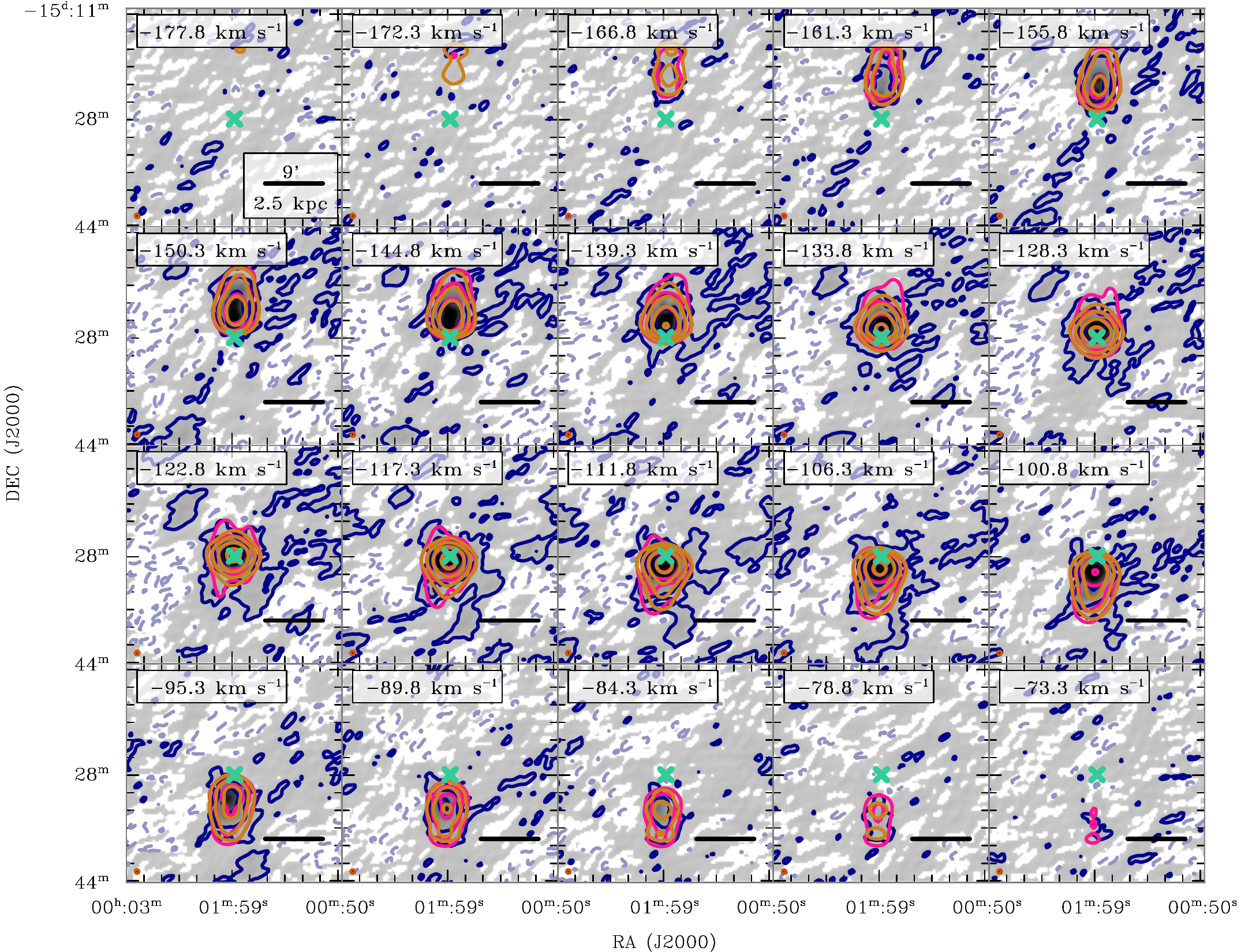}
 \caption{MeerKAT \ion{H}{i} data cube. Contours denote the $-2,2,6,18,54\,-\,\sigma_\mathrm{rms}$-levels,
 where $\sigma_\mathrm{rms}\,=\,5\,\mathrm{mJy}\,\mathrm{beam}^{-1}$. Blue: the observed data cube. Pink:
 the final {\sc TiRiFiC} model. Strong orange: the {\sc TiRiFiC} model allowing for circular motions only.
 Dashed lines represent negative intensities. The crosses represent the kinematical centre of the model,
 the circle to the lower left represents the synthesised beam ($HPBW$).}
 \label{fig:WLM_both_data_cube_20}
\end{figure*}

\begin{figure*}
 \includegraphics[width=\textwidth]{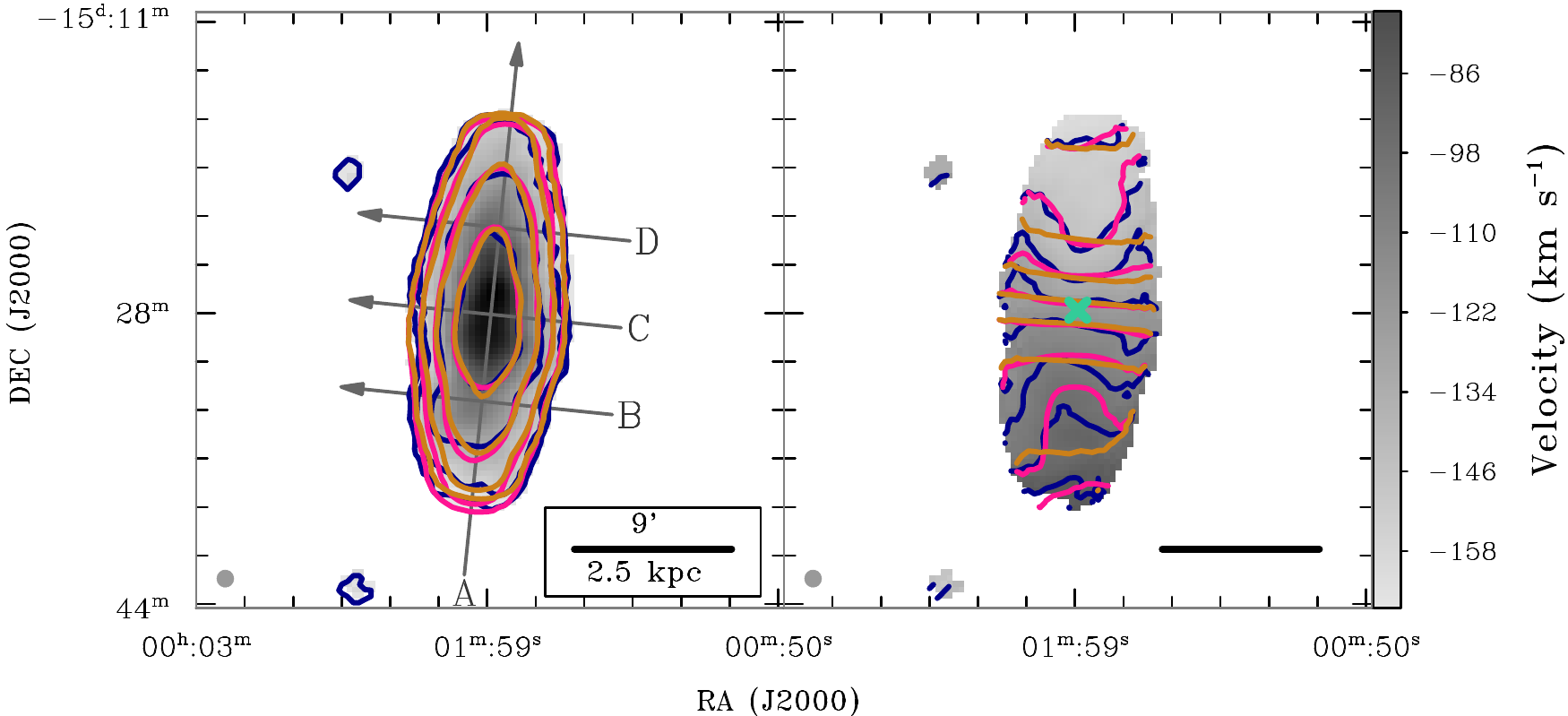}
 \caption{Total-intensity map (left panel) and moment-1 velocity field (right panel) derived from TiRiFiC overlaid onto the observed
 moment maps. Blue: the observed data; strong orange, the TiRiFiC model allowing for circular motions only; pink:
 our final best-fitting TiRiFiC model which include non-circular motions.
 The cross or the intersection between arrows A and C represents the kinematical centre of the model,
 the circle to the lower left shows the synthesised beam ($HPBW$). For the total-intensity maps, contours 
 denote the $\frac{1}{3},1.0,3,9,27\,$-$\,M_\odot\,\mathrm{pc}^{-2}$-levels. Arrows indicate the positions of slices along which
 the position-velocity diagrams in Fig.~\ref{fig:WLM_both_PV-diagrams} were taken.
 For the velocity fields, contours are iso-velocity contours spaced by 10 $\mathrm{km}\,\mathrm{s}^{-1}$ and
 range from -35 $\mathrm{km}\,\mathrm{s}^{-1}$ to 35 $\mathrm{km}\,\mathrm{s}^{-1}$ relative
 to the systemic velocity $v_\mathrm{sys}\,=\,-122\,\mathrm{km}\,\mathrm{s}^{-1}$.}
\label{fig:WLM_both_mom0_mom1_vf}
\end{figure*}
\begin{figure*}
 \includegraphics[width=0.99\textwidth]{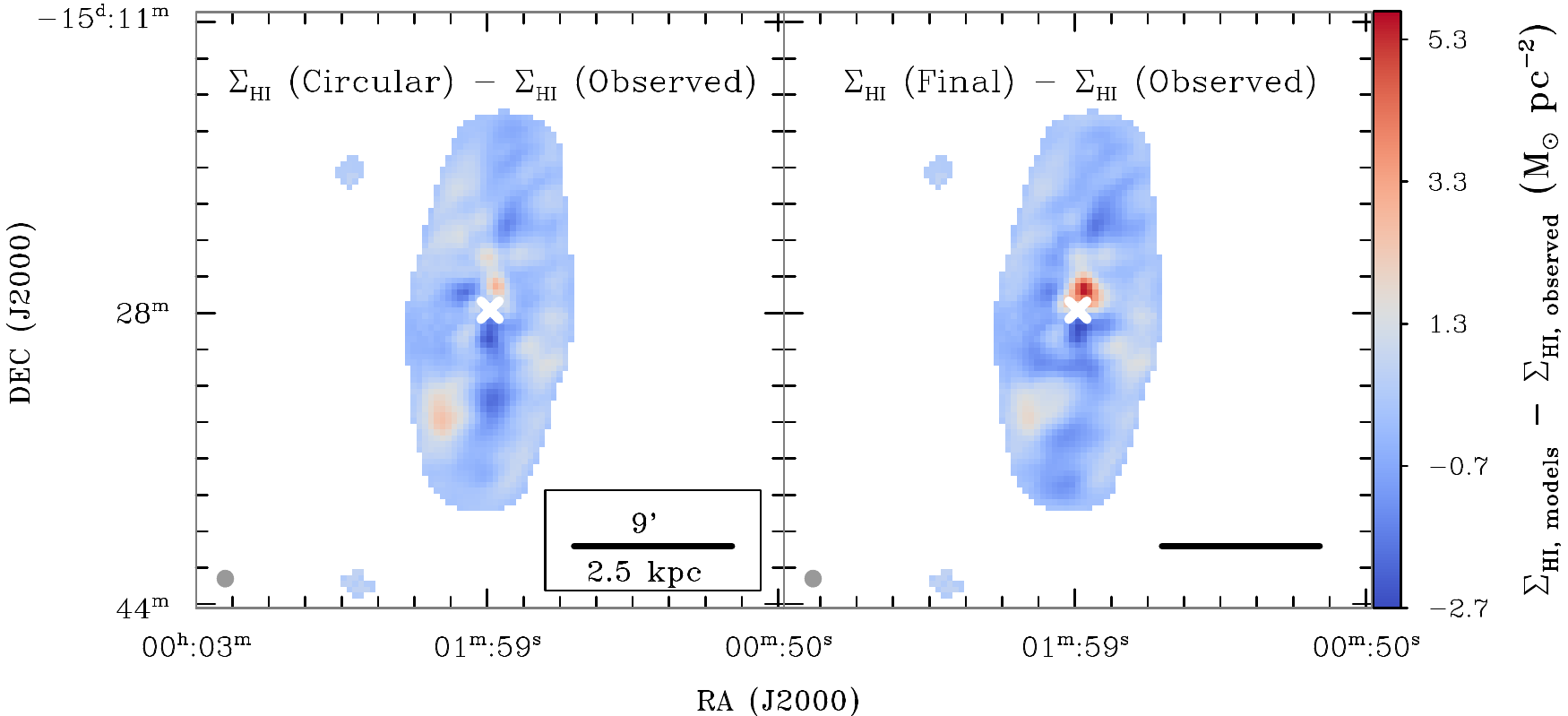}
 \caption{Difference between the surface density maps derived from the TiRiFiC models and the observed data cube.
 Left: TiRiFiC model allowing circular motions only minus the observed data.
 Right: the final TiRiFiC model minus the observed data.}
\label{fig:WLM_mom0diffs}
\end{figure*}
\begin{figure*}
\begin{tabular}{c c}
 \includegraphics[width=0.45\textwidth]{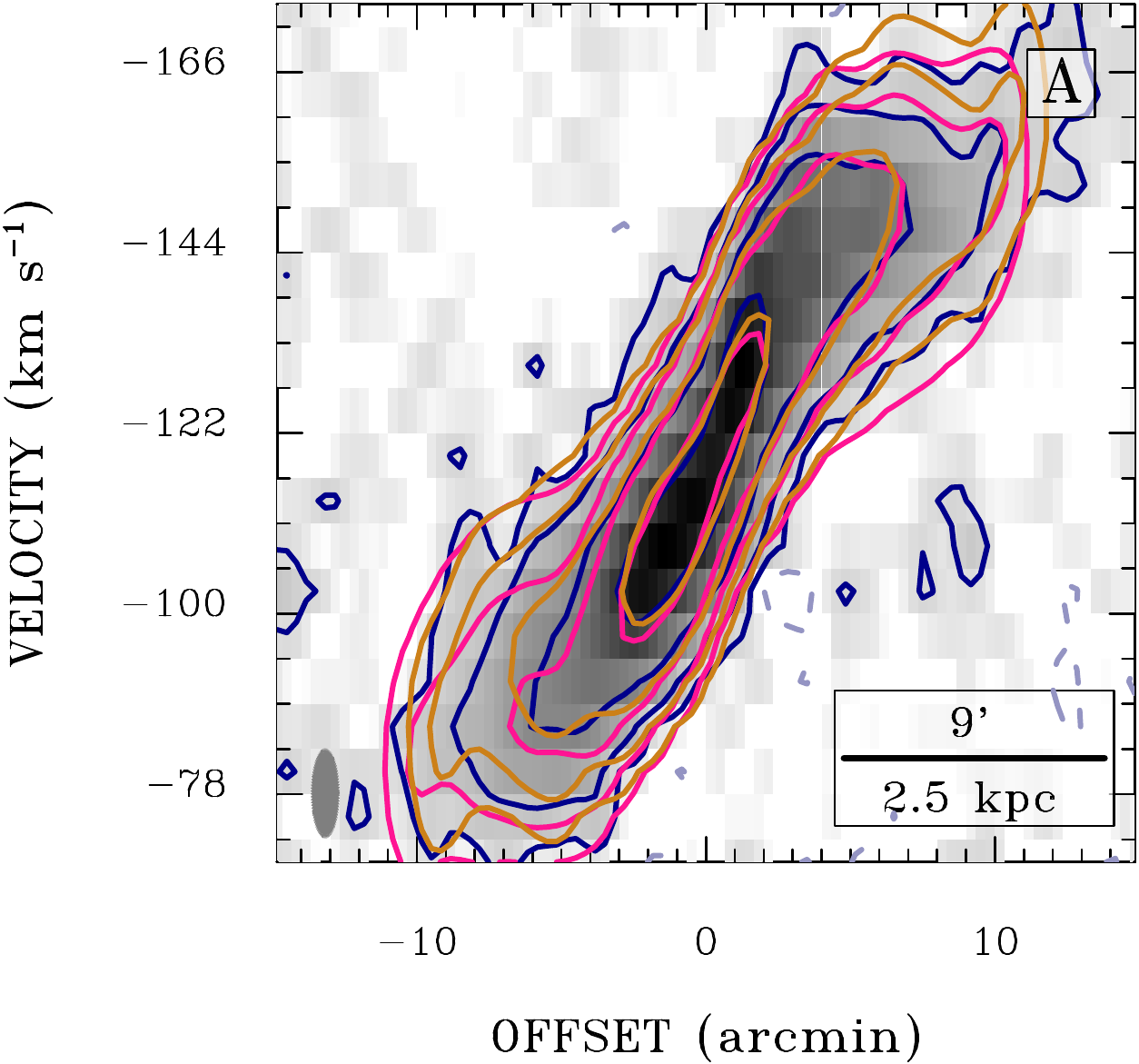}&
 \includegraphics[width=0.45\textwidth]{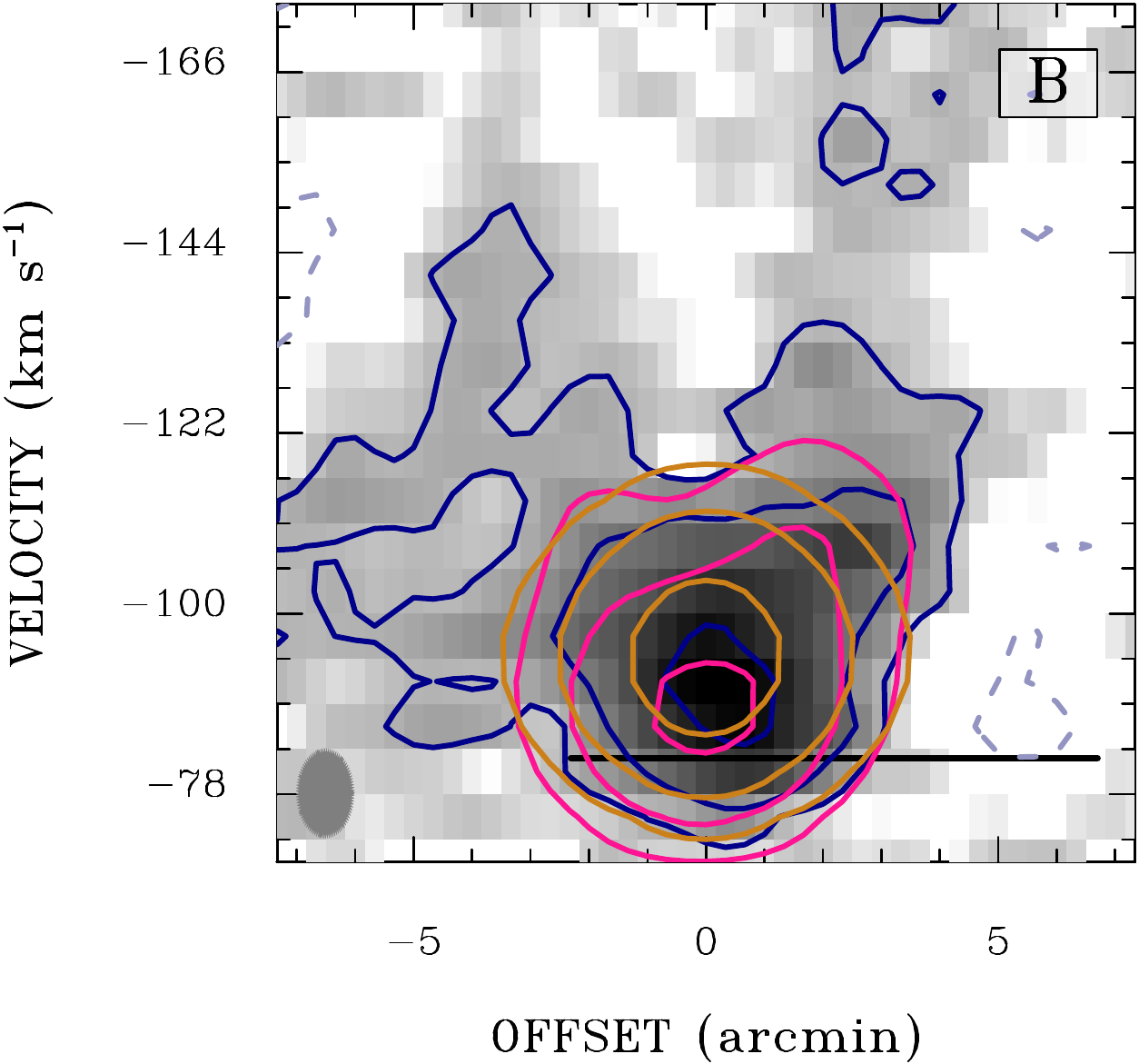}\\\\
 \includegraphics[width=0.45\textwidth]{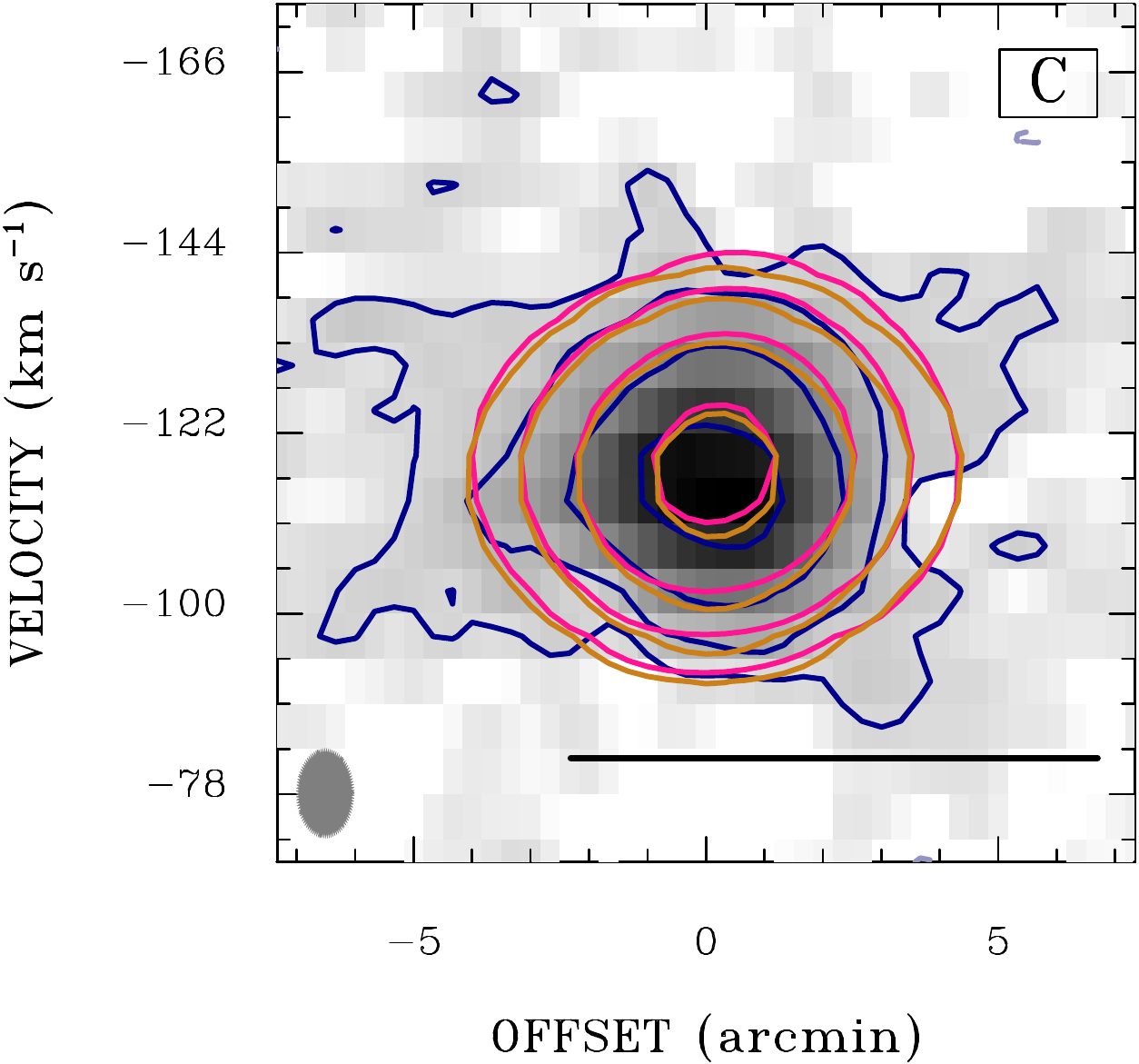}&
 \includegraphics[width=0.45\textwidth]{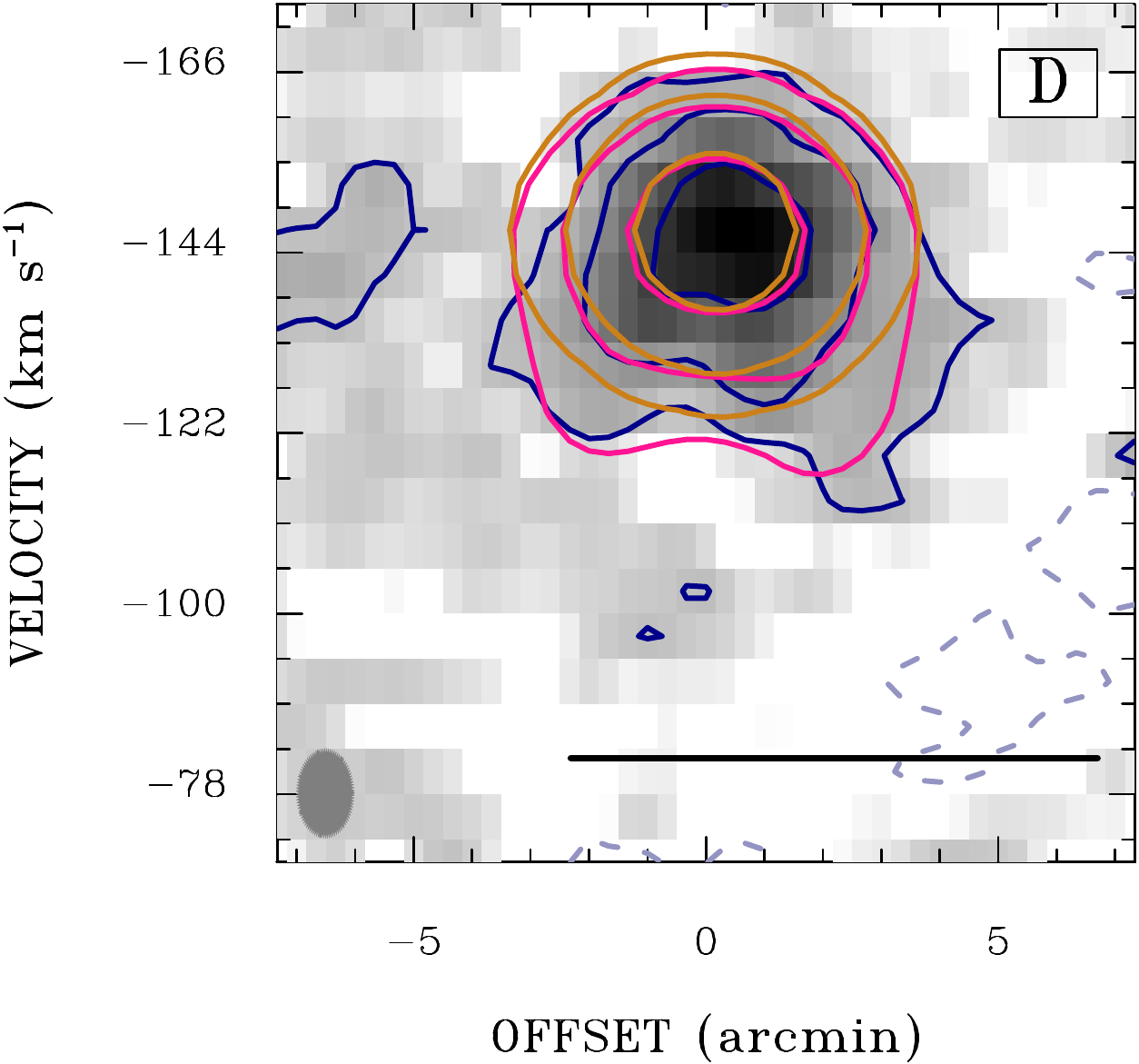}
\end{tabular}
 \caption{Position-velocity diagrams taken along slices shown in Fig.~\ref{fig:WLM_both_mom0_mom1_vf}.
 Contours denote the $-2,2,6,18,54\,-\,\sigma_\mathrm{rms}$-levels, where $\sigma_\mathrm{rms}\,=\,5\,\mathrm{mJy}\,\mathrm{beam}^{-1}$.
 Blue: the observed data cube. Pink: the {\sc TiRiFiC} final model. Green: the {\sc TiRiFiC} model allowing for circular motion only.
 Dashed lines represent negative intensities. The ellipse represents the velocity (2 channels) and the
 spatial resolution ($\sqrt{HPBW_\mathrm{maj}*HPBW_\mathrm{min}}$, where $HPBW_\mathrm{maj}$ and $HPBW_\mathrm{min}$ are the major
 and minor axis half-power-beam-widths).}
\label{fig:WLM_both_PV-diagrams}
\end{figure*}

\begin{figure*}
\begin{tabular}{c c}
\includegraphics[scale=0.4]{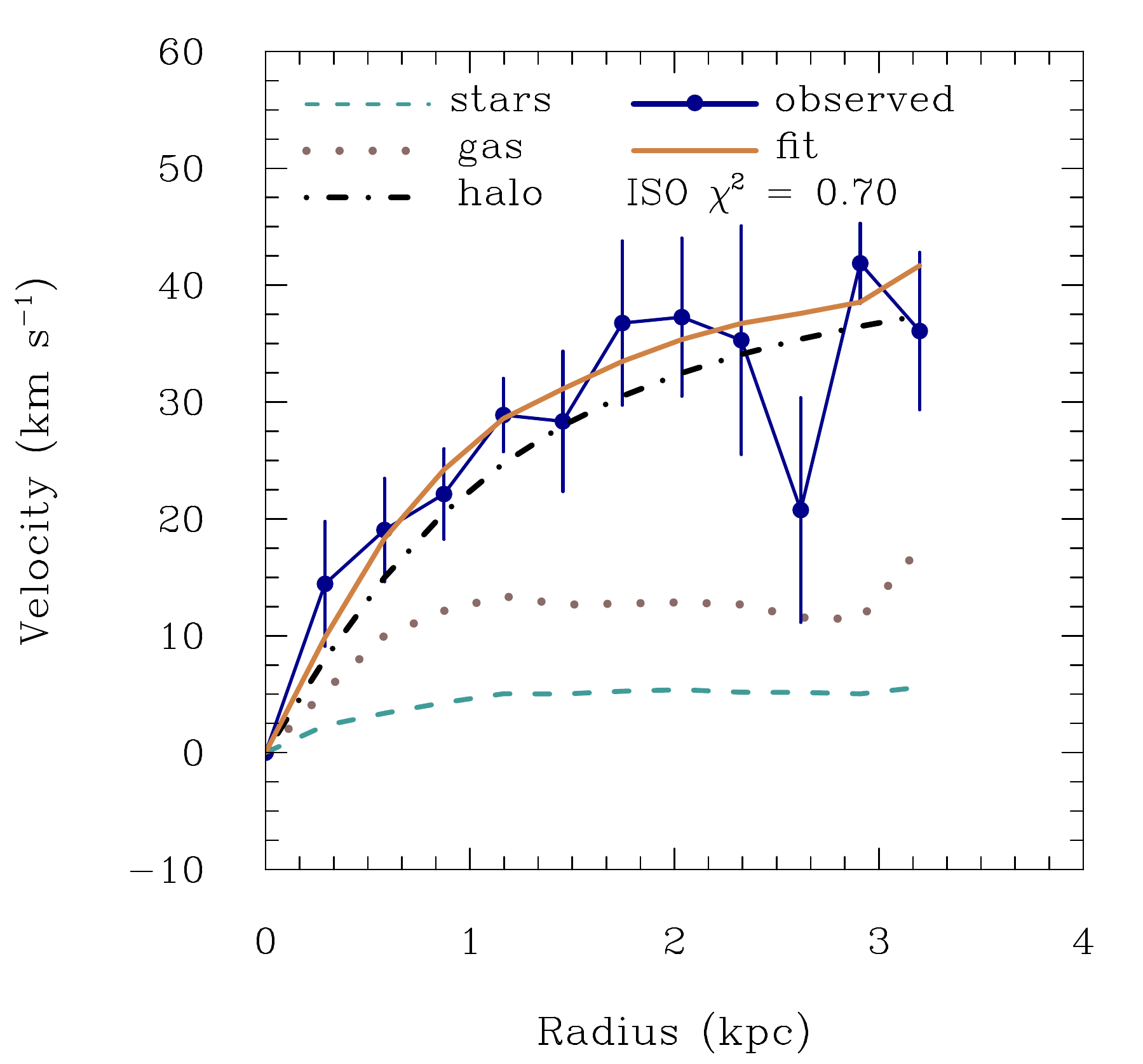} &
\includegraphics[scale=0.4]{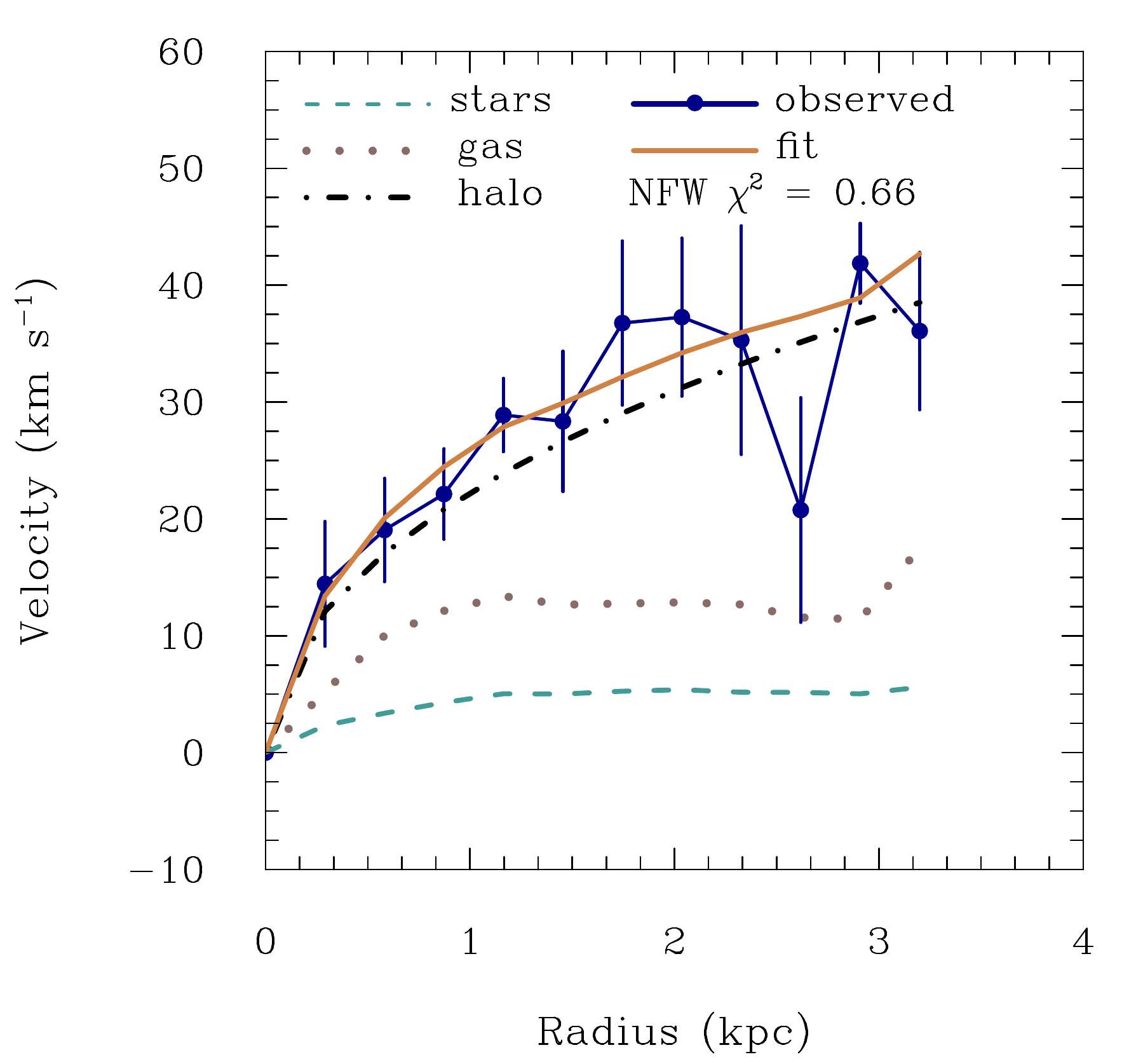}
\end{tabular}
\caption{Mass distribution models of WLM with ISO (left panel) and NFW (right panel) for the model
with only circular motions included.}
  \label{fig:massmodel_circ}
\end{figure*}

\begin{table*}
    \centering
 \begin{small}

    \begin{tabular}{llr} 
       \multicolumn{3}{c}{WLM: Mass models}\\
       \hline
       \hline
        Model & Parameter & $\Upsilon$ = 0.37  \\
    \hline
    ISO & $\rho_{0}$ & 44.21 $\pm$ 15.87 \\
              & $R_{C}$ & 0.98 $\pm$ 0.29 \\
              & $\chi^{2}_{red}$ & 0.70 \\
    \hline
    NFW   &  $R_{200}$ & 133.84 $\pm$ 263.45 \\
                & $C$ & 3.02 $\pm$ 5.98 \\
                & $\chi^{2}_{red}$ &  0.66 \\
    \hline
    \end{tabular}
    \\
      \small
      \textbf{Note.}~~$\rho_{0}$ [$10^{-3} M_{\odot}/pc^3$]: core density of the ISO model.
      $R_{C}$ [kpc]: Core radius of the ISO model. $C$: concentration parameter for the NFW model.
      $R_{200}$~[kpc]: Radius where the mass density contrast with respect to the critical density of the Universe exceeds 200.
      $\chi^{2}_{red}$: Reduced chi-square.
    \end{small}
     \caption{Mass modelling results of WLM using the models with only circular motions.}
    \label{tab:massmodel_c}
 \end{table*}
\bsp	
\label{lastpage}
\end{document}